\documentclass[aps,prd,notitlepage,twocolumn,superscriptaddress,nofootinbib]{revtex4-1}
\usepackage[utf8]{inputenc}
\usepackage[english]{babel}
\usepackage{amsmath}
\usepackage{amsfonts}
\usepackage{amssymb}
\usepackage{listings}
\usepackage{lipsum}
\usepackage{multirow}
\usepackage{datetime}
\usepackage{graphicx}
\usepackage{mathtools}
\usepackage{mathrsfs}
\usepackage{dcolumn}
\usepackage{multirow}
\usepackage{color}   
\usepackage{hyperref}
\hypersetup{
    colorlinks=true, 
    pdfborder = {0 0 0.5 [3 3]},
    anchorcolor=black,
    citecolor=blue,
    linktoc=all,    
    linktocpage=true,
    linkcolor=red,
	urlcolor=blue
}

\begin{document}

\title{Binary boson stars: Merger dynamics and formation of rotating remnant stars}
\author{Nils Siemonsen}
\email[]{nsiemonsen@perimeterinstitute.ca}
\affiliation{Perimeter Institute for Theoretical Physics, Waterloo, Ontario N2L 2Y5, Canada}
\affiliation{Arthur B. McDonald Canadian Astroparticle Physics Research Institute, 64 Bader Lane, Queen's University, Kingston, ON K7L 3N6, Canada}
\affiliation{Department of Physics \& Astronomy, University of Waterloo, Waterloo, ON N2L 3G1, Canada}
\author{William E.\ East}
\affiliation{Perimeter Institute for Theoretical Physics, Waterloo, Ontario N2L 2Y5, Canada}

\date{\today}

\begin{abstract} 
Scalar boson stars have attracted attention as simple models for exploring 
the nonlinear dynamics of a large class of ultra compact and black hole
mimicking objects. Here, we study the impact of interactions in the scalar 
matter making up these stars. In particular, we show the pivotal
role the scalar phase and vortex structure play during the late inspiral, merger,
and post-merger oscillations of a binary boson star, as well as their impact on the properties of the merger remnant. 
To that end, we construct constraint satisfying binary boson star initial data and 
numerically evolve the nonlinear set of Einstein-Klein-Gordon equations.
We demonstrate that the scalar interactions can significantly affect the
inspiral gravitational wave amplitude and phase, and the length of a potential
hypermassive phase shortly after merger. If a black hole is formed after
merger, we find its spin angular momentum to be consistent with similar binary
black hole and binary neutron star merger remnants. Furthermore, we formulate a
mapping that approximately predicts the remnant properties of any given binary
boson star merger. Guided by this mapping, we use numerical evolutions to explicitly demonstrate,
for the first time, that rotating boson stars can form as remnants
from the merger of two non-spinning boson stars. We characterize this new
formation mechanism and discuss its robustness. Finally, we comment on the
implications for rotating Proca stars.
\end{abstract}

\maketitle

\section{Introduction} \label{sec:Intro}
Gravitational wave (GW) and electromagnetic observations from isolated and binary
compact objects can be explained remarkably well by invoking the presence of a
black hole, as predicted by general relativity.  The black hole paradigm
explains physics across several orders of magnitude, from stellar mass compact
binary mergers and X-ray binaries, at smaller scales, to active galactic
nuclei, at larger scales. However, while these observations require
ultracompact central engines, the defining feature of black holes---the event
horizon---remains less well tested. Performing the most stringent tests of the
black hole paradigm necessitates alternative predictions that can be used to
perform model selection with gravitational and electromagnetic observations. 

Such considerations have driven the construction of a large class of exotic ultracompact
objects mimicking many of the observational features of black hole spacetimes
\textit{without} possessing an event horizon (see Ref.~\cite{Cardoso:2019rvt}
for a review). These range from various kinds of fluid stars
\cite{1974ApJ...188..657B,PhysRevD.22.807,Herrera:2004xc}, to string theory-inspired
ultra compact fuzzballs
\cite{Mathur:2005zp,Bena:2007kg,Balasubramanian:2008da}. Lacking horizons,
these objects can exhibit distinct relativistic properties, such as stable light
rings, isolated ergoregions, and super-extremal spins, that are not present in
black hole spacetimes \cite{Cardoso:2019rvt}, and which can lead to new gravitational
phenomenology, including GW echoes
\cite{Cardoso:2016oxy} and ergoregion instabilities \cite{friedman1978},
opening a window into the models of quantum gravity and dark matter from which
these objects emerge. Leveraging GW observations to confront the
black hole hypothesis, however, requires an understanding of
black hole-mimicking objects in the strong-field and highly dynamical regime.
While there are numerous ultracompact object models where stationary solutions have been constructed,
self-consistently evolving these objects is challenging, or even ill-defined, in many cases.

As a result, scalar boson stars (BSs) have received significant attention as proxies for 
generic ultracompact black hole mimickers in the strong-field, dynamical regime. First
constructed in Refs.~\cite{Kaup:1968zz,Ruffini:1969qy,Seidel:1993zk},
BSs have been studied extensively in isolation in a variety of
models with various scalar self-interactions \cite{Seidel:1991zh,Alcubierre:2003sx,Schunck:2003kk,Liebling:2012fv,Visinelli:2021uve}. Analogous
to fluid stars, a family of solutions in a particular model contains a
single, or pair, of stable branches
\cite{Friedberg:1986tq,Balakrishna:1997ej,Schunck:1999zu}. Stability studies in
the linear regime have demonstrated the robustness of non-rotating BS solutions
to perturbations \cite{Sorkin:1981jc, Gleiser:1988rq,Gleiser:1988ih,
Lee:1988av,Kusmartsev:1990cr,Guzman:2004jw,Sanchis-Gual:2021phr}. Spherically
symmetric BSs are consistently formed dynamically from coherent
\cite{Seidel:1993zk,Guzman:2003kt} and incoherent
\cite{Amin:2019ums,Levkov:2018kau,Veltmaat:2018dfz,Arvanitaki:2019rax} diffuse
initial states, as well as through binary BS mergers in head-on
\cite{Lai:2004fw,Choptuik:2009ww,Paredes:2015wga,Bernal:2006ci,Schwabe:2016rze,Palenzuela:2006wp,Mundim:2010hi,Bezares:2017mzk,Helfer:2021brt}
and orbiting mergers
\cite{Palenzuela:2007dm,Palenzuela:2017kcg,Bezares:2022obu,Bezares:2018qwa,Bezares:2017mzk},
further solidifying their robustness (for a review, see
Ref.~\cite{Liebling:2012fv}). Analogous analyses have shown similar dynamical
behavior in vector stars, called Proca stars (PSs)\footnote{Here, we  
will use BS to refer exclusively to stars made from scalar matter, and PS to refer
to those formed from a vector field.
}
\cite{Brito:2015pxa,Sanchis-Gual:2022mkk,Sanchis-Gual:2018oui,DiGiovanni:2018bvo,Sanchis-Gual:2017bhw}.
While the existence, stability, and formation of spherically symmetric stars
from isolated or binary solutions is well-studied, the role of angular momentum
in these systems is still poorly understood. It was noticed early on that
spherical stars cannot rotate perturbatively \cite{Kobayashi:1994qi} (i.e.,
they are not continuously connected to their non-rotating counterparts; notable
exceptions exist in the Newtonian limit \cite{Kling:2020xjj}). Later it was
found that rotating BS (and PS) solutions with quantized angular momentum exist
\cite{Kleihaus:2005me,Kleihaus:2007vk,Brito:2015pxa}. However, only stars in
scalar models with self-interactions are perturbatively stable against a
non-axisymmetric instability
\cite{Sanchis-Gual:2019ljs,DiGiovanni:2020ror,Siemonsen:2020hcg,Dmitriev:2021utv}.
Lastly, despite numerous efforts to form rotating BSs dynamically
\cite{Palenzuela:2017kcg,Bezares:2022obu,Bezares:2018qwa,Bezares:2017mzk,Croft:2022bxq},
even in those models with rotating stars without known instabilities
\cite{Siemonsen:2020hcg}, no rotating BSs has been formed from the merger of two
non-spinning stars.
If BSs mergers cannot form a (non-black-hole) rotating remnant,  
that would seem to place a serious impediment on their ability 
to mimic black holes without invoking horizons.

\begin{figure}[t]
\includegraphics[width=0.49\textwidth]{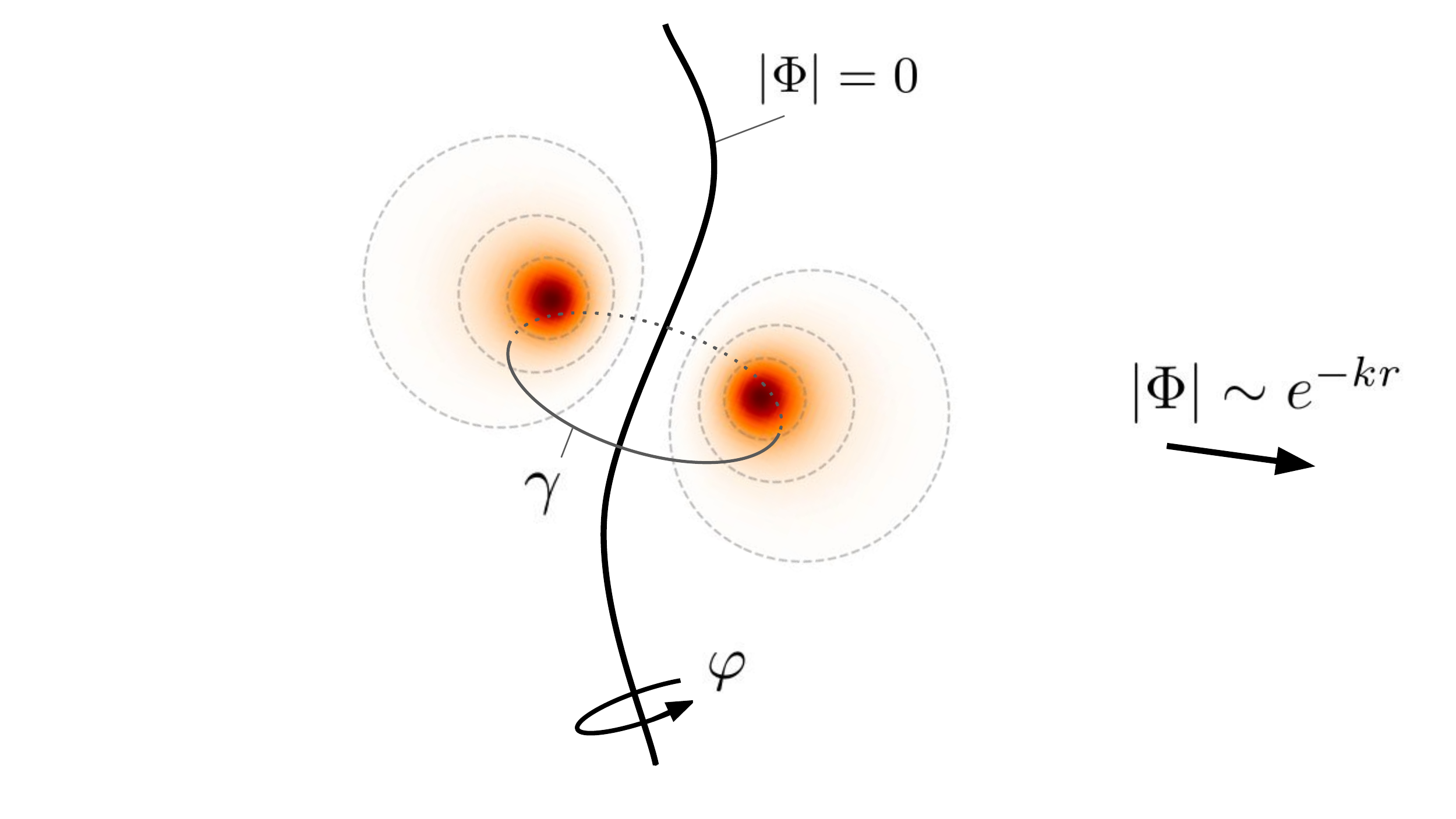}
\caption{We show schematically an axial slice through a rotating BS at a fixed time. The scalar field magnitude $|\Phi|$ (red/yellow color) vanishes along the central vortex line, attains a maximum value some distance from the vortex line, and drops off exponentially towards large distances $\sim e^{-kr}$, with some $k>0$. Surfaces of constant scalar field magnitude are indicated as gray dashed lines. Integrating the gradient of the scalar phase $\text{Arg}(\Phi)=\psi$ along the path $\gamma$ around the vortex in the azimuthal direction $\varphi$ gives the vortex index $q$ of the rotating BS, as defined in \eqref{eq:vortexdef}.}
\label{fig:vortex}
\end{figure}

BSs are stationary, non-topological solutions of a complex massive scalar field
$\Phi$ with a global $U(1)$ symmetry minimally coupled to gravity (the
generalization of Q-ball solutions to include self-gravity \cite{Friedberg:1976me,Coleman:1985ki}). The tendency of a
localized wave solution to disperse is balanced by gravitational attraction,
while the gravitational decay of BSs (dissipation through gravitational radiation) is precluded by their conserved
$U(1)$-charge $Q$,
counting the number of bosons in the system. Generally, the scalar field of
an isolated BS takes the form
\begin{align}
\Phi=\phi e^{i(\omega t-m\varphi)},
\label{eq:scalaransatz}
\end{align}
with magnitude $\phi$, frequency $\omega$, integer azimuthal index $m$, time
coordinate $t$, and azimuthal coordinate $\varphi$. For spherical solutions, the
scalar field magnitude is maximum at the origin, and exponentially 
decays towards spatial infinity. Their rotating counterparts with
non-zero angular momentum $J=mQ$, on the other hand, exhibit vanishing scalar field
magnitude at their centers and are classified by the index $|m|\geq 1$, leading
to toroidal surfaces of constant scalar field magnitude. Therefore, rotating BSs can
also be understood as \textit{vortices} with gravitationally bound scalar
matter. This is shown schematically in \figurename{ \ref{fig:vortex}}.
Within a spacelike slice of a spacetime with scalar field $\Phi=\phi
e^{i\psi}$, we define a vortex to be the line $L$ through the hypersurface
such that the integral of the gradient of the scalar phase $\psi$ along the
oriented loop $\gamma$ in a sufficiently small neighborhood around $L$ is a
non-zero integer $q$,
\begin{align}
\frac{1}{2\pi}\oint_\gamma d\ell_i D^i\psi= q ,
\label{eq:vortexdef}
\end{align}
called the vortex index. Here, $D_i$ is the covariant derivative in the
hypersurface. Since for non-spinning BSs the phase $\psi$ is constant in space, i.e., $D^i\psi=0$, 
applying this definition gives $q=0$. In the
case of isolated rotating BSs, the gradient $D^i\psi$ is non-trivial. In fact,
the vortex line is the line of vanishing $|\Phi|$ through the center of mass of
these solutions (as shown in \figurename{ \ref{fig:vortex}}). The vortex index
$q$ is exactly the azimuthal index $m$ of the star, $|q|=m$. Hence, in the
context of these rotating BS solutions, the connection between angular momentum and
the vortex index is manifest. Lastly, reversing the orientation of $\gamma$ implies
$q\rightarrow -q$.

The appearance and role of vortices in various contexts have been the subject
of extensive study for decades. In particular, quantized vortices are generic
features in Bose-Einstein condensates \cite{PRA.65.023603}, superfluids
\cite{PRL.71.1375,PRL.91.135301,PRB.38.2398}, wave-like dark matter
\cite{Yu:2002sz,Sikivie:2009qn,Kain:2010rb,Rindler-Daller:2011afd}, or cosmic
strings \cite{Kibble:1976sj}; all of which are described by scalar models similar (or
identical) to the scalar theories with BS solutions considered here. In these
contexts, vortices appear dynamically, for instance, as the result of symmetry
breaking phase transitions \cite{Kibble:1976sj,Zurek:1985qw,delCampo:2013nla},
or the introduction of angular momentum into the system \cite{Rindler-Daller:2011afd,PRA.65.023603}. Therefore,
from vortex lattices in trapped superfluids to cosmic string networks and
spinning dark matter halos, vortices are important in a wide variety of
phenomena. Here, we show that vortices, their relation to angular momentum, and
the evolution of the phase of the scalar field are also crucial ingredients in 
understanding and predicting the merger dynamics of binary BSs.

In this work, we numerically evolve the nonlinear Einstein-Klein-Gordon system
of equations in 3D to study the role of the scalar field during the late
inspiral, merger, and ringdown of spinning and non-spinning binary BSs in
different nonlinear scalar models with a global $U(1)$ symmetry. As the scalar
interactions are exponentially enhanced with decreasing binary separation, we
find that the scalar phase plays a crucial role during the later inspiral and
merger of binary BSs. We illustrate some cases where the nature of the endstate of a binary
merger is determined by the relative phase of the stars at early
times. Secondly, we provide a mapping that approximately predicts the outcome
of any given binary (or multi) BS and PS merger. Utilizing this mapping
to guide the choice of parameters, we show, for the first time, using numerical evolutions, cases where rotating BSs form
dynamically from a non-spinning binary BS merger. We provide a set of
necessary conditions for the formation of these rotating BS and
PS remnants and identify the regions in parameter space where this formation
channel is expected to be active. 

The remainder of the paper is organized as follows. In
Sec.~\ref{sec:mergerdynamics}, we briefly review the role of scalar
interactions in the dynamics of binary Q-balls, then proceed to apply these results
to binary BS inspirals, first qualitatively, and then systematically, in the case
of a non-spinning binary inspiral, and finally construct a mapping to
approximately predict the remnant properties of any given binary BS inspiral.
In Sec.~\ref{sec:formationofrotremn}, we begin by listing the necessary
conditions for the formation of a rotating remnant BS from the merger of a
non-spinning binary, explicitly demonstrate the formation of a rotating BS
remnant by numerically evolving a suitable system, discuss the robustness and
other characteristics of this formation channel, and study the remnant resulting from 
the merger of a spinning BS with a non-spinning BS.
In Sec.~\ref{sec:conc}, we further discuss our findings and conclude. Finally, in the appendices, we revisit a non-axisymmetric instability present in rotating BSs in light of our findings, and provide further details on our numerical setup. We use geometric units with $G=c=1$ throughout.

\section{Merger dynamics} \label{sec:mergerdynamics}

During the merger of two BSs in a nonlinear
scalar model, scalar interactions play an important role along with gravitational interactions.
Due to the exponential fall-off of the scalar field amplitude at
large distances from an isolated star, the scalar forces are also exponentially
suppressed for a binary at large separations. Conversely, the scalar field interaction is exponentially enhanced
during the later inspiral and merger of two BSs, and is crucial to understanding the merger dynamics.

In Sec.~\ref{sec:scalarmodels}, we introduce the nonlinear complex scalar
theories considered in this work. Continuing in
Sec.~\ref{sec:scalarinteractions}, we briefly review known results for Q-balls
and BSs, and conjecture how these results can be translated to the inspiral and
merger dynamics of binary BSs. We apply this intuition to the inspiral of a
representative binary BS in Sec.~\ref{sec:inspmergdynamics}, and study the
scalar interaction dependence on the scalar phase systematically in the context
of a binary BS inspiral in Sec.~\ref{sec:gwimprints}. Lastly, in
Sec.~\ref{sec:remnantmap}, we conjecture a mapping that identifies the remnant
of any multi BS encounter and illustrate its utility in the context of an
inspiraling spinning binary BS. (We also comment on the implications for PSs.)

\subsection{Scalar models} \label{sec:scalarmodels}

In this work, we mainly consider a complex scalar field $\Phi$ whose potential $V(|\Phi|)$ includes
both a quadratic term corresponding to a mass $\mu$, as well as higher order contributions from self-interactions. We also
consider massive complex vector fields $A_\mu$ in select cases in order to 
compare to the behavior of scalar theories; for details on this and the PS
solutions in these models, see Appendix~\ref{app:bssolutions}. The scalar
dynamics occurs in a dynamical spacetime $g_{\mu\nu}$ of Ricci curvature
scalar $R$. Therefore, the globally $U(1)$ invariant action is
\begin{align}
S=\int d^4x\sqrt{-g}\left[\frac{R}{16\pi}-g^{\mu\nu}\nabla_{(\mu}\bar{\Phi}\nabla_{\nu)}\Phi-V(|\Phi|)\right].
\label{eq:action}
\end{align}
Here, and in the following, overbars denote complex conjugation. Scalar self-interactions are classified into \textit{attractive} and \textit{repulsive} potentials. Attractive (repulsive) potentials are those with negative (positive) first correction beyond the mass-term, 
which means the increasing nonlinear self-interactions, e.g. in a binary BS with decreasing separation, reduce (enhance) the potential energy. We focus on one type of potential for each class and note that similar behavior is expected generally for self-interactions of the same class. The solitonic potential \cite{Friedberg:1986tq}
\begin{align}
V_{\rm sol}(|\Phi|)=\mu^2|\Phi|^2\left[1-\frac{2|\Phi|^2}{\sigma^2}\right]^2,
\label{eq:solitonic}
\end{align}
is characterized by the coupling constant $\sigma$. In the
$\sigma\rightarrow\infty$ limit, the self-interactions reduce to the simple massive
scalar case with \textit{attractive} higher order corrections: $V_{\rm
sol}=\mu^2|\Phi|^2-4\mu^2|\Phi|^4/\sigma^2+\mathcal{O}(|\Phi|^6)$. In this
scalar model, there are ultracompact solutions with stable light rings, even in spherical
symmetry~\cite{Palenzuela:2017kcg,Boskovic:2021nfs}. In the flat
spacetime limit, this scalar theory admits exact, non-topological, solitonic 
solutions called Q-balls, which are continuously connected to their
self-gravitating cousins, i.e., BSs \cite{Kleihaus:2005me}. The repulsive
scalar interactions are represented by
\begin{align}
V_{\rm rep}(|\Phi|)=\mu^2|\Phi|^2+\lambda|\Phi|^4,
\label{eq:repulsive}
\end{align}
with $\lambda>0$. Both the scalar solitonic and repulsive models (as well as
the massive vector model) admit non-trivial, stationary, axisymmetric solutions
describing a gravitationally-bound relativistic scalar (or vector) condensate.
The details of how these scalar and vector star solutions are obtained from the
action \eqref{eq:action}, and constructed numerically, can be found in
Appendix~\ref{app:bssolutions} (for PSs) and in, e.g., Ref.~\cite{Siemonsen:2020hcg}
(for BSs). Here, we simply note that in the scalar case the ansatz
\eqref{eq:scalaransatz} with asymptotically flat boundary conditions yields an
infinite set of families of BS solutions each labeled by their azimuthal number
$m$, and parameterized by their frequency $\omega<\mu$. The spherically symmetric $m=0$ family of
solutions is non-rotating,
while the $m\neq0$ families consist of non-perturbatively rotating
stars with \textit{quantized} angular momentum $J=mQ$, where $Q$ is the
global $U(1)$-charge. We define the radius of BS solutions as the circular radius
$R$, within which 99\% of the solution's energy is contained, and the
compactness $C$ as the ratio of the ADM mass $M$ to radius, $C=M/R$.

\subsection{Scalar interactions} \label{sec:scalarinteractions}

On a flat background, a Q-ball can be decomposed, as in
\eqref{eq:scalaransatz}, into a spatial profile $\phi$ and a phase $\omega t-m\varphi\rightarrow\psi$. 
In the case of
an isolated, non-spinning (i.e., $m=0$) Q-ball, the spatial profile peaks at the
center of the solution and decays exponentially as $\phi\sim e^{-kr}$ at large
distances $r$ from the center, while the complex phase exhibits a harmonic time
dependence $\psi=\omega t$, together with the arbitrary constant phase-shift
$\psi\rightarrow \psi+\alpha$, under which the model is symmetric. A binary
Q-ball with separation $|\textbf{x}_1-\textbf{x}_2|= D\gg 1/k_i$, with $i=1$ and $2$,
is approximately given by the scalar field profile
\begin{align}
\Phi\sim e^{i\omega_1 t}\phi_1(\textbf{x}_1)+e^{i(\omega_2 t+\alpha)}\phi_2(\textbf{x}_2).
\label{eq:binaryqball}
\end{align}
The scalar self-interactions lead to momentum exchange---a scalar
force---between the two solitons
\cite{Axenides:1999hs,Battye:2000qj,Bowcock:2008dn}. This force ultimately
originates from the internal non-stationarity of the complex phase of the
scalar field. For the binary defined in \eqref{eq:binaryqball}, the scalar
force has a dependence given by \cite{Axenides:1999hs,Bowcock:2008dn}
\begin{align}
\tilde{F}\sim \cos[(\omega_1-\omega_2)t+\alpha]e^{-D(k_1+k_2)},
\label{eq:scalarforce}
\end{align}
for $1/D \ll k_1,k_2$.  The magnitude of $\tilde{F}$ is exponentially suppressed by the
distance $D$ between the solitons, while the sign of $\tilde{F}$ is determined by the
phase evolution of the binary. This spatial dependence and importance of the
complex phase is applicable also in the BS case, as we see below.  In the limit
of equal frequency, $\omega_1=\omega_2$, the temporal dependence vanishes, and
the sign of $\tilde{F}$ is determined solely by the phase-offset $\alpha$. In the
general case of $\omega_1\neq \omega_2$, a breathing motion appears in response
to the harmonically oscillating force applied on each Q-ball
\cite{Axenides:1999hs}, independent of the constant offset $\alpha$. Therefore,
the complex phase dynamics determine the sign of the effective force applied,
while the magnitude is exponentially suppressed by the distance between the
solitons. 

The dynamics of the complex phases $\psi_1$ and $\psi_2$ is non-trivial in
the presence of nonlinear scalar interactions \cite{Battye:2000qj}. Integrating
out the spatial degrees of freedom, assuming a large separation, and identical
solitons, the evolution follows $\ddot{\psi}_1+\ddot{\psi}_2=0$ and
$\ddot{\psi}_1-\ddot{\psi}_2\sim \varepsilon^2\sin(\psi_1-\psi_2)$,
with the overlap $\varepsilon\sim e^{-D(k_1+k_2)}$. For the special case of
$\psi_1=\psi_2+n\pi$, with $n\in\mathbb{Z}$, the evolution trivializes,
i.e., $\dot{\psi}_1,\dot{\psi}_2=$ const., and the soliton's phase-evolution
is set by a single frequency. However, in general, an initial phase-offset
$\alpha=(\psi_1-\psi_2)|_{t=0}\in(0,\pi)$, implies
$\ddot{\psi}_1,\ddot{\psi}_2\neq 0$. Therefore, the phases start evolving
towards $\dot{\psi}_1>\dot{\psi}_2$, i.e., towards a state of different
frequencies. The frequency evolution implies a change in charge of the two
solitons, since the frequency uniquely parameterizes the family of Q-ball
solutions of (in general) different charge. Hence, a non-trivial complex phase
evolution ensues due to nonlinear scalar interactions, implying charge 
transfer between the two Q-balls, and altering the nature of the force \eqref{eq:scalarforce}.

Furthermore, in the presence of gravity (i.e., in the case of BSs), no
nonlinear scalar potential is required for non-trivial interactions to occur.
In Ref.~\cite{Palenzuela:2006wp}, an effective repulsion was observed in the
collision of two mini BSs (stars in linear scalar models) when comparing the
case when the complex scalar field is \textit{anti-symmetric} under star
interchange, i.e., in \eqref{eq:binaryqball}, if $\alpha=\pi$, $\phi_1=\phi_2$,
and $\omega_1=\omega_2$, then $\Phi\leftrightarrow -\Phi$ when
$\textbf{x}_1\leftrightarrow \textbf{x}_2$, to the symmetric case. Related to
this, static spacetimes corresponding to BSs with two (or more) peaks
have been constructed in the linear scalar model, which can be thought of as
corresponding to two non-spinning BSs kept in equilibrium by the same
relative phase difference~\cite{Yoshida:1997nd,Herdeiro:2021mol,Cunha:2022tvk}.
In the following, we will refer to this equilibrium state as the dipolar BS
(DBS) solution.

Lastly, angular momentum, present in rotating Q-balls, spinning BS solutions,
or in inspiraling binary BSs, plays an important role in the evolution of the
phase, since angular momentum and the appearance of vortices in a massive
complex scalar theory are tightly connected. To illustrate this, we restrict to
a flat spacetime, where we can use the symmetries of the background to define local
notions of angular momentum. We define the angular momentum density $\rho_A$
with respect to the Killing field $\eta_A$ associated with the rotational
symmetry in the $A$-th direction as $\rho_A= T_{0j} \eta^j_A$, in Cartesian
coordinates.
From the $U(1)$ complex scalar stress-energy tensor, we know that
$T_{0i}=-2\dot{\phi}\partial_i\phi-2\dot{\psi}\phi^2\partial_i\psi$, when
decomposing $\Phi=\phi e^{i\psi}$. Therefore, the vorticity in the $A$-th
direction, which we define as $\nu_A:=\eta_A^j\partial_j\psi$, is related to
the associated angular momentum density by
\begin{align}
\rho_A\sim \phi^2 \nu_A\partial_t\psi,
\label{eq:angmomdensity}
\end{align}
where we dropped the $\psi$-independent term. In the case of rotating Q-balls,
with field profile $\psi=\omega t-m\varphi$, we find that the vorticity in all
directions, excluding the spin-direction, vanishes. The latter is simply given
by $\rho_{\rm spin}=-2\omega\phi^2 \nu_{\rm spin}=2m\omega \phi^2$. In the case
of the rotating BS, the expression is modified only by curvature corrections.
This makes the connection between vorticity and angular momentum manifest, as
the total angular momentum in the rotating Q-ball (and BS) solution is
quantized by the vortex index $m$, as $J=mQ$. Note, a vortex in the $U(1)$
nonlinear scalar models considered here does \textit{not}, by itself, contain
energy. Hence, vortices can be created and destroyed spontaneously, as long as
angular momentum is conserved. This is in stark contrast to, for instance,
cosmic strings in global $U(1)$-Higgs or Higgs-Abelian models, where the string
itself is a non-zero energy solution that is topologically protected.

Ultimately, the evolution of the complex phase of the scalar field determines
the type of self-interaction-induced momentum and charge transfer between two
Q-balls, as well as the angular momentum present in the solution, while these
effects are exponentially suppressed by the distance between the solitons.

\subsection{Scalar interactions in a binary BS inspiral} \label{sec:inspmergdynamics}

Let us translate the findings discussed in the previous section from the flat
spacetime case, to the self-gravitating BS case in the context of a binary
inspiral. For most of the early inspiral of a binary BS, the scalar interactions
are exponentially suppressed, and hence, sub-dominant to the gravitational
interactions. At this stage of the inspiral, the scalar phase of each star
grows linearly according to the star's stationary frequency. In the late
inspiral, i.e., when $D\sim 1/k_i$, scalar interactions increase in importance,
and the phases of each star start to differ from the stationary prediction,
i.e., $\ddot{\psi}_{1,2}\neq 0$. At this stage, linear predictions for the
evolution of the scalar phase break down, and nonlinear simulations are
necessary. Many of the effects present in Q-balls cannot be easily quantified
in the context of dynamical gravity due to a lack of background symmetries and
well-defined local quantities. Therefore, in the following we attempt to
provide intuition for the scalar interaction-driven physical processes active
during the later inspiral and merger of binary BS, and leave a more rigorous
and systematic study to the later sections.

\begin{figure*}
\includegraphics[width=0.99\textwidth]{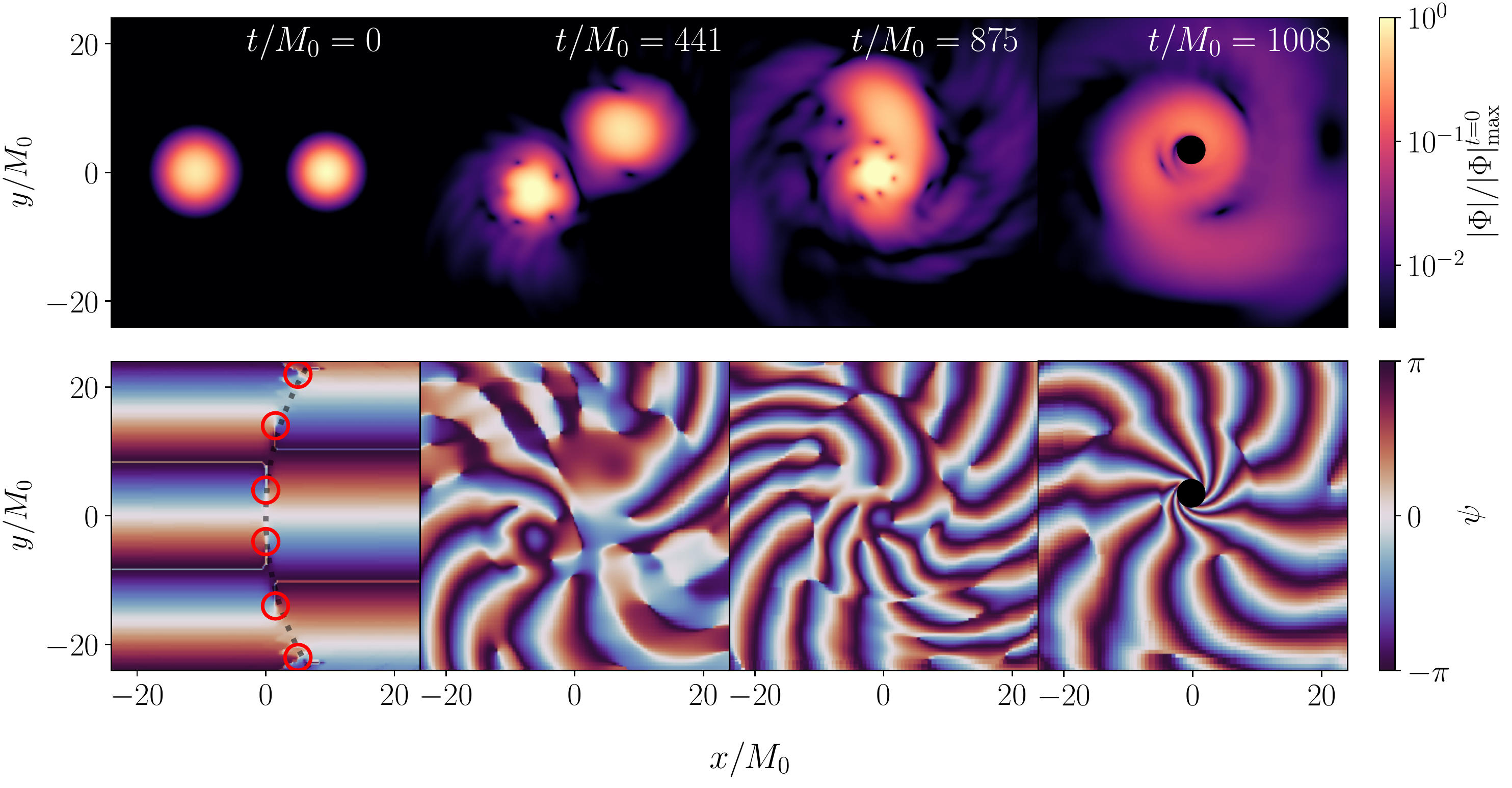}
\caption{We show four snapshots of the equatorial plane during the merger of a
    non-spinning binary BS with frequencies $\omega_1/\mu=0.9$ and
    $\omega_2/\mu=0.86$, with no initial phase-offset $\alpha=0$, total ADM
    mass $M_0$, an initial coordinate separation $D=20M_0$, and in the repulsive
    scalar model \eqref{eq:repulsive} with $\lambda/\mu^2=10^3$. The
    compactness of the higher and lower frequency star is $C=0.08$ and $C=0.12$, respectively. The total
    angular momentum points into the page, and the orbit has Newtonian eccentricity $e=0.13$. The final black hole parameters are shown in \figurename{
        \ref{fig:representativebbsinspiralbhgw}}. \textit{(top row)} We show
    the magnitude of the scalar field $|\Phi|$, normalized by the maximum
    magnitude of the scalar field in the initial time-slice $|\Phi|_{\rm
    max}^{t=0}$. \textit{(bottom row)} We show the complex phase $\psi\in
    (-\pi,\pi)$ at the corresponding times in the equatorial plane. At
    $t/M_0=0$, we indicate the locations of the $q=1$ vortices by red circles,
    while we indicate the surface of $\phi_1\sim\phi_2$ defined below
    \eqref{eq:complexphaseinitialtimeslice} with a black dashed line. Notice,
    the white lines in the first panel of the bottom row are interpolation
    artifacts and correspond to $\psi=\pm \pi$.}
\label{fig:representativebbsinspiral}
\end{figure*}

In order to understand the phase evolution within a binary BS during the late
inspiral, it is instructive to begin by studying the phase of a single, boosted,
non-spinning BS with field profile
\begin{align}
\Phi=\phi(\textbf{x})e^{i(\omega t+\alpha)}
\end{align}
in the stationary frame. With the primed coordinates $(t',\textbf{x}')$ denoting a boosted frame, defined by the boost vector $\beta^i$, the complex phase of the boosted BS in the $t'=0$ slice is
\begin{align}
\psi(\textbf{x}')=\omega \beta^i x_i' +\alpha \mod 2\pi.
\end{align}
Therefore, due to the mixing of temporal and spatial degrees of freedom in the
Lorentz boost, the complex phase of the scalar field is a monotonic function of
\textit{space} in the boosted frame, with slope determined by the star's
frequency $\omega$\footnote{The magnitude $\phi$ is simply Lorentz contracted
in the boost direction.}. We now move to the case of an inspiraling binary BS.
The construction of binary BS initial data satisfying the Hamiltonian and momentum constraints is outlined in Appendix~\ref{app:idandnumevo}, with further details in
\cite{inprep}. Here, we simply note that the binary's scalar field is a
superposition of two isolated stars, boosted with velocity $v_1,v_2$ along
$\beta_1^j,\beta^j_2$. Explicitly, the constraint-solving scalar field profile
in the initial time-slice in the center of mass frame is
\begin{align}
\Phi_{\rm BBS}=\phi_1 e^{i(\omega_1 \beta_1^ix'_i)}+\phi_2 e^{i(\omega_2\beta_2^ix'_i+\alpha)}.
\end{align}
Therefore, in the center of mass frame, the spatial dependence of the complex phase $\psi_{\rm BBS}$ of $\Phi_{\rm BBS}$ is
\begin{align}
\psi_{\rm BBS}\approx
\begin{cases} 
\omega_1 \beta_1^ix'_i, & \phi_1 \gg \phi_2, \\
\omega_2 \beta_2^ix'_i+\alpha, & \phi_2 \gg \phi_1.
\end{cases}
\label{eq:complexphaseinitialtimeslice}
\end{align}
In the regions with $\phi_1\sim\phi_2$, an infinite set of $|q|=1$ vortices
appear. In fact, if the scalar field initial data of the binary BS is constructed by superposing the individual star's fields, then these vortices cannot be removed, unless all angular momentum is removed from the system (e.g., in a head-on collision). We find that a subset of these vortices becomes dynamically important in the binary evolutions discussed in the remainder of this work.

\begin{table}[b]
\begin{ruledtabular}
\begin{tabular}{c c|c c | c c c}
Frequency & $C$ & $x_0/M_0$ & $v_x$ & $J_0/M_0^2$ & $\alpha$ & $e$ \\
\hline \hline
$\omega_1/\mu=0.90$ & $0.08$ & $-10.69$ & $0.12$ & \multirow{ 2}{*}{0.9} & \multirow{ 2}{*}{0} &\multirow{ 2}{*}{$0.13$} \\ 
$\omega_2/\mu=0.86$ & $0.12$ & $9.31$ & $-0.10$ & & & \\ 
\end{tabular} 
\end{ruledtabular}
\caption{The properties of the non-spinning binary BS initial data discussed in the main text. The two stars have frequencies $\omega_{1,2}$, with initial phase offset $\alpha$, are positioned at coordinate locations $x_0$ and $y_0/M_0=0$, have boost velocities $v_x$ and $v_y=0$ (with Newtonian eccentricity $e$), compactness $C$, and mass-ratio $\tilde{q}=1.13$. Here $M_0$ is the ADM mass, and $J_0$ is the similarly defined global angular momentum, the latter given by eq. (7.63) of Ref.~\cite{Gourgoulhon:2007ue}. Note, this definition of angular momentum, albeit commonly used, is not free from gauge-dependency (for further details, see Ref.~\cite{Gourgoulhon:2007ue}).}
\label{tab:introBBSidproperties}
\end{table}

To illustrate this, and the subsequent evolution, we consider a non-spinning
binary BS in the repulsive scalar model \eqref{eq:repulsive} with coupling
$\lambda/\mu^2=10^3$. The stars of the binary are prepared with frequencies
$\omega_1/\mu=0.9$ and $\omega_2/\mu=0.86$ and initial phase-offset $\alpha=0$.
The initial coordinate separation is $D=20M_0$, with binary ADM mass $M_0$ (see
\tablename{ \ref{tab:introBBSidproperties}} for details on the parameters of
the initial data). In the language of the previous section, this choice enables
phase evolution and breathing behavior. We comment on the former below, while
the latter is likely a small effect on timescales larger than the orbital
timescales of this binary. In \figurename{
    \ref{fig:representativebbsinspiral}}, we show the magnitude $\phi_{\rm
BBS}$ and phase $\psi_{\rm BBS}$ of the equatorial plane of the binary BS at
four different times throughout the evolution. We focus first on the initial
time-slice. The less compact of the two stars, i.e., the star with smaller
central scalar field magnitude, is the star with smaller charge and larger
frequency, $\omega_1/\mu=0.9$. The complex phase of this binary, shown in the
bottom row of \figurename{ \ref{fig:representativebbsinspiral}}, illustrates
the structure discussed above. The phase is a monotonically increasing function
of space in the boost direction on either side of the black dashed line,
indicating where $\phi_1\sim\phi_2$. The slope of the phase is different on
either side of $\phi_1\sim\phi_2$ in the equatorial plane, since $\omega_1\neq
\omega_2$. The surface of $\phi_1\sim\phi_2$ is a 2-dimensional surface in the
spatial hypersurface separating regions of $\phi_2>\phi_1$ from those with
$\phi_2<\phi_1$. Lastly, in the equatorial plane, along the line of
$\phi_1\sim\phi_2$ there is a set of $q=1$ vortices indicated with red circles. 

The chosen boost parameters do not lead to a quasi-circular inspiral, but
rather to an orbit with non-zero eccentricity. Therefore, the stars go through
an initial orbital phase, during which their (coordinate) distance increases
over time. During this time, the scalar interactions are exponentially
suppressed, and the dynamics are largely dominated by gravitational
interactions of the stars. Throughout this phase, the complex phase of each
star increases approximately linearly, and independently, in time. After passing the
apoapsis, the coordinate distance between the stars decreases, leading to
enhanced scalar interactions. The closest approach is achieved around
$t/M_0=350$. During the periapsis passage, there is significant overlap between the two
scalar field profiles. The evolution of $\psi$ is no longer approximately
linear in time. The enhanced scalar interaction, in conjunction with the
presence of the vortices in $\psi$ between the two stars, leads to the transfer
of these vortices onto \textit{both} of the BSs. In the second snapshot of
\figurename{ \ref{fig:representativebbsinspiral}}, the vortices can be seen
just after the closest approach. After the periapsis, the vortices orbit around
each individual star in the same direction as the overall orbital motion.
Qualitatively, the orbiting vortices indicate angular momentum transfer from
the orbit to the spin angular momentum
that can be assigned to each star. While stationary BSs cannot rotate
perturbatively \cite{Kobayashi:1994qi}, time-dependent solutions may. During
the second close encounter, the scalar interactions dominate the dynamics. The
star with the larger frequency transfers large amounts of scalar charge onto
its heavier companion. This process is shown in the third snapshot of
\figurename{ \ref{fig:representativebbsinspiral}}. As the heavier star accretes
scalar matter, its charge and central scalar field magnitude increase.
Ultimately, the lighter (and less compact) star is completely tidally disrupted
and accretes rapidly onto the heaver companion. The mass of this companion
increases beyond the maximum mass the star can stably support, and it collapses into a black hole. This situation is
shown in the fourth panel of \figurename{ \ref{fig:representativebbsinspiral}}.
The remnant black hole moves through the surrounding residual scalar matter, continuing to accrete it. 

\begin{figure}[t]
\includegraphics[width=0.48\textwidth]{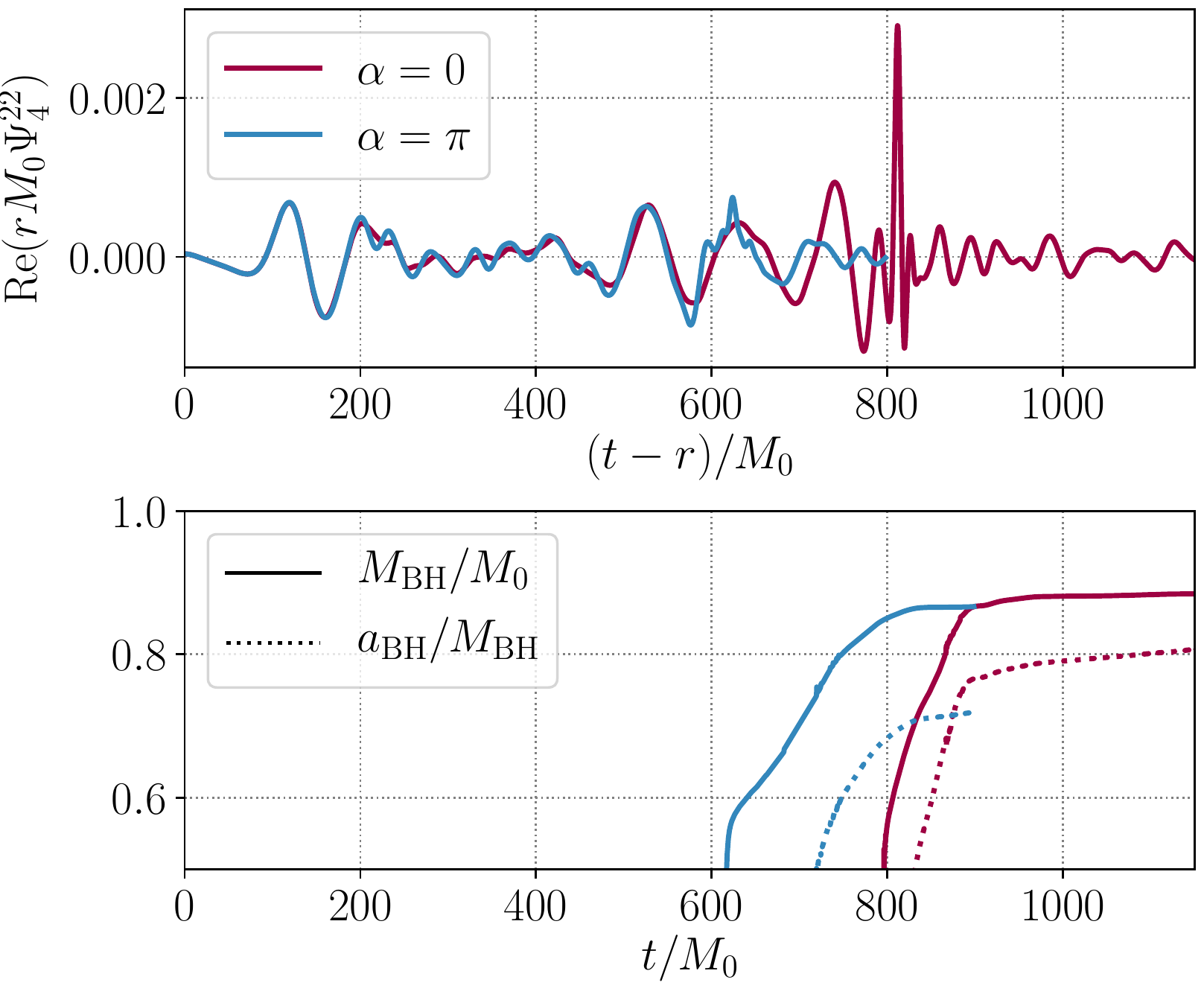}
    \caption{\textit{(top panel)} We show the $(\ell,m)=(2,2)$ mode of the $s=-2$-
    weighted spherical harmonic components of the Weyl Newman-Penrose scalar $\Psi_4$
(extracted at coordinate radius $r=100M_0$) emitted during the binary BS
inspiral of the case with initial phase-offset $\alpha=0$ shown in \figurename{
    \ref{fig:representativebbsinspiral}} and discussed in the main text. 
We compare this to the GWs from the same binary inspiral with initial
phase-offset $\alpha=\pi$. \textit{(bottom panel)} The 
mass $M_{\rm BH}$ (solid) and dimensionless spin parameter $a_{\rm
BH}$ (dashed) measured from the apparent horizons of the remnant black holes formed in the inspiral of the top
panel. We evolve the $\alpha=\pi$ case for only roughly $900M_0$.}
\label{fig:representativebbsinspiralbhgw}
\end{figure}

The gravitational radiation, as well as the final black hole parameters, are
shown in \figurename{ \ref{fig:representativebbsinspiralbhgw}}. There, we also
compare to an evolution of the same binary BS initial data, except with initial
phase-offset of $\alpha=\pi$. This case has two important features
that distinguish it from the $\alpha=0$ case: \textit{(i)} The $\alpha=\pi$ case collapses
to a black hole roughly $200M_0$ earlier than the $\alpha=0$ case, and
\textit{(ii)} the gravitational waveforms differ in amplitude 
from $t=200M_0$ to $t=500M_0$. The high-frequency features in the waveform
shown in the top panel of \figurename{ \ref{fig:representativebbsinspiralbhgw}}
originate from time-dependent features on scales smaller than the
sizes of the stars. Comparing the orbital period of the vortices
around the center of each star with the frequency of the small scale features
in the emitted GWs, we can identify the orbiting vortices as the source of the
high-frequency GWs. The differing amplitudes indicate that the vortices in the
$\alpha=\pi$ case are surrounded by larger $|\Phi|$ values. Consulting
\eqref{eq:angmomdensity}, this implies also a locally enhanced angular momentum
density. This is consistent with finding \textit{(i)}. Larger angular momentum
transfer from the orbit to the spin implies earlier merger times, and hence, a
more rapid transfer of charge to the heavy star and subsequent black hole
formation. The final black hole parameters are, however, roughly independent of
the initial phase offset. In fact, the spin angular momentum of the remnant
black hole is comparable to that formed by the merger of a binary neutron star
or binary black hole. Hence, beside the residual scalar matter, the remnant
black hole retains little memory of the type of binary it was made from. 

A few remarks are in order. Many of the statements above are purely qualitative
in nature, and are mainly made to provide intuition. In particular, the transfer
of orbital to spin angular momentum deserves a more rigorous analysis (e.g., using 
techniques developed in Refs.~\cite{Clough:2021qlv,Croft:2022gks}). 
The evolution of the
scalar phase during the second encounter is nonlinear, making it challenging to
gain intuition from applying linear methods discussed in the previous section.
In the following section, we return to the dependence of the GWs on the 
initial phase and a more systematic analysis of the phase evolution. 
Furthermore, the use of vortices as a tool to understand the nonlinear phase
evolution and to predict the remnant is the subject of
Secs.~\ref{sec:remnantmap} and \ref{sec:formationofrotremn}.

\subsection{Gravitational wave imprints} \label{sec:gwimprints}

\begin{figure*}[t]
\centering
\includegraphics[width=0.485\textwidth]{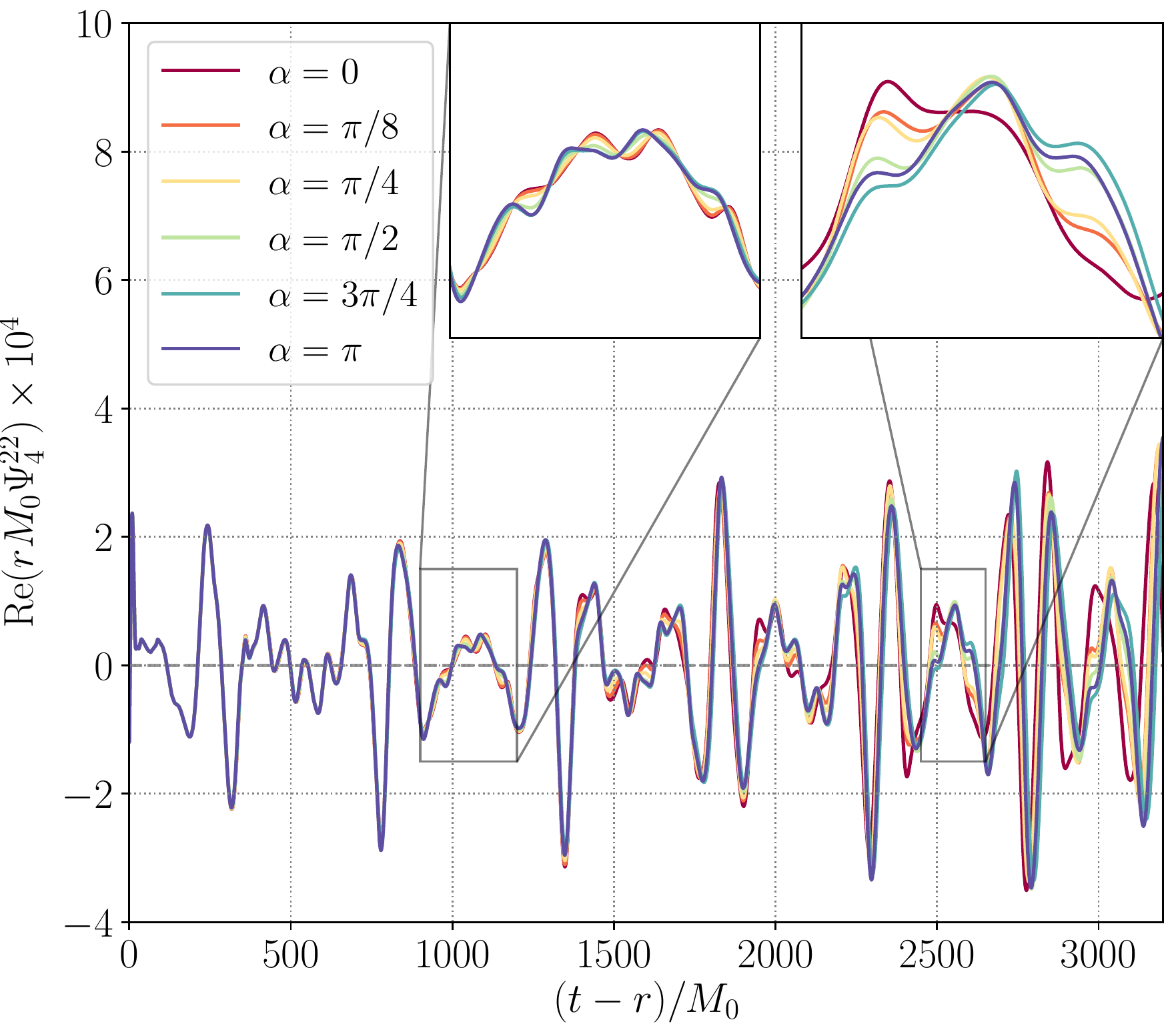}
\hfill
\includegraphics[width=0.485\textwidth]{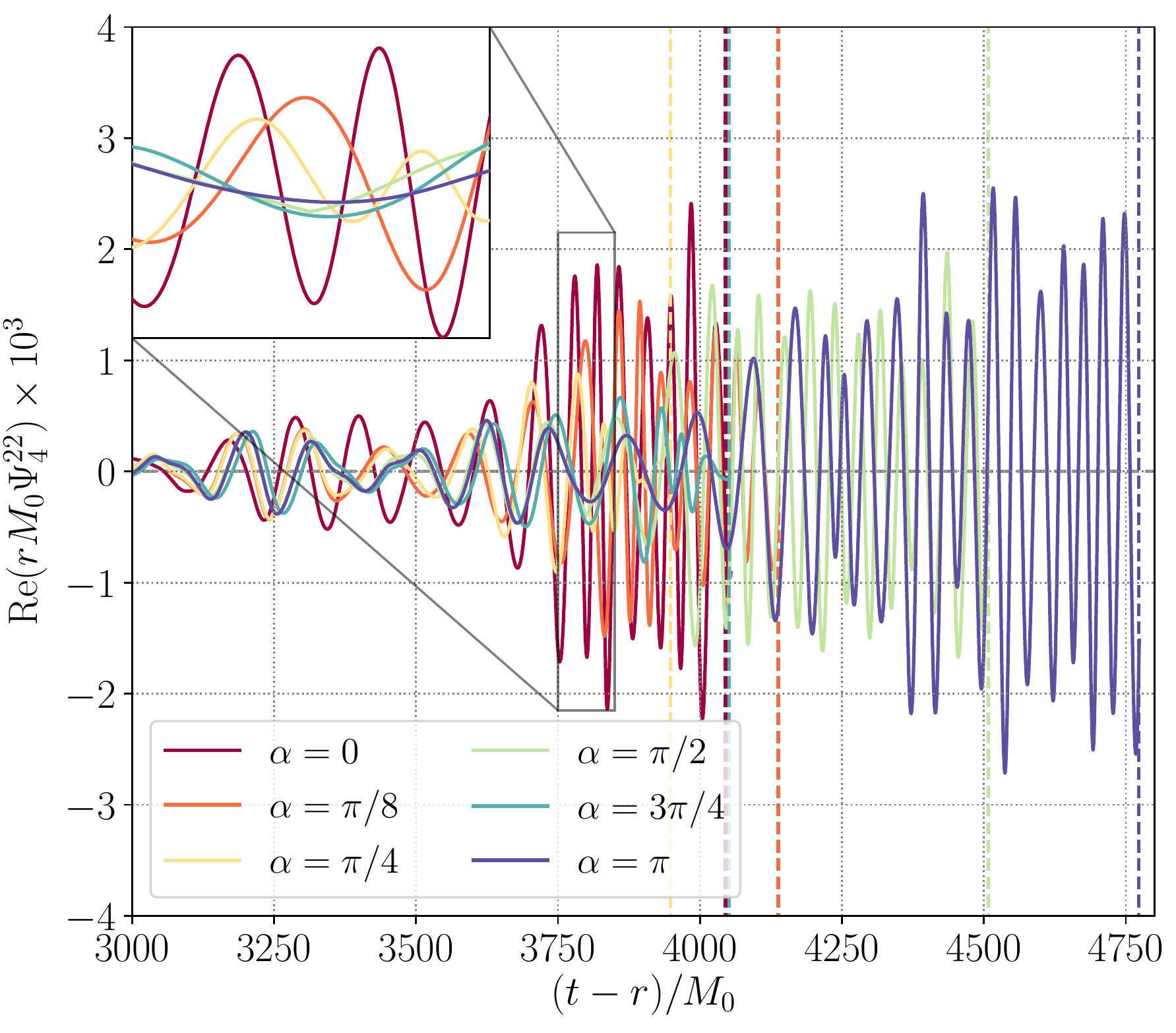}
\caption{The GWs (extracted at a coordinate radius $r/M_0=100$) during the
    inspiral \textit{(left)} and merger \textit{(right)} of the binary BS described in the main text.  The different cases
    correspond to identical initial binary systems, except with different
    values of the phase offset $\alpha$ in the range $0$ to $\pi$.
    Specifically, we show the $(\ell,m)=(2,2)$ $s=-2$-weighted spherical
    harmonic component of the Newman-Penrose scalar $\Psi_4$ as a function of
    retarded time $t-r$. 
    The differences between the various $\alpha$-cases are
    due to the enhancement of nonlinear scalar effects during each close
    encounter of the binary and towards merger, as the separation between the
    stars shrinks. Notice the different scale used on the left and the right. After merger, the waveforms are terminated around 
    the time when the system
        collapses to a black hole, which is indicated by a vertical dashed line.
    We show the GWs after gravitational collapse of
    the $\alpha=0$ and $\pi$ cases in \figurename{ \ref{fig:gwringdown}}.} \label{fig:gwinspiral}
\end{figure*}

In the previous two sections, we first reviewed the importance of the scalar
phase evolution for the nonlinear dynamics of scalar solitons in the absence of gravity, and then
qualitatively applied some of these concepts to the case of an inspiraling
non-spinning binary BS. In this section, we investigate the role of the scalar
phase in the inspiral and merger of a non-spinning binary BS and the emitted
GWs more systematically. The variation of the scalar phase in binary BS mergers
has been studied only in the case of head-on collisions in
Refs.~\cite{Lai:2004fw,Palenzuela:2006wp,Palenzuela:2007dm,Cardoso:2016oxy,Evstafyeva:2022bpr}
(see Ref.~\cite{Sanchis-Gual:2022mkk} for a study in the PS case). Here, we
consider the impact of the scalar interactions on  the \textit{inspiral} of a
binary BS and connect our observations directly to the physical intuition
provided in Sec.~\ref{sec:scalarinteractions}, for the first time. We find
several significant differences between the head-on collisions studied in
\cite{Lai:2004fw,Palenzuela:2006wp,Palenzuela:2007dm,Cardoso:2016oxy,Evstafyeva:2022bpr}
and the inspirals considered here: \textit{(i)} the effect of the scalar interactions
accumulate secularly throughout the inspiral, eventually resulting in strong de-phasing and
modulations of the emitted GW amplitudes; \textit{(ii)} vortices appear due
to the orbital angular momentum and drive dominant high-frequency
features in the emitted GWs at late times during the merger; and \textit{(iii)} the time
to collapse to a black hole post-merger depends sensitively on the scalar interactions,
while the remnant's properties are mostly insensitive to the scalar
interactions driving the proceeding dynamics.

In order to illustrate the role of the scalar phase and vortex structure during
the inspiral, merger, and ringdown, we consider a non-spinning binary BS in the
repulsive scalar model given by \eqref{eq:repulsive} with
$\lambda/\mu^2=10^3$, and focus on an equal frequency case with
$\omega_1=\omega_2=0.9\mu$. With a compactness of $C=0.08$ for each star, the
impact of the scalar interactions is enhanced compared with a system consisting
of highly compact constituent BSs (and $\omega\ll\mu$), since at fixed
separation of the stars' center of mass, the overlap of the two stars' scalar fields is larger. The
constraint satisfying initial data is constructed using the methods outlined in
Appendix~\ref{app:idandnumevo}. In the initial time slice, the binary system
has ADM mass $M_0$, the stars' coordinate positions are at $x_0/M_0=\pm 10$ and
$y_0/M_0=0$ (hence, initial coordinate separation $D=20M_0$), and the initial
coordinate velocities are $v_x=0$ and $v_y=\mp 0.12$ (with these parameters the
Newtonian eccentricity is $e=0.15$). These initial velocities were chosen to
result in an eccentric orbit, with multiple periapses before final merger. This
allows us to observe the effects of repeatedly enhanced and suppressed scalar
self-interactions on the GWs. In order to study the impact of the scalar field phase
on the inspiral, we vary the initial phase-offset $\alpha$ between the BSs 
considering the values $\alpha\in\{0,\pi/8,\pi/4,\pi/2,3\pi/4,\pi\}$.
Note that in this
strong-coupling regime of the repulsive scalar model, stable rotating $m=1$ BSs
were demonstrated to exist in Ref.~\cite{Siemonsen:2020hcg} (and in Newtonian gravity in Ref.~\cite{Dmitriev:2021utv}).
However, we choose binary BS parameters that are expected to result in the formation of a rotating black
hole since the sum of the charges of the constituent stars is larger than the maximum charge
of \textit{both} the $m=0$ and $m=1$ families of solutions, and indeed we find collapse postmerger.

In \figurename{ \ref{fig:gwinspiral}}, we show the GWs emitted during the
inspiral of this binary BS for each initial phase offset $\alpha$. Focusing
first on the $\alpha=0$ case, the non-negligible eccentricity in the binary is
reflected in the gravitational waveform as periodic peaks around
$(t-r)/M_0\approx 250$, 750, 1300, 1800, 2300, and  $2750$, corresponding to the close
encounters of the stars. In between these close encounters, the GW signal is
characterized by a high-frequency component emerging from spurious oscillation
modes excited in the individual stars due to the way the initial data is constructed. (Though 
we do not do so here, these spurious
oscillations can be alleviated by modifying the superposition of the two isolated boosted stars and utilizing a different scaling for the conformal kinetic energy of the scalar field, as shown in Ref.~\cite{inprep}.) 
Though the cases
with different values of $\alpha$ initially have essentially indistinguishable
orbits and GWs, following the first encounter around $(t-r)/M_0\approx 250$,
the differences in the GW amplitude $|\Psi_4^{22}|$ grow \footnote{All cases considered here were obtained with
identical numerical setups both to construct the initial data and to evolve it,
suggesting that the \textit{relative} differences in the GWs of the evolutions
shown in \figurename{ \ref{fig:gwinspiral}} are driven by scalar interactions,
rather than being due solely to numerical truncation error. However, as discussed in
Appendix~\ref{app:idandnumevo}, the estimated truncation error in the GW amplitude
is comparable to the difference between the cases with different values of $\alpha$.} (see also the two
insets).
During the periods of large separation
between the stars, the scalar phase of each star evolves approximately linearly
in time. However, scalar interactions are exponentially enhanced during each close encounter,
and become more significant as the orbital period and periapse distance shrink due to GW emission
during the course of the merger.
Since the nonlinear interactions depend on, and affect the scalar phase of the star,
as discussed in Sec.~\ref{sec:scalarinteractions},
they secularly affect the evolution and GWs of the binary BS system.

These differences in \figurename{ \ref{fig:gwinspiral}} are primarily
differences in the GW amplitude. However, as can already be seen there, at late
times, there is also a non-negligible \textit{de-phasing} between the
waveforms.  This can be clearly seen in \figurename{ \ref{fig:gwinspiral}}, where
we show the GWs emitted during the late inspiral and merger of the binary
systems. Though the exact point of contact for the two stars is not
well-defined, As the stars begin to merge, the GW amplitude increases
significantly---by up to an order of magnitude---as can be seen comparing
the left and right panels of \figurename{ \ref{fig:gwinspiral}} (note
the difference in scales). While the exact point of contact for the two stars
is not well-defined, one can loosely determine the merger time by this sudden
increase in amplitude. It is clear from the right panel of \figurename{ \ref{fig:gwinspiral}}
that this time depends strongly on the initial phase offset $\alpha$. For
instance, the $\alpha=0$ case merges around $(t-r)/M_0\approx 3750$, while the
binary with initial phase offset $\alpha=\pi$ merges at $(t-r)/M_0\approx
4150$. After the merger, the system enters into a transient state
consisting of a dynamical remnant star that is temporarily prevented from collapsing
due to excess energy and/or angular momentum,
analogous to what may happen post-merger in a binary neutron star.
A series of scalar vortices (some of which still present from the initial data) rapidly
orbit around the center of mass of the remnant indicating large perturbative
angular momentum of the \textit{hypermassive BS}. 
\footnote{In analogy with the neutron star case (see, e.g., Ref.~\cite{Baumgarte:1999cq}),
here we use hypermassive to refer to an object with total mass above the maximum
stable $m=1$ rotating BS within the same scalar model.
We use hypermassive BS to refer to the merger remnants even when these
are highly perturbed, and far from equilibrium solutions.
}
These vortices are small scale features orbiting on
the scale of the original constituent stars, which lead to high-frequency GW
emission (similar to what we described in Sec.~\ref{sec:inspmergdynamics}).
This is reflected in the sudden increase in the GW frequency after the merger
of the stars, as shown in \figurename{ \ref{fig:gwinspiral}}, which matches the
orbital frequency of the vortices around the center of mass of the hypermassive
BS. The length of this hypermassive state depends on the initial phase-offset,
and hence, on the nonlinear scalar dynamics. For instance, the hypermassive
phase in the $\alpha=\pi/4$ case only lasts $t/M_0\approx 150$, while for the
$\alpha=\pi$ case, it lasts $t/M_0\approx 600$. The latter is longer lived,
since the symmetry gives rise to a vortex at the center of mass throughout the
evolution, which acts to delay the gravitational collapse of the hypermassive
remnant.  However, in all the cases with different values of $\alpha$, we find
eventual collapse to a black hole, at times indicated by the vertical dashed
lines in \figurename{ \ref{fig:gwinspiral}}.  We cannot identify a clear trend in
the dependence of the collapse time on initial scalar phase offset $\alpha$.
This may be explained by the fact that the intrinsic BS inverse frequency
$\omega_{1,2}^{-1}\approx 1.4 M_0$ is much shorter than the time to merger
$T_m\sim\mathcal{O}(10^3M_0)$, so that the repeated scalar interactions,
operating on timescales $\sim 1/\omega$, accumulate differences nonlinearly
throughout the inspiral of length $T_m$, leading to a significantly different
states entering the hypermassive phase and subsequent gravitational collapse.
Hence, any consistent $\alpha$-trend present at early times appears lost in the
accumulated nonlinear shift.

\begin{figure}[t]
\centering
\includegraphics[width=0.485\textwidth]{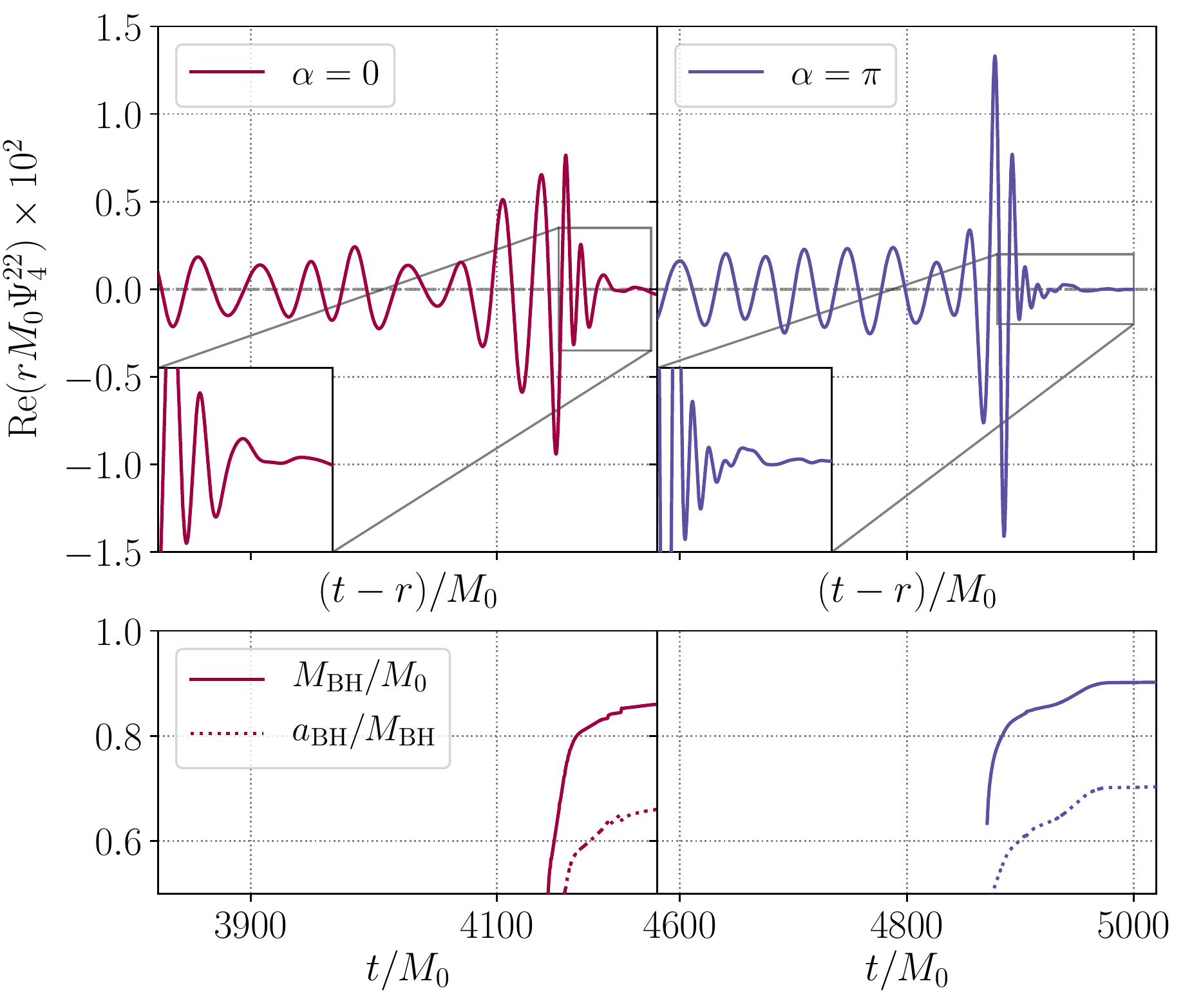}
\caption{\textit{(top panels)} We show the GWs emitted around the time of black hole formation by the binary BS systems discussed in the main text and shown in \figurename{ \ref{fig:gwinspiral}}, for initial phase offsets $\alpha=0$ and $\pi$. Notice the difference in scale compared with \figurename{ \ref{fig:gwinspiral}}. \textit{(bottom panels)} The mass $M_{\rm BH}$ and spin parameter $a_{\rm BH}$ of the remnant black holes formed as functions of \textit{coordinate time} $t$ (corresponding to the retarded time $t-r$ in the top panels).}
\label{fig:gwringdown}
\end{figure}

In \figurename{ \ref{fig:gwringdown}}, we show the GWs through the collapse of
the hypermassive BS state to the final remnant black hole in the case of
$\alpha=0$ and $\alpha=\pi$. As pointed out above, the time of collapse is
significantly different due to the accumulated nonlinear scalar (and
gravitational) interactions. However, the GWs radiated during the collapse to
a black hole are qualitatively similar. The amplitude rapidly increases
by a factor of a few and then decays. This sudden increase in amplitude
is likely driven by the vortices in the hypermassive remnant being forced onto
tighter orbits with higher orbital frequency just before horizon appearance.
The subsequent ringdown differs somewhat from a spinning black hole in vacuum
due to the residual scalar matter (of mass $\sim 0.1 M_{\rm BH}$) orbiting the
black hole.  As can be seen in the insets in the top panel of \figurename{
    \ref{fig:gwringdown}}, this has the effect of washing out the exponential
decay of the GW amplitude.  The mass and dimensionless spin parameters of the
remnant black hole are only slightly smaller, in the $\alpha=0$ case compared
to the $\alpha=\pi$ case. Therefore, despite the quantitatively different
inspiral and merger dynamics, depending sensitively on the initial scalar phase
configuration of the stars, the remnant black holes show only little memory of
the initial BS binary from which they emerged. In fact, the dimensionless spin
of the remnant black holes are roughly $a_{\rm BH}/M_{\rm BH}\approx 0.7$,
which is consistent with the quasi-circular merger of a non-spinning equal-mass
binary black hole, and the lower range of the values of $a_{\rm BH}/M_{\rm
BH}\approx 0.7$--$0.8$ found for prompt collapse following a binary neutron star
merger (see, e.g., Ref.~\cite{Bernuzzi:2013rza,Cokluk:2023xio}).

In summary, we find that, during the
late inspiral of a binary BS, scalar interactions, which depend on the scalar phase offset
between the stars, have a significant, cumulative effect, resulting in modulations of
the amplitude and de-phasing of the GWs, and affecting the merger time by  hundreds of light crossing times
of the system. The two stars merge into a hypermassive BS remnant that is
characterized by a series of vortices orbiting rapidly around its center of
mass, resulting in emitted high frequency GWs. Lastly, the hypermassive remnant
collapses to a remnant black hole with mass and spin which, despite the large
difference in the inspiral dynamics, we find to be largely insensitive to the
nonlinear scalar dynamics prior to collapse.

Here, we considered the special case of an equal frequency binary BS.
Therefore, the scalar interactions in the initial binary BS are completely
characterized by the scalar phase offset $\alpha$ between the stars. In a more
complex scenario with $\omega_1\neq \omega_2$, the different linear evolutions
of both stars' phases would lead to oscillations in the nature of the scalar
interactions at characteristic frequencies $\omega_1\pm\omega_2$.  The
oscillatory nature of the interaction accumulates nonlinearly during the later
inspiral, with the phase offset it leads to at the point of contact depending
on the length of the inspiral. The value of this phase offset just before
merger will likely have a strong effect on the qualitative behavior of the
system after merger (when nonlinear interactions drive the dynamics), as it
does in the scenario studied here. In fact, in
Refs.~\cite{Sanchis-Gual:2022mkk,Evstafyeva:2022bpr}, it was noted that in the
case of head-on collisions, the GW emission is predictably dependent on the
phase-offset at early times and the point of contact. 
Our results, however, suggest that in the
case of a BS inspiral, the nonlinear interactions prior to merger, which
accumulate secularly, render a reliable prediction of the phase-offset at the
point of contact challenging.

\subsection{Remnant map} \label{sec:remnantmap}

In the previous section, we demonstrated the complex dependence of the scalar
and gravitational dynamics during the merger process on the binary BS
parameters, in particular the scalar phase of the BSs.  In order to provide a
more straightforward understanding of these systems, here we outline and motive
a small set of criteria, and an associated mapping, in order to guide
predicting the outcome of a given merger of a binary BS, i.e., to determine
whether a spinning BS, non-spinning BS, or a black hole will be formed.

\subsubsection{Details of the remnant map}

Recall, within a given scalar (or vector) model, there exists a set of
one-parameter families of BS (or PS) solutions $\mathbb{B}_{m}$ indexed by the
azimuthal index $m$ (excluding ``excited" states with a higher number of radial
nodes).  
A representative $\mathcal{B}\in\mathbb{B}_m$ of the
one-parameter family of star solutions (for fixed $m$) is identified, in
general, by its mass $M$, charge $Q$, frequency $\omega$, and angular momentum
$J$ ($J=0$ for spherical stars). Crucially however, if one restricts to the
\textit{stable} branches of a family of solutions, a particular BS (or PS)
solution is \textit{uniquely} identified by the charge alone $\mathcal{B}(Q)$.

During a binary BS (or PS) merger, energy and
angular momentum will come both from the constituent stars, as well as from the
orbit, and some fraction of these quantities will be carried away both by
gravitational and scalar radiation.  The total boson number (i.e., the
$U(1)$-charge of the system), on the other hand, will only be affected by
scalar radiation. We expect this to be subdominant to gravitational radiation
since the scalar field is massive, and higher energy processes are required to
elevate a bound boson into an asymptotically free state.
Therefore we will make the approximation that the scalar charge is conserved
during the merger here, but return to
the implications of scalar particle loss below. Hence, a core assumption of the mapping from the binary's constituent BSs (formally, $\mathcal{B}_1\in\mathbb{B}_{m_1}$ and $\mathcal{B}_2\in\mathbb{B}_{m_2}$) with charges $Q_{1,2}$ into the remnant solution $\mathcal{B}_r$ of charge $Q_r$ is charge conservation $Q_1+Q_2=Q_r$.

Combining the assumption of charge conservation, as well as restricting to the stable branches of the one-parameter families of solutions $\mathbb{B}_{m_{1,2}}$, enables us to introduce $\mathcal{R}_F$, which maps all properties of the merging binary BS (or PS), i.e., properties of $(\mathcal{B}_1,\mathcal{B}_2)$ into those of the remnant $\mathcal{B}_r$, formally written as
\begin{align}
\mathcal{R}_F(\mathcal{B}_1,\mathcal{B}_2)=\mathcal{B}_r.
\label{eq:map}
\end{align}
Since $\mathcal{B}_{1,2}$ are uniquely identified by any pair of the
constituents properties (e.g., the frequencies $\omega_{1,2}$), the map
$\mathcal{R}_F$ can take various explicit forms. For example, the explicit 
map of the frequencies of the inspiraling binary
$(\omega_1,\omega_2)$ into the frequency of the remnant $\omega_r$ is
$\mathcal{R}_F^\omega(\omega_1,\omega_2)=Q_r^{-1}[Q_1(\omega_1)+Q_2(\omega_2)]=\omega_r$,
using charge conservation and the inverse $Q^{-1}_r$ of $Q_r(\omega_r)$ for a chosen family of solutions.

So far, we have not specified into which family of stationary solutions
$\mathcal{B}_r\in F$ the remnant map $\mathcal{R}_F$ maps (this freedom is
indicated by the subscript of $\mathcal{R}_F$). In principle, $F$ could be
\textit{any} set of stationary solutions allowed in the scalar (or vector)
theory at hand, i.e., \eqref{eq:action} and \eqref{eq:vectoraction}, that
satisfies the charge conservation assumption; for instance, a non-rotating BS,
a DBS solution, a spinning BSs, etc. However, many
of these possibilities can be rejected using a series of conditions which the
remnant has to approximately satisfy. 
Of course, it is possible that are no suitable remnant star solutions for a
given binary. For example, the total charge of the binary may be above the
maximum charge of any remnant non-vacuum family, $Q_1+Q_2> Q^{\rm max}_r$, in
which case one may expect a black hole to form.
As we show below, the combination of
these conditions and the remnant map \eqref{eq:map} can be used to understand
the outcome of even complex merger scenarios. In the following, we introduce
the conditions first, and then return to a practical application of the formal
map \eqref{eq:map}.

The first set of conditions, we call the \textit{kinematic conditions}. The masses and angular momentum of the constituents of the binary, as well as the remaining orbital angular momentum at merger $J_{\rm orb}$, should satisfy
\begin{align}
\begin{aligned}
1. & \ \ M_1+M_2\gtrsim M_r,\\
2. & \ \ J_1+J_2+J_{\rm orb.}\gtrsim J_r,
\end{aligned}
\label{eq:kinematicconditions}
\end{align}
at the point of contact of the stars. Here, $M_r$ and $J_r$ are the remnant 
solution's total mass and angular momentum. This condition holds only for
aligned-spin scenarios and here we are neglecting the correction from the
orbital energy as being small. 

Secondly, the \textit{stability condition}, states that the remnant 
stationary solution must be free of any linear or nonlinear instability:
\begin{align}
3. \quad \mathcal{B}_r \text{  is stable.}
\label{eq:stabilitycondition}
\end{align}

Lastly, the \textit{vortex condition} concerns the scalar phase dynamics. We conjecture that the vortex number $m_r$ of the final remnant of the binary merger is given by the vortex number of a closed loop $\Gamma$ enclosing the center of mass at the point of contact of the stars as well as  
all significant vortex lines of the merging binary:
\begin{align}
\begin{aligned}
4. & \ \ m_r=\frac{1}{2\pi}\oint_\Gamma d\ell_i D^i\psi.
\end{aligned}
\label{eq:vortexcondition}
\end{align}
By significant vortex lines, we mean to exclude, for example, those that may arise as
perturbations to the constituent BSs during the inspiral (as in Fig.~\ref{fig:representativebbsinspiral}),
though we will return to this below. 
Here, $\Gamma$ is negatively oriented with respect to the total angular momentum\footnote{Note, paired with the choice of $J\rightarrow -J$ and $m\rightarrow -m$ (or $Q\rightarrow -Q$), the negative orientation of $\Gamma$ with respect to the
total angular momentum of the remnant is a convention.} $J_{\rm ADM}^\mu$.  
These conditions are approximate, and we explicitly show below the degree to which 
they must be satisfied in example mergers.

The procedure to determine the family of stationary solutions $F$
and identify the remnant $\mathcal{B}_r$ is as follows: \textit{(i)} Construct
$\mathcal{R}_F$ for a plausible family of remnants $F$, \textit{(ii)}
reject this mapping if it violates one of the four conditions listed above, and
\textit{(iii)} begin at step \textit{(i)} with a different plausible family.
Therefore, the map \eqref{eq:map} is used to eliminate possibilities until the
correct remnant remains. A priori, considering all plausible families
$F$ appears to be a daunting task. However, in practice, we find
that only focusing on the families of non-spinning and $m=1$ rotating isolated
stars (within the considered model) suffices, with the vortex condition
\eqref{eq:vortexcondition} efficiently discriminating between the two. For example, a DBS solution can be approximated by the sum of two non-spinning stars, particularly in the Newtonian limit (see Ref.~\cite{Cunha:2022tvk}).

In Ref.~\cite{Bezares:2022obu}, a similar set of criteria was considered (in
particular conditions 1.--3.). However, crucially, here we add the vortex
condition \eqref{eq:vortexcondition}, and point out its vital role in the
binary merger process below. 
Furthermore, while the ``point of contact" is an ill-defined notion in the case
of BSs (and PSs), the more relativistic the stars become, the easier it becomes to identify
the point of contact. Similarly, the vortex condition should be regarded
somewhat approximately, as we will show using examples of un-equal mass mergers
of a rotating and a non-rotating BS. 
In the special case of a head-on collision (i.e., with exactly vanishing
angular momentum), no vortex exists. Lastly, this map may be extended to PSs as
well by the addition of another condition, which governs the relative
\textit{direction} of the vector fields of a given PS binary. We return to
applying this remnant map to PS in Sec.~\ref{sec:rotremnformationcrit}, but
leave a more detailed analysis to future work.

\subsubsection{Applying the remnant map}

\begin{figure*}
\includegraphics[width=0.99\textwidth]{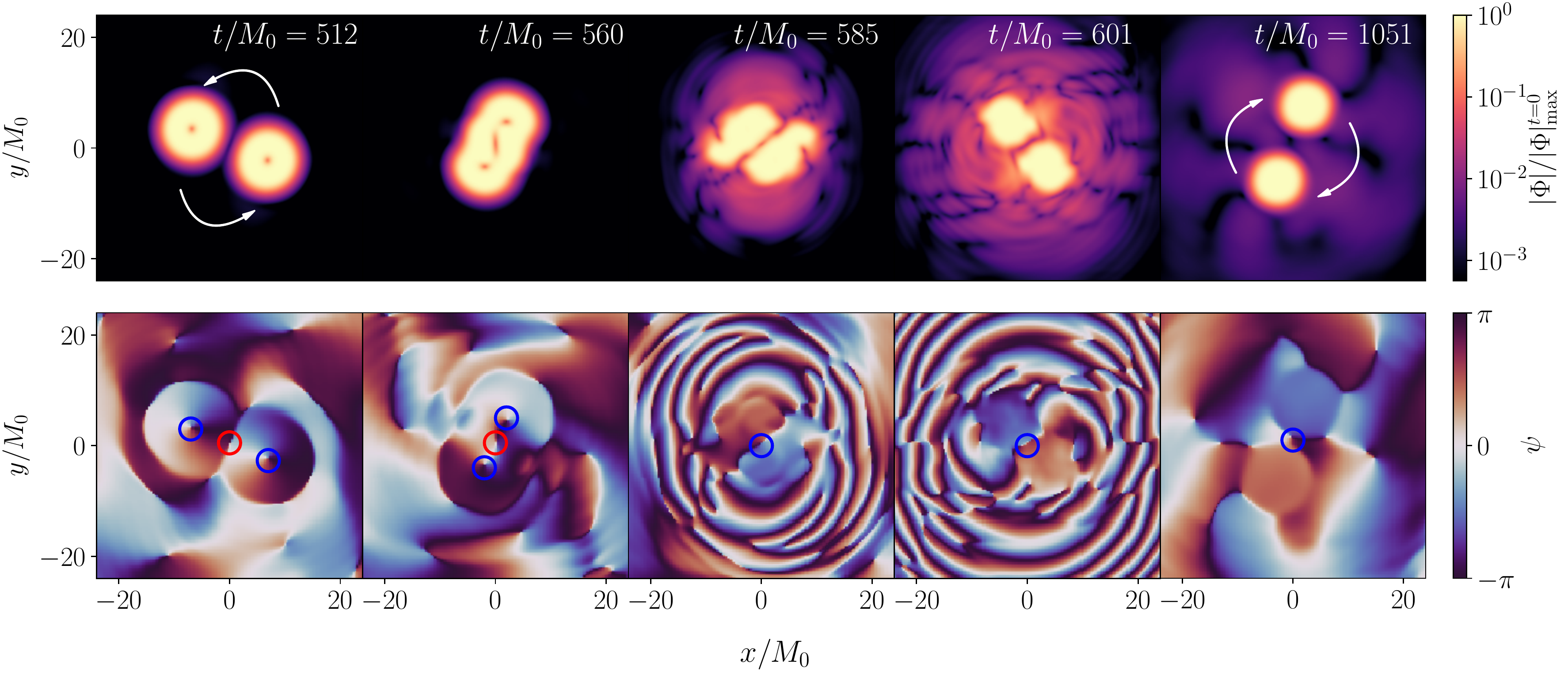}
\caption{Five snapshots of the equatorial plane of the spinning binary BS
    inspiral discussed in Sec.~\ref{sec:remnantmap}. \textit{(top row)} We show
    the magnitude of the scalar field in the equatorial plane, normalized by the
    initial maximum of the magnitude $|\Phi|_{\rm max}^{t=0}$. \textit{(bottom
    row)} We present the corresponding scalar phase $\psi\in(-\pi,\pi)$. We
    indicate the location of all relevant $q=-1$ vortices by red circles, and
    all relevant $q=1$ vortices by blue circles. (All vortex indices are
    measured with respect to the total angular momentum). The arrows indicate
    the direction of rotation of the inspiraling binary and the binary endstate. }
\label{fig:rotbbsexample}
\end{figure*}

We now illustrate the importance and utility of the above considerations with an 
example, in particular, of a binary BS where a single remnant is not formed at 
merger. We consider a spinning binary BS inspiral of identical $m_1=m_2=1$ BSs with
$\omega_1=\omega_2=0.4\mu$ in the solitonic scalar model with $\sigma=0.05$
[see \eqref{eq:solitonic}], of compactness $C=0.12$, at an initial coordinate
distance of $D=10M_0$ and initial phase-offset of $\alpha=0$. The orbit of the
binary is set up such that $|J_{\rm orb.}|<|J_1+J_2|$, with $J_{\rm orb.}$
anti-aligned with $J_1$ and $J_2$. Lastly, we note that, if the spinning binary
BS initial data is prepared with a vanishing phase-offset initially, the
symmetry of the binary fixes the location of a $q=-1$ vortex at the center of
mass of the system throughout the inspiral. 
Notice that each \textit{individual} spinning star has vortex number
$m_{1,2}=1$ (with respect to the total angular momentum). 

We can now employ the remnant map \eqref{eq:map}, and consider possible
outcomes of this inspiral. First, because of the vortex condition
\eqref{eq:vortexcondition} with $m_r=q+m_1+m_2=1$, no single non-spinning star can be
formed, as that would require $m_r=0$. Secondly, the total angular momentum of
the inspiraling system, $J_{\rm total}=J_1+J_2+J_{\rm orb.}$, is
\textit{smaller} than the angular momentum of a remnant spinning BS in the
$|m_r|=1$ family of solutions, as can be check explicitly using \eqref{eq:map}.
Therefore, using the remnant map with \eqref{eq:vortexcondition} and conditions \eqref{eq:kinematicconditions}, 
we can already rule out that the remnant
$\mathcal{B}_r$ is a single non-spinning BS or a $m_r \geq 1$ rotating BS 
[ $|m_r|>1$ would violate both the vortex condition and, likely, the stability condition
\eqref{eq:stabilitycondition}]. Hence, if the merger does not result in a black hole,
the only option is that the remnant
consists of at least \textit{two} stars. The reflection symmetry of the system
with respect to the center of mass (due to the choice of identical stars with
vanishing phase-offset in the initial data) suggests that the remnant is made up of 
an integer number of identical stars. Furthermore, as we show in Appendix~\ref{app:revisitNAI}, a single
isolated non-spinning BS is energetically favorable to a $m=1$ spinning BS of the same charge. Therefore,
two non-spinning remnant BSs are energetically favored over two spinning BSs.
Finally, due to the $m_r=1$ vortex of the remnant, the scalar phases of
each non-spinning stars will be exactly out of phase with respect to
each other. This implies they will be ``bouncing" off each other due to the
effective repulsion associated with this phase difference reviewed in Sec.~\ref{sec:scalarinteractions}, analogous
to the state found in \cite{Yoshida:1997nd,Palenzuela:2006wp}. Lastly, the two
stars (in a configuration similar to the DBS solution) will move around the
common center of mass (and the $m_r=1$ vortex) now with orbital angular
momentum $J_{\rm total}$, i.e., they orbit in the \textit{opposite} direction
to the inspiraling orbit. Hence, using \eqref{eq:map} we obtain a final state
for the system that satisfies all of the above conditions. Therefore, using the remnant
map \eqref{eq:map} and the four conditions listed above, we were able to
qualitatively predict large portions of the \textit{nonlinear} dynamics of the
binary system as well as the final remnant and its properties. Of course, we
cannot rule out that the final state is composed of more than two stars, which,
however, could be addressed in principle by considering, whether four stars as
the remnant $\mathcal{B}_r$ is favored over two stars. 

We confirm this picture by numerically evolving this spinning binary BS system,
as shown in \figurename{ \ref{fig:rotbbsexample}}. 
Before entering the regime of strong scalar interactions,
i.e., for times $t/M_0<500$, the phases of each spinning star oscillate around
the respective central vortex (marked in blue) roughly at the star's internal
frequency. Notably, a $q=-1$ vortex is present at the center of mass by
construction of the binary initial data (marked in red). From $t/M_0\approx
550$ to $t/M_0\approx 650$, the two stars interact nonlinearly, both
gravitationally and through the scalar self-interaction. During this
interaction, the two $m_{1,2}=1$ vortices of the two spinning stars merge with
the $q=-1$ vortex of the orbital angular momentum at the center of mass to form a
single $m_r=q+m_1+m_2=1$ remnant vortex fixed at the center of mass.
Furthermore, the nonlinear interaction of the two spinning stars result in the
formation of \textit{two} (approximately) non-spinning BSs. This addition of
vortex numbers can equivalently be understood using angular momentum
conservation discussed in Sec.~\ref{sec:scalarinteractions}: The spin angular
momenta of the two merging stars add to the \textit{oppositely oriented}
orbital angular momentum during the merger. The remnant stars are void of
spin-angular momentum, such that the remaining vortex $m_r$ must aligned with
the total (and now only orbital) angular momentum. Finally, the $m_r=1$ vortex
at the center of mass remains and dictates that the two remnant stars are
precisely out of phase. The latter can be seen in the last three snapshots of
\figurename{ \ref{fig:rotbbsexample}}. The outcome of the merger is a DBS solution
\cite{Yoshida:1997nd,Palenzuela:2006wp}, however, with non-zero orbital angular
momentum (implying the presence of the central $m_r=1$ vortex). Hence, we find
that this DBS state orbits around the central vortex with the remaining angular
momentum (the arrows in \figurename{ \ref{fig:rotbbsexample}} indicating the
sense of rotation about the center of mass). It is plausible that the final stationary state is a DBS solution with vanishing angular momentum (i.e., the system radiates the appropriate amount of energy and angular momentum to migrate towards a stationary DBS solution). 

This explicitly demonstrates both the utility of the remnant
map together with the kinematic, stability, and vortex conditions in predicting
the outcome of highly nonlinear mergers, as well as the important role
the scalar interactions, in particular, the scalar phase evolution and
vortex structures, play during the merger of binary BSs.

\section{Formation of rotating boson stars} \label{sec:formationofrotremn}

Finding a scenario where a rotating BS forms dynamically from the merger of two
non-rotating stars has been an open problem. In the past, numerous attempts and 
configurations were considered to form a rotating BS remnant from a binary inspiral,
most notably in Refs.~\cite{Palenzuela:2017kcg,Bezares:2022obu} (see also Refs.~\cite{Palenzuela:2007dm,Bezares:2017mzk,Sanchis-Gual:2018oui}). 
Here, we argue, that all of
these attempts violate one or several of the conditions needed to form rotating
BSs outlined in the previous section. In particular, we point out that 
the scalar phase and
vortex structure play a pivotal role in forming a persistent rotating BS
remnant. 

To show this explicitly, we proceed by applying the remnant map and set of conditions from the
previous section to this formation scenario in
Sec.~\ref{sec:rotremnformationcrit}, and demonstrate by numerical evolutions in
Sec.~\ref{sec:formationdynamics} that some appropriately chosen initial data
satisfying these conditions do in fact lead to a rotating remnant.  Hence,
for the first time, we find rotating BS remnants that form dynamically from the merger of
two non-spinning BSs. We then study the robustness of this formation mechanism
to variations in angular momentum and scalar phase in
Sec.~\ref{sec:formationrobustness}, discuss the implications of these
conditions on the characteristics of this new formation channel in
Sec.~\ref{sec:rotremnformationchannel}, and finally, in
Sec.~\ref{sec:partiallyspinning} analyze the merger remnant of a binary where
one BS is spinning and the other is non-spinning, showing that whether it is
rotating or non-rotating depends on the binary's mass-ratio. 

\subsection{Formation criteria \& parameter space} \label{sec:rotremnformationcrit}

\begin{figure*}[t]
\includegraphics[width=0.306\textwidth]{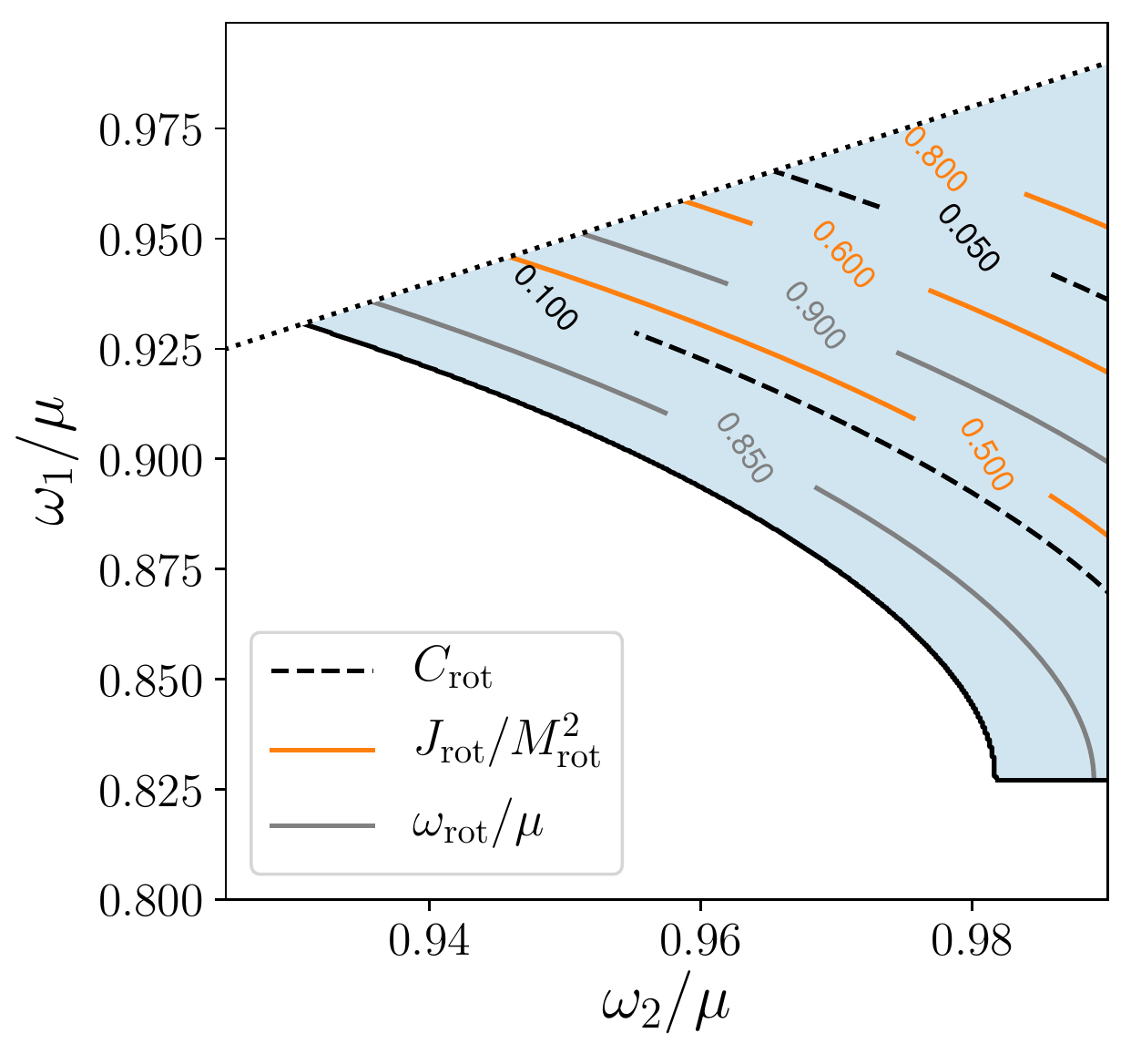}
\includegraphics[width=0.28\textwidth]{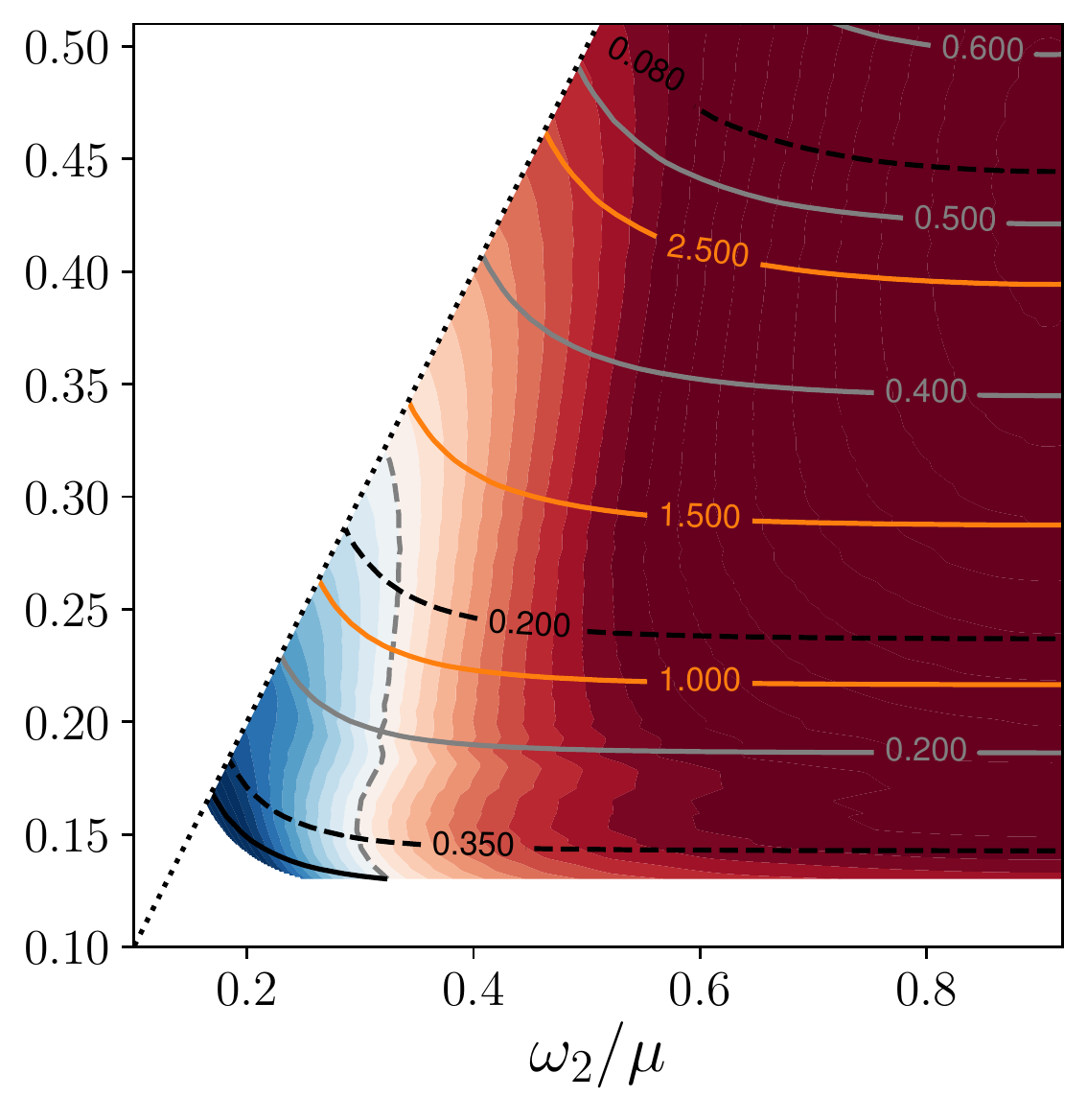}
\includegraphics[width=0.3635\textwidth]{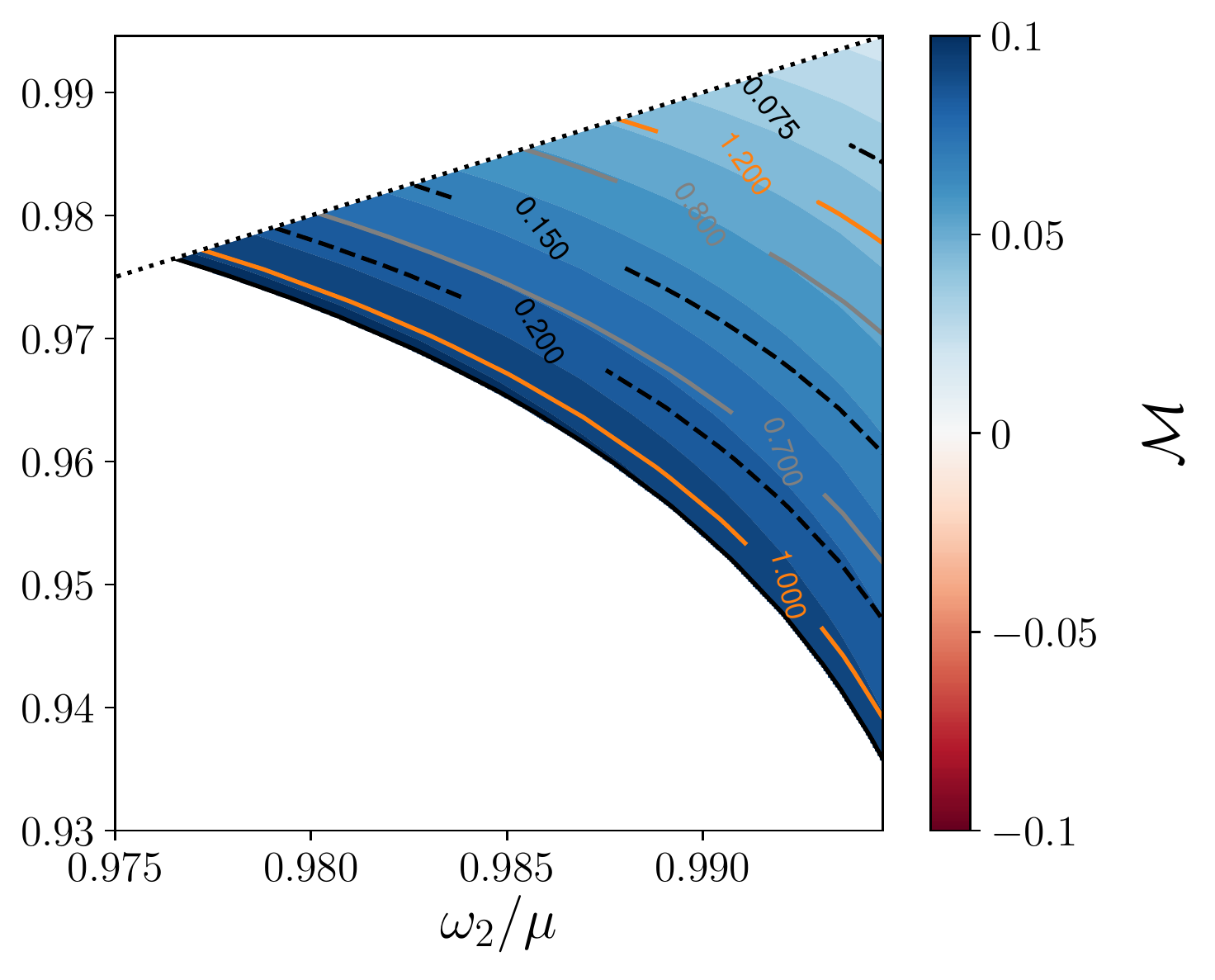}
\caption{We plot the properties (dimensionless angular momentum $J_{\rm
    rot}/M^2_{\rm rot}$, frequency $\omega_{\rm rot}$, and compactness $C_{\rm
    rot}=M_{\rm rot}/R_{\rm rot}$) of a $m=1$ rotating remnant star, assuming
    the remnant map \eqref{eq:map}, as a function of the initial
    non-spinning BS frequencies $\omega_1$ and $\omega_2$ (\textit{contour
    lines}). In addition, we show the normalized mass difference
    $\mathcal{M}=(M_1+M_2-M_{\rm rot})/(M_1+M_2)$ across the binary star
    parameter space (\textit{contour plot}). Notice, the plot is symmetric
    under the interchange $\omega_1\leftrightarrow \omega_2$, and we only consider
    the regime where $Q_1+Q_2<Q_{\rm rot}^{\max}$. \textit{(left)} The
    binary parameter space in the repulsive scalar model with
    $\lambda/\mu^2=10^3$, \textit{(middle)} the solitonic self-interactions
    with $\sigma=0.05$, and \textit{(right)} in the massive vector model
    without self-interactions. We explicitly restrict to only the radially
    stable Newtonian branches in the \textit{left} and \textit{right} panels, and
    the radially stable relativistic branch in the \textit{middle} panel. In the
    \textit{middle}, the dashed gray line indicates where $\mathcal{M}=0$.
    Notice, for a $\sigma=0.1$ solitonic scalar theory, no region with
    $\mathcal{M}>0$ exists. The non-axisymmetric linear instability found in
    Ref.~\cite{Sanchis-Gual:2019ljs} is likely absent in the \textit{right} panel;
    however, it is present in the \textit{middle} for all
    $\omega_{\rm rot}/\mu>0.5$, and may be present in the \textit{left} panel for 
    some solutions with $\omega_{\rm rot}/\mu<0.9$, as shown in Ref.~\cite{Siemonsen:2020hcg}.}
\label{fig:Parameterspace}
\end{figure*}

We begin by considering how the remnant map \eqref{eq:map} relates  
two non-spinning BS solution to a $m=1$ rotating BS solution, and in
what range the kinematic and stability conditions, defined in \eqref{eq:kinematicconditions} and \eqref{eq:stabilitycondition} (discussed in Sec.~\ref{sec:remnantmap}), respectively,
will be satisfied. We return to a discussion of the vortex condition below.
To that end, we consider three
different families of non-rotating stars: \textit{(i)} a repulsive family of
BSs with $\lambda/\mu^2=10^3$, \textit{(ii)} a solitonic set of BSs with
coupling $\sigma=0.05$, and \textit{(iii)} a family of non-rotating PSs
(without vector self-interactions). For each of these families of solutions, a
corresponding family of $m=1$ rotating stars exists. We label the frequencies
of the constituents of the initial non-spinning binary by $\omega_1$ and
$\omega_2$ and choose, without loss of generality, that $\omega_2\geq\omega_1$.
For each combination of stationary stars $(\omega_1,\omega_2)$ 
we use the charge conservation mapping \eqref{eq:map} combined with the one-to-one charge-frequency
relation, $Q_r(\omega_r)\rightarrow Q_{m=1}(\omega_{\rm rot})$ for
corresponding $m=1$ family of rotating BSs (and PSs) in each of the three
considered models to map all properties of the two stars in the
initial binary system into the properties of a single \textit{potential} $m=1$
rotating remnant star. 

In \figurename{ \ref{fig:Parameterspace}}, we show the results of the above
constructed mapping for all three theories. Using this, we are able to apply
the kinematic conditions to isolate the parts of the initial non-spinning
binary parameter space that are suitable for the formation of a rotating remnant BS.
First, we consider condition 1. in \eqref{eq:kinematicconditions} (the
kinematic condition). In order for the formation of a rotating remnant to be
energetically favorable from a non-spinning binary, we must require
\begin{align}
\mathcal{M}=\frac{M_1+M_2-M_{\rm rot}}{M_1+M_2}>0.
\label{eq:Mconditionforrotremn}
\end{align}
Here $M_1$, $M_2$, and $M_{\rm rot}$ are the ADM masses of the stationary
isolated non-spinning stars and the rotating star, respectively. From
\figurename{ \ref{fig:Parameterspace}}, we can deduce that for models
\textit{(i)} and \textit{(iii)}, the entire initial binary parameter space
favors the formation of a rotating remnant on energetic grounds alone, while
only the highly relativistic regime of the binary parameter space of model
\textit{(ii)} has $\mathcal{M}>0$. This narrows down the possible initial
binary configurations that might lead to the formation of a rotating remnant.
Furthermore, for the formation, we must require that the orbit of the inspiral
contains sufficient angular momentum, i.e., condition 2. in
\eqref{eq:kinematicconditions} dictates that
\begin{align}
|J_{\rm orb.}|>|J_{\rm rot}|.
\label{eq:Jconditionforrotremn}
\end{align}
This restricts the binary orbit to a subset of all possible inspirals. In fact,
as we discuss below in Sec.~\ref{sec:formationrobustness}, it is not necessary for 
\eqref{eq:Jconditionforrotremn} to be strictly satisfied, and furthermore
there is also an upper bound on $J_{\rm orb.}$ for the successful
formation of a rotating remnant star, which we determine
empirically below for an example inspiral. 

According to the stability condition \eqref{eq:stabilitycondition}, 
the remnant rotating BS (or PS) must be a linearly (and non-linearly) stable solution.
Despite the recent progress in understanding the stability properties of these
rotating solutions, this is a subtle point, and we simply state that so far, no linear
or nonlinear instability is known to exist in the parts of the rotating BS and PS
parameter space we are interested in (see \figurename{ \ref{fig:Parameterspace}} 
for an indication where known instabilities may be active). Finally, the last condition---the vortex
condition---defined in \eqref{eq:vortexcondition}, dictates that at the point
of contact of the non-spinning binary, there must exist a $|q|=1$ vortex in the 
phase in the vicinity of the center of mass of the system. This will be determined
by the relative phases of the binary constituents. With these restrictions in
hand, we are now able to explicitly determine whether a rotating remnant star is
formed dynamically, when all the above conditions are met, in the next section.

\subsection{Formation dynamics} \label{sec:formationdynamics}

\begin{table}[b]
\begin{ruledtabular}
\begin{tabular}{c|c c c c c c c}
Model & $\omega_{\rm sph}/\mu$ & $J_0/Q_0$ & $x_0/M_0$ & $y_0/M_0$ & $v_x$ & $v_y$ & $\alpha$ \\
\hline \hline
repulsive & $0.95$ & $2.12$ & $\pm 15$ & $\pm 10$ & $\mp 0.06$ & $\pm 0.02$ & $\pi$ \\ 
\hline 
solitonic & $0.25$ & $1.16$ & $\pm 20$ & $\pm 3$ & $\mp 0.3$ & $0$ & $\pi$ \\ 
\end{tabular} 
\end{ruledtabular}
\caption{The properties of non-spinning binary BS initial data leading to the
    formation of a rotating BS remnant. The two stars are identical, i.e.,
    $\omega_1=\omega_2=\omega_{\rm sph}$, with initial phase offset $\alpha$,
    are positioned at coordinate locations $x_0$ and $y_0$ (upper signs refer
    to the first star), and have boost velocities $v_x$ and $v_y$. The initial
    orbital angular momentum is $J_0$, the ADM mass is $M_0$, and $Q_0$
    refers to the initial $U(1)$-charge of the binary. The couplings are
    $\lambda/\mu^2=10^3$ and $\sigma=0.05$ for the repulsive and solitonic
    scalar models, and the compactness of these stars is $C=0.037$ and $C=0.13$
    in each of the models, respectively.}
\label{tab:rotBBSidproperties}
\end{table}

\begin{figure*}
\includegraphics[width=0.99\textwidth]{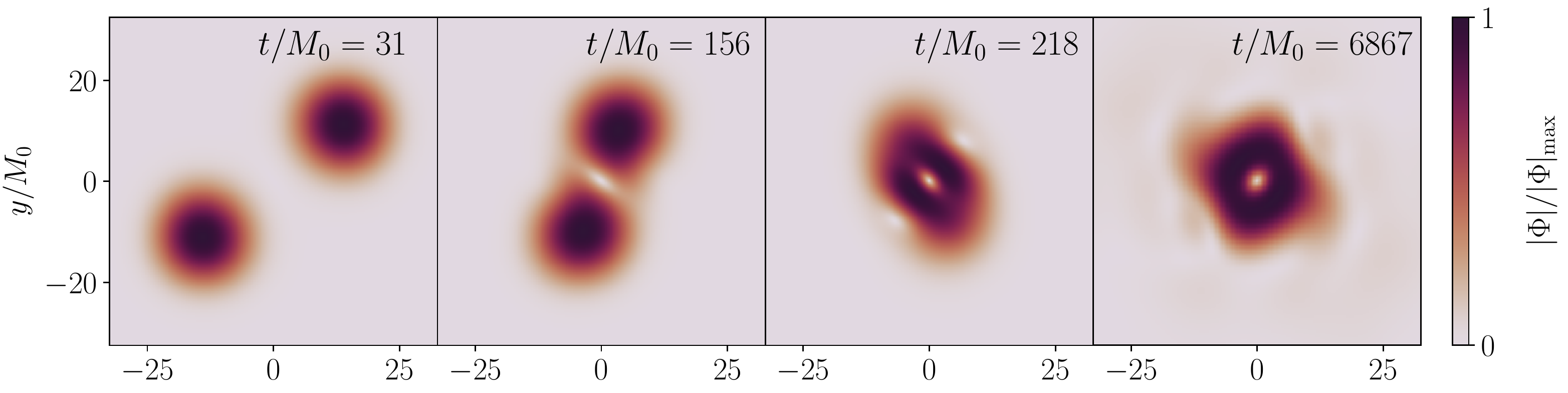}
\includegraphics[width=0.99\textwidth]{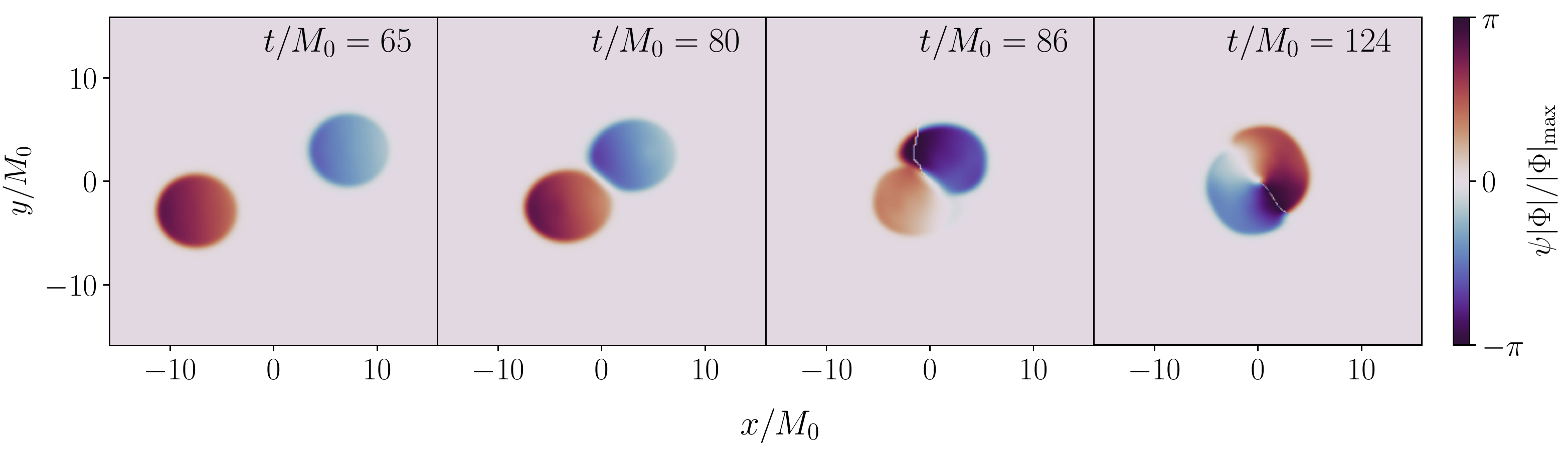}
\begin{flushleft}
\includegraphics[width=0.465\textwidth]{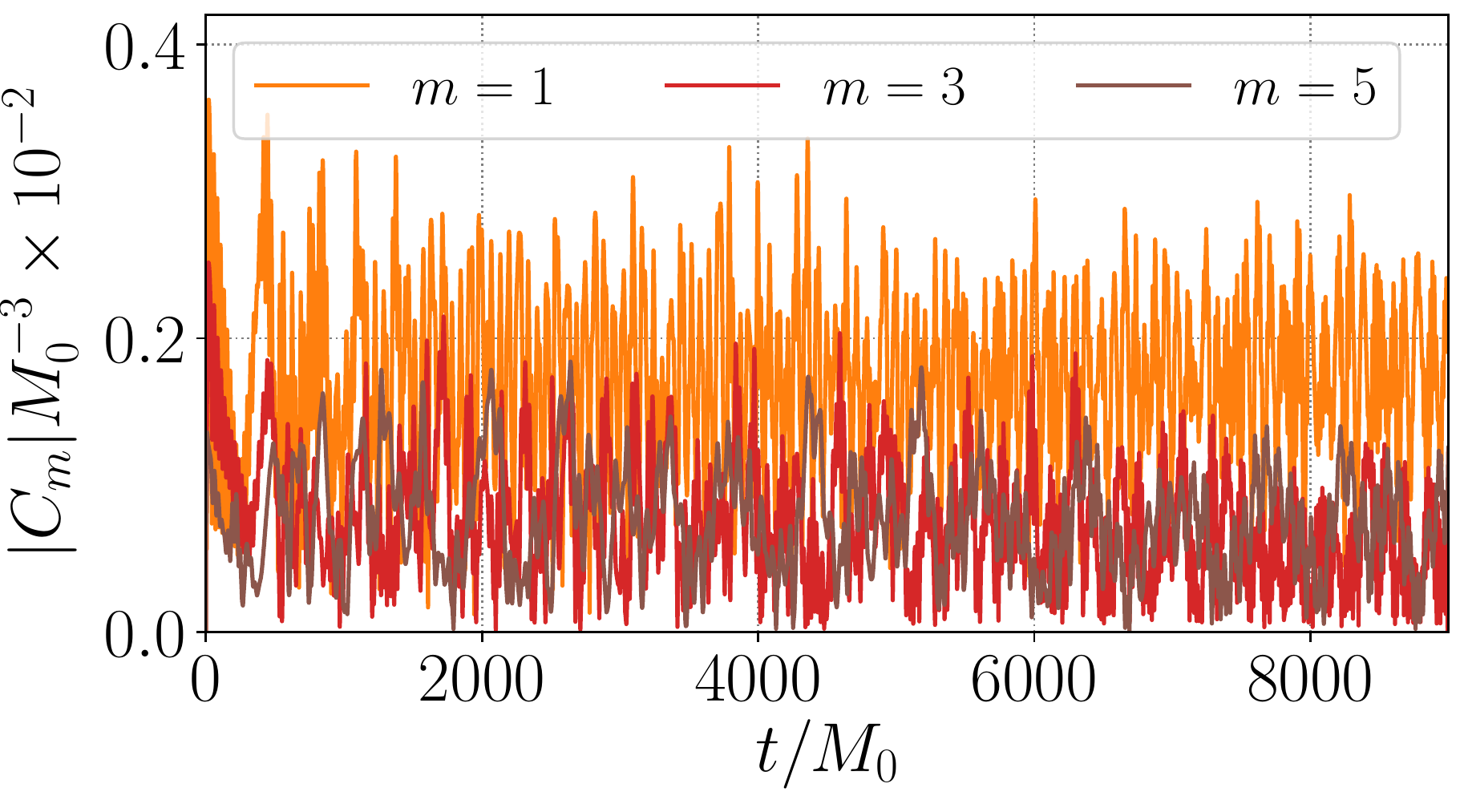}
\includegraphics[width=0.465\textwidth]{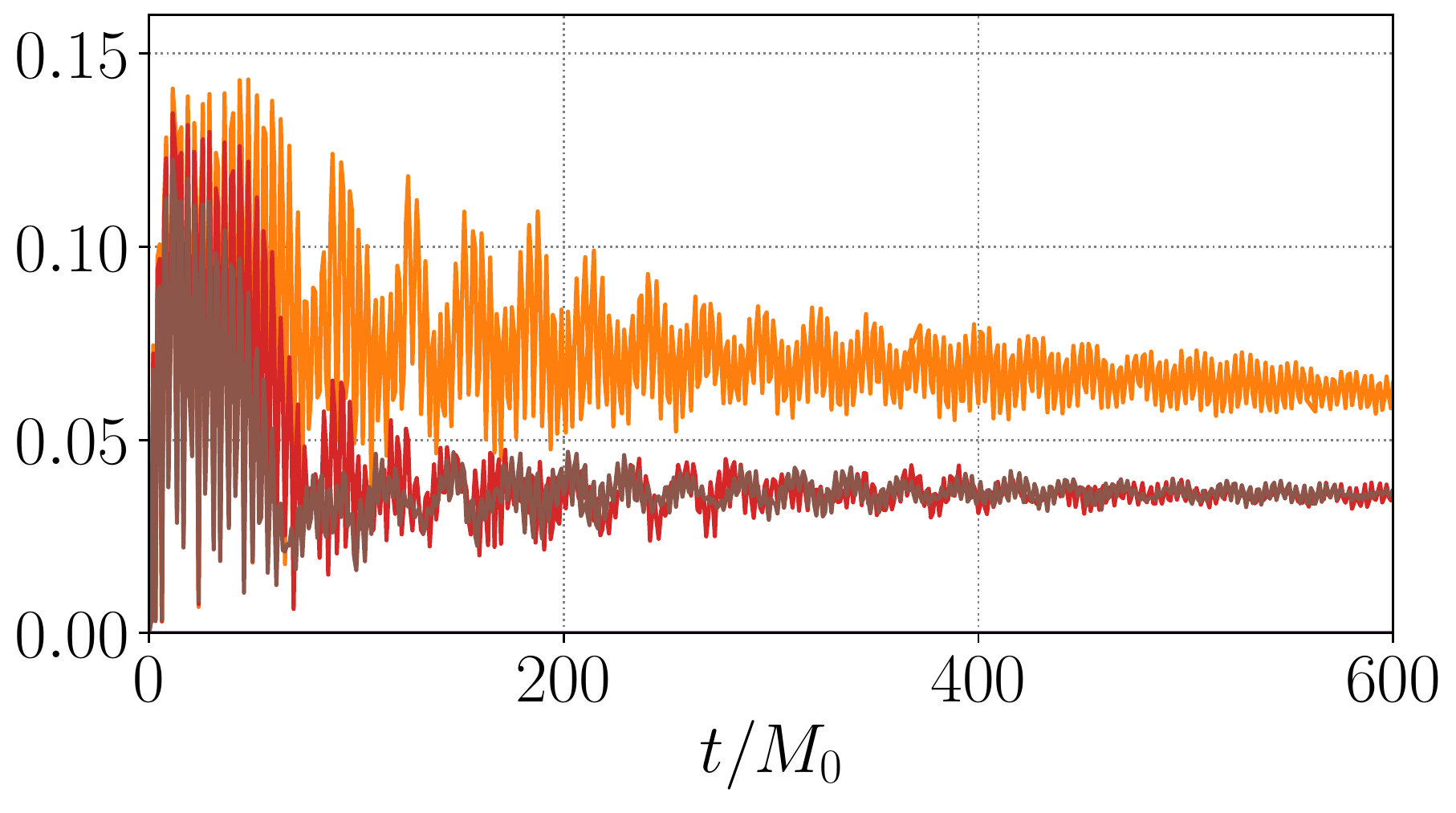}
\end{flushleft}
\caption{We illustrate the dynamics leading to the formation of $m=1$ rotating
BS remnants from the merger of two non-spinning BSs. \textit{(top row)} Here, we
focus on the binary star in the repulsive scalar model described in \tablename{
\ref{tab:rotBBSidproperties}}. We show the magnitude of the scalar field
normalized by the maximum magnitude $|\Phi|/|\Phi|_{\rm max}$ in the equatorial
plane at four different times during merger. \textit{(middle row)} We plot the
product of the scalar phase $\psi$ and the maximum normalized scalar field
magnitude $\psi|\Phi|/|\Phi|_{\rm max}$ in the equatorial plane at four
different times during the merger. Here, we show the binary in the solitonic
scalar theory with properties given in \tablename{
\ref{tab:rotBBSidproperties}}. \textit{(bottom row)} The evolution of odd-$m$
components of the azimuthal decomposition \eqref{eq:azimuthaldecomp} of both binary
mergers (the repulsive binary on the left, and the solitonic binary on the
right). Notice, in the case of the solitonic binary, we show in
Appendix~\ref{app:idandnumevo} that the $m=3$ and $5$ modes are dominated by
truncation error at late times and converge to zero. The even--$m$ modes are
negligible throughout the evolution.}
\label{fig:Formationofrotremn}
\end{figure*}

Consulting the remnant map shown in \figurename{ \ref{fig:Parameterspace}}, we
begin by constructing non-spinning binary BS initial data that satisfies the
conditions of the previous section. Here, we focus exclusively on the repulsive
and solitonic scalar models, with $\lambda/\mu^2=10^3$ and $\sigma=0.05$,
respectively, and return to a discussion of the vector case below. In order to
ensure that the vortex condition \eqref{eq:vortexcondition} is satisfied at the point of contact, we
restrict to a binary made up of identical stars, i.e.,
$\omega_1=\omega_2=\omega_{\rm sph}$, but with opposite initial phase,
$\alpha=\pi$. The remaining characteristics of the initial data are summarized
in \tablename{ \ref{tab:rotBBSidproperties}}. The frequency of the non-spinning
stars are chosen to satisfy the kinematic and stability conditions.
Specifically, comparing the properties presented in \tablename{
    \ref{tab:rotBBSidproperties}} with \figurename{ \ref{fig:Parameterspace}},
it is evident that these binaries satisfy $\mathcal{M}>0$, and hence condition
\eqref{eq:Mconditionforrotremn}, and are in regions of the rotating BS
parameter space without any known instabilities. The initial positions and
velocities are set up in order to force the binary on a highly elliptical orbit
and merge during the first encounter. The boost velocities are chosen to
achieve sufficient angular momentum to form the rotating remnant star predicted
by the remnant map: $J_{\rm rot}/M_{\rm rot}^2=0.56$ and $J_{\rm rot}/M_{\rm
rot}^2=0.95$ for the stars in the repulsive and the solitonic scalar models,
respectively. This ensures that condition \eqref{eq:Jconditionforrotremn} is
satisfied at the point of contact. Finally, the center of mass of these
binaries exhibits a $q=1$ vortex throughout the evolution (measured with
respect to the orbital angular momentum), by virtue of $\omega_1=\omega_2$ and
initial phase-offset $\alpha=\pi$, hence, satisfying the vortex condition
\eqref{eq:vortexcondition}. The details of the numerical construction of the
constraint satisfying initial data are given in Appendix~\ref{app:idandnumevo}.

In \figurename{ \ref{fig:Formationofrotremn}}, we show a few snapshots from the numerical evolution of the two sets of binary BSs constructed above, as well as the time evolution of the azimuthal mode decomposition of the real part, $\Phi_R=\text{Re}(\Phi)$, of the scalar field around the center of mass, and along the angular momentum direction
\begin{align}
C_m=\int d^3x \Phi_R e^{im\varphi}.
\label{eq:azimuthaldecomp}
\end{align}
In the early stages of the evolution, $t/M_0<80$ and $t/M_0<150$, for the
solitonic and repulsive cases, respectively, the phases of each stars evolve approximately
linearly in time. At the point of contact, the scalar and gravitational
interactions are highly nonlinear. During this merger phase, the scalar matter
attaches to the vortex line at the center of mass, and the system rings down to
a state characterized by a single $m=1$ mode in the scalar field---a rotating
BS remnant. All other modes decay away over time. From the harmonic
time-dependence of $\text{Re}(C_1)$, we obtain the late-time rotating remnant
frequency of $\omega_{\rm rot} M_0=0.78$ in the solitonic case. This is well
approximated by the remnant map shown in \figurename{
    \ref{fig:Parameterspace}}, which predicts $\omega_{\rm
rot}=0.75/M_0=0.22\mu$, with the difference being attributable to  the
(Richardson-extrapolated) charge loss of $\Delta Q/Q_0=0.04$ occurring during
merger, as shown explicitly in Appendix~\ref{app:idandnumevo}.

This explicitly demonstrates that, indeed, rotating BSs can be formed from the
merger of two non-spinning BSs, given the conditions in
Sec.~\ref{sec:rotremnformationcrit} are met. In particular, in the following
section, we show explicitly that (at least approximately) satisfying the vortex condition is crucial for
successfully forming the rotating remnant. In the case of the formation of
rotating PSs (instead of BSs), the vector phase may play a similarly
important role. However, these solutions also possess an intrinsic
preferred direction (due to the vector field), and hence, the merger dynamics
might not only be governed by the vector phase, but also the vector direction.

\subsection{Robustness of formation mechanism} \label{sec:formationrobustness}

Having demonstrated the dynamics leading to
the formation of rotating BSs from the merger of two non-rotating stars in several cases,
we now analyze the robustness of this formation channel to variations in both
the total angular momentum and the phase offset of the binary. To that end, we
focus on the binary BS in the solitonic scalar model we found to form a
rotating BS remnant in the previous section, and vary its initial angular
momentum and phase offset, while keeping all other parameters, in particular,
the $U(1)$-charge, fixed. We show that there exists a set of binary BS initial
data with non-zero measure that form a rotating BS remnant, 
demonstrating the robustness of the formation mechanism.

\subsubsection{Variation of angular momentum}

\begin{figure}[t]
\includegraphics[width=0.48\textwidth]{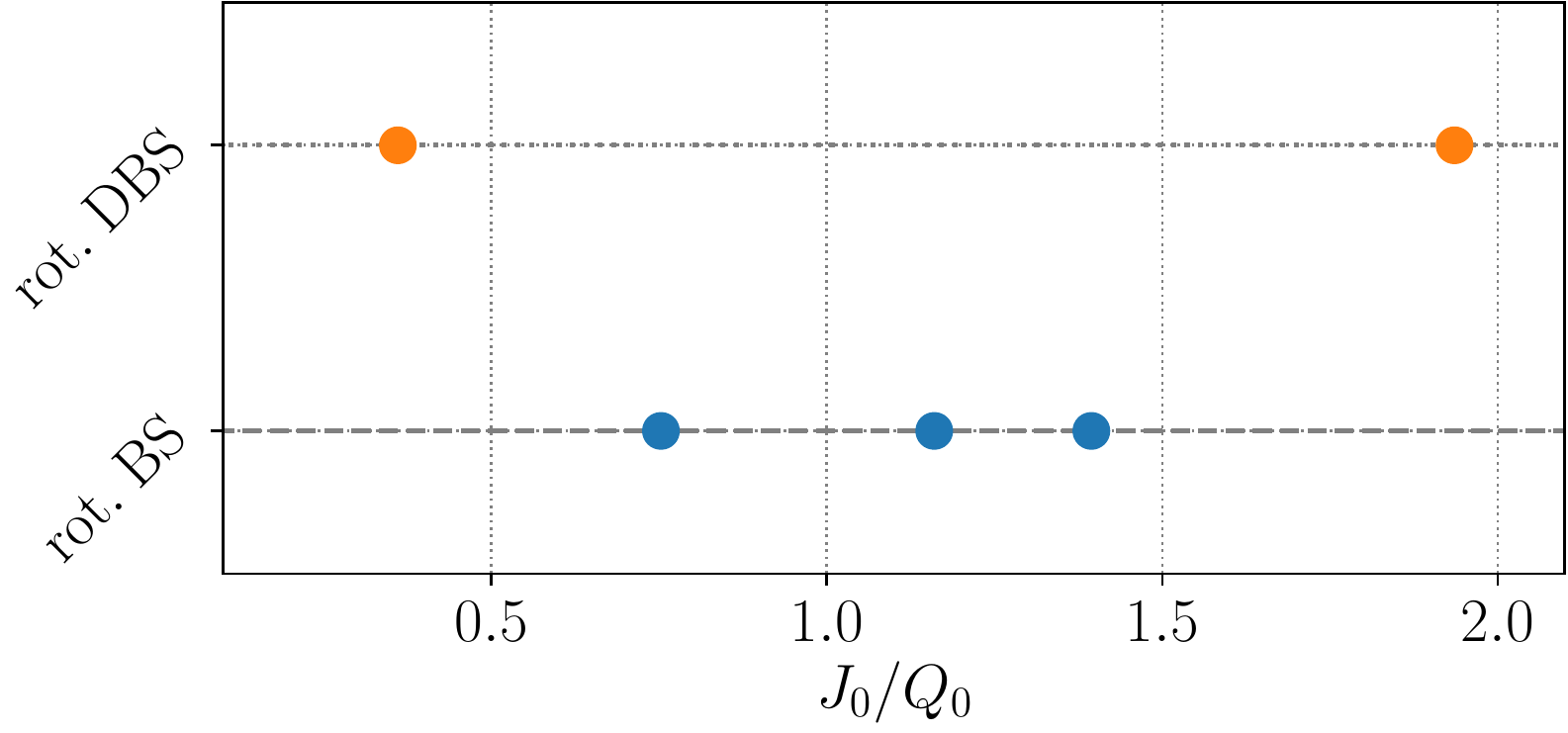}
\caption{We classify the remnants of the mergers of a non-spinning binary BS
    into rotating DBS and rotating BS solutions as a function of the total
    (initial orbital) angular momentum $J_0$. The angular momentum is
    normalized by the $U(1)$-charge of the binary $Q_0$.
    }
\label{fig:angularmomrobustness}
\end{figure}

To quantify the robustness of the formation channel to variations in the binary
angular momentum, we perform a series of simulations of the solitonic binary
BS specified in \tablename{ \ref{tab:rotBBSidproperties}} with varying initial
boosts $|v_x|\in\{0.1,0.25,0.3,0.35,0.45\}$. This changes the initial angular
momentum $J_0$ away from the value of the binary specified in \tablename{
    \ref{tab:rotBBSidproperties}}. We
evolve these sets of initial data through merger (see
Appendix~\ref{app:angularmomentumvariation} for snapshots of the evolution). In
\figurename{ \ref{fig:angularmomrobustness}}, the remnant of the merger is
classified as either a rotating DBS solution (consisting of two orbiting, non-spinning stars, as described in
Secs.~\ref{sec:scalarinteractions} and~\ref{sec:remnantmap}), or a $m=1$ rotating
BS solution. While both remnant classes posses a $|q|=1$ vortex at the center
of mass, the scalar field magnitude morphology is distinct; rotating BS
solutions are stationary and exhibit toroidal surfaces of constant scalar field magnitude, in
contrast to rotating DBS solutions, which exhibit two disconnected surfaces of
constant scalar field magnitude\footnote{A more robust method to identify the
remnants as rotating BSs (opposed to rotating DBSs) is to, for instance, 
explicitly check the consistency of $J_{\rm rem}=mQ_{\rm rem}$ of the remnant
solution at late times (satisfied only by rotating BSs).}. Notice also, the rotating DBS
remnant is \textit{not} a stationary solution, as it continues to radiate
energy and angular momentum. Consulting \figurename{
    \ref{fig:angularmomrobustness}}, a variation of the initial angular
momentum of up to $\Delta J_0/Q_0\approx \pm 50 \%$ still leads to the prompt formation of a rotating BS remnant\footnote{Note, the initial
angular momentum $J_0$ is determined using the \textit{gauge-dependent} ``ADM"
angular momentum of the initial data.}. However, if $|\Delta J_0|/Q_0$ is above some critical threshold (shown in \figurename{
    \ref{fig:angularmomrobustness}}), then the system settles temporarily into a rotating DBS solution
with long orbital period if $\Delta J_0<0$, and short orbital period if $\Delta
J_0>0$. If both the initial angular momentum $J_0$ and total charge $Q_0$ of
the binary system were conserved through merger, then only the initial data
satisfying $J_0=Q_0$ could form a $m=1$ rotating remnant. However, due to
scalar and gravitational radiation, both angular momentum and $U(1)$-charge may
be carried away from the system resulting in the formation of a rotating BS
remnant for initial data with a range of initial angular momenta. For instance, based on the scalar field morphology, we find the binary configuration with $J_0/Q_0\approx 0.75$ (shown in \figurename{ \ref{fig:angularmomrobustness}}) to settle into rotating BS after merger (see also \figurename{ \ref{fig:angmomvariationevolution}}). This highly perturbed remnant plausibly continues to emit residual energy and $U(1)$-charge, approaching $J/Q\approx 1$ at late times. Furthermore, the rotating DBS remnant with $J_0/Q_0\lesssim 0.5$ may settle into a non-rotating DBS solution as suggested in Sec.~\ref{sec:remnantmap}, while the spinning DBS remnant with $J_0/Q_0\gtrsim 1.5$ could plausibly settle into either a non-rotating DBS solution or a rotating BS as the system radiates angular momentum. Ultimately, the
formation mechanism is robust against variations of binary orbital angular
momentum to the degree shown in \figurename{ \ref{fig:angularmomrobustness}}.

\subsubsection{Variations of scalar phase}
Thus far, we have tested the robustness of the formation of a rotating
BS from merger by varying the initial angular momentum of the binary BS's orbit, while
keeping the initial scalar phase-offset between both stars fixed, i.e.,
$\alpha=\pi$. In this section, we fix $J_0/Q_0=1.16$ and instead vary the
initial phase-offset between the two solitonic stars $\alpha$.
Hence, we considering binary mergers in the solitonic scalar model with
parameters given in \tablename{ \ref{tab:rotBBSidproperties}}, but with different initial
phases $\alpha/\pi\in [0,63/64]$. We demonstrated in
Sec.~\ref{sec:formationdynamics} that the $\alpha/\pi=1$ case results in the
formation of a $m=1$ rotating BS remnant. The $\alpha=0$ binary promptly
leads to the formation of a \textit{non-rotating} BS remnant, shedding all 
its orbital angular momentum. Hence, there exists a 
critical value, $0<\alpha_{\rm crit}<\pi$, for which the remnant is 
marginally either a non-spinning or a rotating BS remnant.

\begin{figure}
\includegraphics[width=0.484\textwidth]{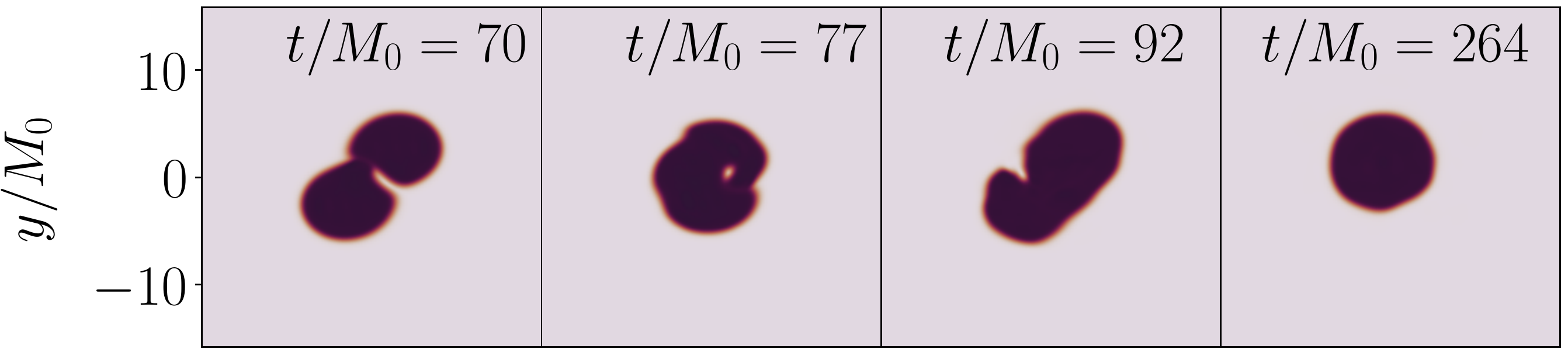}
\includegraphics[width=0.482\textwidth]{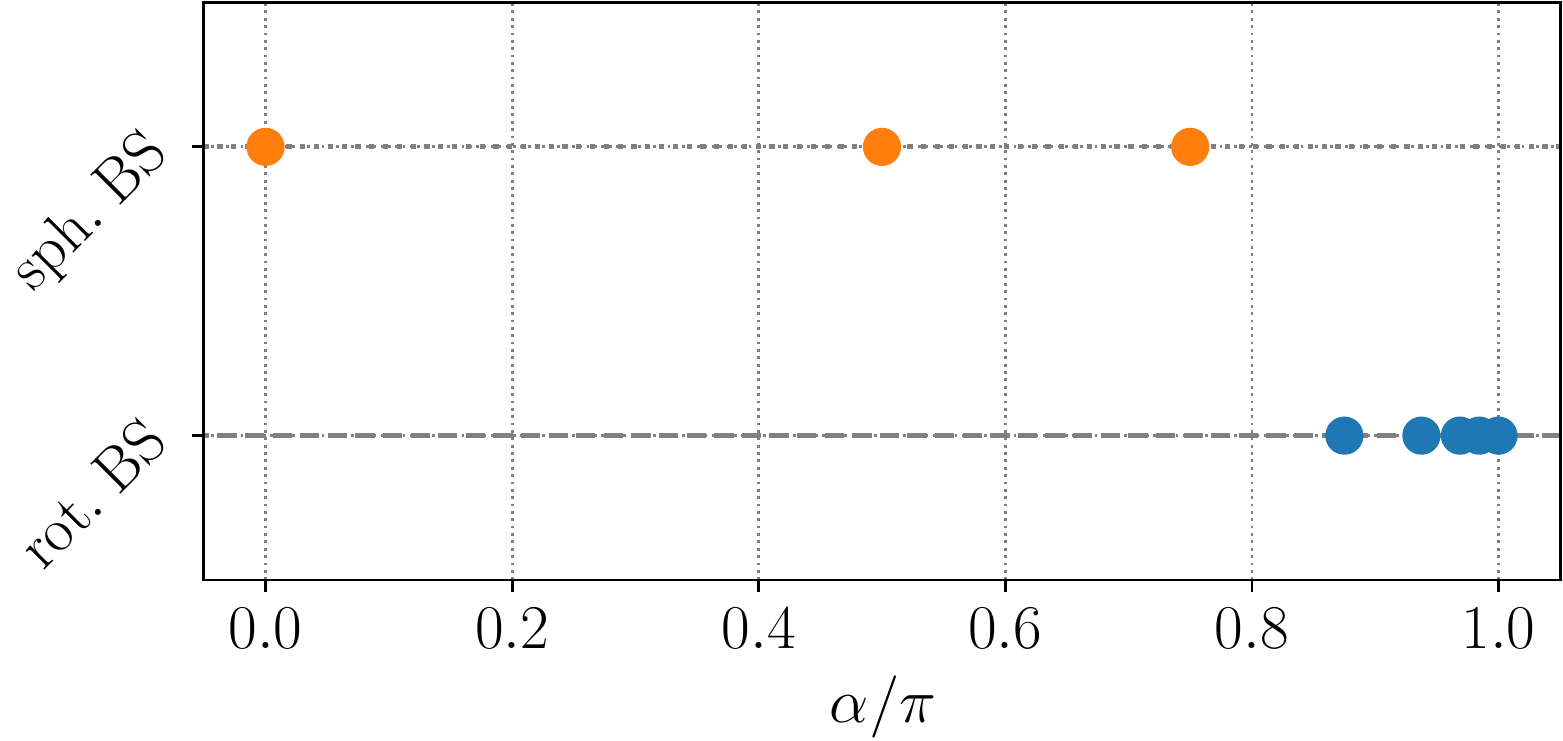}
\caption{\textit{(top)} We show the normalized magnitude of the scalar field $|\Phi|/|\Phi|_{\rm \max}^{t=0}$ in the equatorial plane at four different times during the evolution of the binary BS merger with parameters given in \tablename{ \ref{tab:rotBBSidproperties}}, but with initial phase-offset $\alpha/\pi=3/4$. \textit{(bottom)} We classify the remnant solution of binary BSs with different values of the initial phase $\alpha$ into spherical or rotating ($m=1$) BSs. We consider $\alpha/\pi\in\{1,63/64,31/32,15/16,7/8,3/4,1/2,0\}$.}
\label{fig:phasevariation}
\end{figure}

In the bottom panel of \figurename{
\ref{fig:phasevariation}}, we classify the remnants of the merger of the
sequence of these binary BSs with varying initial phase into
spherical and rotating BSs. We find all binaries with $\alpha/\pi\geq 7/8$
form a $m=1$ rotating remnant at late times, while all binaries with
$\alpha/\pi\leq 3/4$ result in a non-rotating BS
remnant\footnote{As another illustration of this type of behavior, we also considered the binary BS in the repulsive
scalar model characterized in \tablename{ \ref{tab:rotBBSidproperties}}, with
initial phase-offset $\alpha/\pi=7/8$; this binary tends towards a non-rotating
BS remnant at late times after merger.}. 
We find all remnant rotating
stars persist, without further sign of instability, for the remainder of the
evolutions of length $t\gtrsim 1000 M_0$.\footnote{We note 
that at lower numerical resolutions, the rotating remnant formed
when $\alpha/\pi\geq 7/8$ 
appears to have a growing perturbation that eventually ejects the central vortex from the star.
However, this behavior disappears at sufficiently high resolutions. Details 
can be found in Appendix~\ref{app:vortexejinstability}.
}
Therefore, the formation scenario discussed in
Sec.~\ref{sec:rotremnformationcrit} does not require the phase offset of the
merging binary to be fine-tuned, and is robust up to a phase variation of
$|\Delta\alpha|/\pi\approx0.12$ around $\alpha/\pi=1$. In the top panel of \figurename{ \ref{fig:phasevariation}}, we show a few
snapshots of the evolution of the binary with $\alpha/\pi=3/4$. In the first
panel, the vortex is located to the right of the center of mass during the
initial contact of the stars. The vortex then traverses the merging object in
the second panel and is ejected in the third panel. The final state (fourth
panel) is a perturbed non-rotating BS. This behavior is representative of all cases with $1/2\leq \alpha/\pi\leq 3/4$. In the $\alpha=0$ limit, any vortex present due to the non-vanishing orbital angular momentum stays \textit{outside} the remnant star altogether.

\subsection{Formation channel} \label{sec:rotremnformationchannel}

In the previous sections, we demonstrated that rotating BSs can be 
formed from the merger of two non-spinning stars without fine tuning of the binary
parameters for several examples. In this section, we consider more generally
for what systems the approximate conditions for the formation of a
rotating BS remnant listed in Sec.~\ref{sec:rotremnformationcrit} are satisfied,
and its implications for different formation channels for rotating BSs.
We focus exclusively on binary mergers, and do not discuss, for example, formation
of a rotating star from the collapse of coherent rotating cloud (as in, e.g., Ref.~\cite{Sanchis-Gual:2019ljs}).

First, from 
\figurename{ \ref{fig:Parameterspace}}, it is clear that the energy condition \eqref{eq:Mconditionforrotremn}
is not restrictive for the repulsive scalar model and PSs (since
$\mathcal{M}>0$ across the parameter space). However, this condition does restrict 
the solitonic scalar theory to a small
$M_1/M_2\sim\mathcal{O}(1)$ region in the highly relativistic part of the stable
branch of the families of stars. Therefore, while in the former two models, the 
formation of a rotating star is possible in the entire parameter space,
including the Newtonian limit, the situation is more complex in the solitonic
scalar model. There, a spherical star would first have to be formed in the
relativistic part of the family of BSs (i.e., in the region of parameter space
in \figurename{ \ref{fig:Parameterspace}} where $\mathcal{M}>0$). If a star is formed on the Newtonian branch of the non-spinning solutions within the solitonic scalar model, then it can move towards the relativistic branch by gaining mass (e.g., through a sequence of binary BS mergers), which generically leads to a more compact, i.e., more relativistic, remnant. Once these stars have migrated to the relativistic branch, the formation of a rotating BS remnant is possible from a relativistic binary with mass-ratio close to unity.

Secondly, the stability condition \eqref{eq:stabilitycondition} is the most
restrictive, as all known $|m|>1$ rotating BSs are linearly unstable
\cite{Sanchis-Gual:2019ljs,Siemonsen:2020hcg,Dmitriev:2021utv}, allowing only
the formation of $m=1$ rotating BS remnants in scalar models \textit{with}
self-interactions (i.e., in the limit of vanishing scalar self-interactions,
all $m=1$ BSs are likely linearly unstable). Furthermore, in the case of attractive
scalar self-interactions (i.e., the solitonic scalar model), even the $m=1$ BSs
are linearly unstable in the Newtonian limit \cite{Dmitriev:2021utv} and
unstable to the ergoregion instability in the relativistic limit
\cite{friedman1978,Moschidis:2016zjy}, implying that in the solitonic model
\textit{only} those non-spinning stars in the relativistic part of the
parameter space may form rotating BS remnants \textit{without} ergoregions
during merger. In the case of the repulsive scalar model (with sufficiently
strong self-interactions), all $m=1$ rotating BSs below the maximum mass of the
family of solutions are linearly stable
\cite{Siemonsen:2020hcg,Dmitriev:2021utv}. The $m=1$ rotating PSs were shown to
be linearly stable using numerical simulations in \cite{Sanchis-Gual:2019ljs},
suggesting that also rotating PSs may be formed across a large part of the
parameter space. The $m=1$ rotating solutions in the repulsive scalar models
and the massive vector theory exhibit ergoregions only near the maximum mass
and in the strong coupling limit, only marginally affecting the parameter space in which the
formation of a rotating remnant is possible.

\begin{figure}
\includegraphics[width=0.49\textwidth]{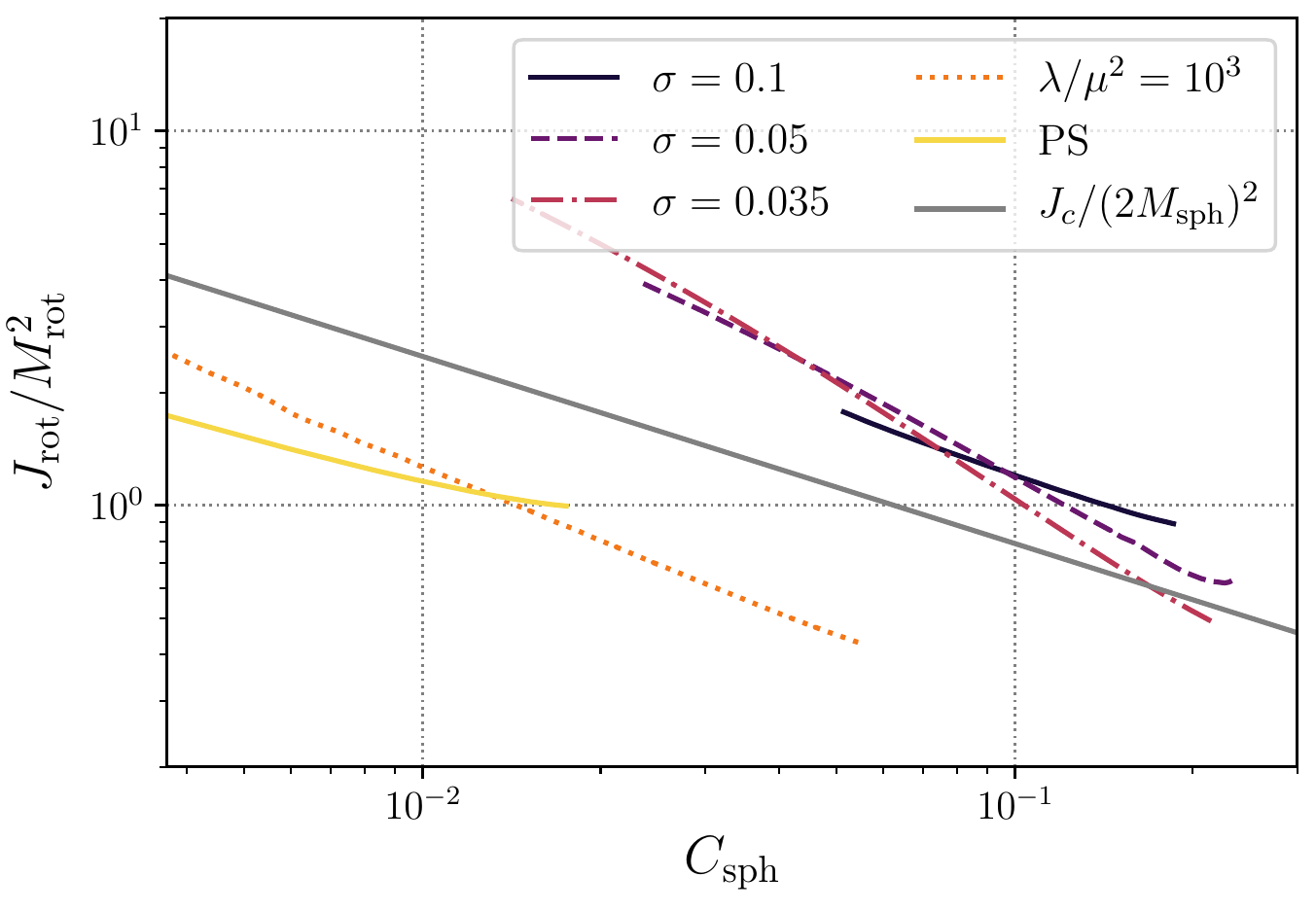}
\caption{As a function of the compactness $C_{\rm sph}$ of two identical
    non-rotating stars in various models and couplings, we compare the
    dimensionless angular momentum of the corresponding $m=1$ rotating BS and PS
    solution obtained using the remnant map \eqref{eq:map}. In particular, we
    compare three families of stars in the solitonic scalar model of coupling
    strength $\sigma$ with the family of PSs, scalar BSs in the repulsive
    scalar model (labelled with its coupling $\lambda$), and the Newtonian
    quasi-circular angular momentum at the point of contact $J_c$, derived in
    the main text. We focus only on the solution branches below the maximum
    mass. Notice, however, that not all the solitonic cases that are plotted lie in the stable
    part of the $m=1$ rotating BS parameter space.
    }
\label{fig:angularmomofrotremn}
\end{figure}

Lastly, the angular momentum condition \eqref{eq:Jconditionforrotremn},
restricts the types of orbits that could lead to a rotating remnant BS. In the
following, we focus entirely on the equal mass limit of the inspiral binary for
simplicity. In the Newtonian limit, a compact binary, composed of two point masses
$M_{\rm sph}$ on circular orbits with separation $d$ and orbital frequency
$\Omega=\sqrt{2M_{\rm sph}/d^3}$, possesses orbital angular $J_{\rm orb}=\nu\Omega
d^2$ with reduced mass $\nu=M_{\rm sph}/2$. Furthermore, if each compact object
has a radius $R_{\rm sph}$ and compactness $C_{\rm sph}=M_{\rm sph}/R_{\rm
sph}$, then the angular momentum of the binary, at the point of contact of the
two objects, is $J_c=M^2_{\rm sph}/\sqrt{C_{\rm sph}}$. For a typical
non-spinning neutron star, $C_{\rm sph}=0.125$, the dimensionless orbital
angular momentum at contact is $J_c/(2M_{\rm sph})^2=0.7$, which is roughly
consistent with non-linear simulations~\cite{Bernuzzi:2013rza,Cokluk:2023xio}. In
\figurename{ \ref{fig:angularmomofrotremn}}, we compare this quasi-circular
orbit estimate $J_c$ to the spin angular momenta of four families of $m=1$ BSs
and the family of $m=1$ PSs as a function of the compactness of the equal mass
non-spinning binary they emerge from  using the remnant map \eqref{eq:map}.
From \figurename{ \ref{fig:angularmomofrotremn}}, it is clear that the rotating
non-relativistic stars (those in the repulsive scalar model and the PSs)
require far less angular momentum than a quasi-circular orbit would provide,
while on the other hand, the rotating relativistic stars (those in the scalar
theory with solitonic potential) require far more angular momentum than a
quasi-circular orbit could provide. While our estimate is purely Newtonian, and
compares only the dimensionless spin at the point of contact (i.e., neglects
 radiation emitted during the merger process), it provides a rough estimate for
the parameter space, where rotating BSs may form through quasi-circular
inspirals. Therefore, from \figurename{ \ref{fig:angularmomofrotremn}} we
conclude that in the solitonic and repulsive scalar models, as well as in the case
of PSs, a quasi-circular orbit is unlikely to lead to the formation of a rotating
remnant BS, with exceptions only in isolated and small parts of the parameter
space. 

Due to the non-relativistic nature of this estimate, there may be highly
relativistic rotating solitonic BSs that could be formed through quasi-circular
orbits, or the possibility that scalar interactions between diffuse repulsive
stars radiate sufficient angular momentum during the late inspiral and merger 
leading to a rotating BS remnant. For instance, the binary inspiral in the repulsive scalar model shown to result in a rotating remnant BS in Sec.~\ref{sec:formationdynamics} possessed an oversupply of angular momentum, i.e., $J_0/Q_0=2.12$, and yet formed a (highly perturbed) $m=1$ rotating BS remnant. This may similarly hold for merging PSs. However, it appears that the types of 
orbits that may robustly lead to the formation of a rotating remnant are 
relativistic, highly
eccentric encounters of non-spinning relativistic stars in the solitonic scalar
model, while only those stars with non-relativistic velocities and mild impact
parameters may merge into rotating remnant PSs and BSs in the repulsive scalar
model. 

\subsection{Merger of mixed spinning-non-spinning BS binaries} \label{sec:partiallyspinning}

\begin{figure}[t]
\centering
\includegraphics[width=0.49\textwidth]{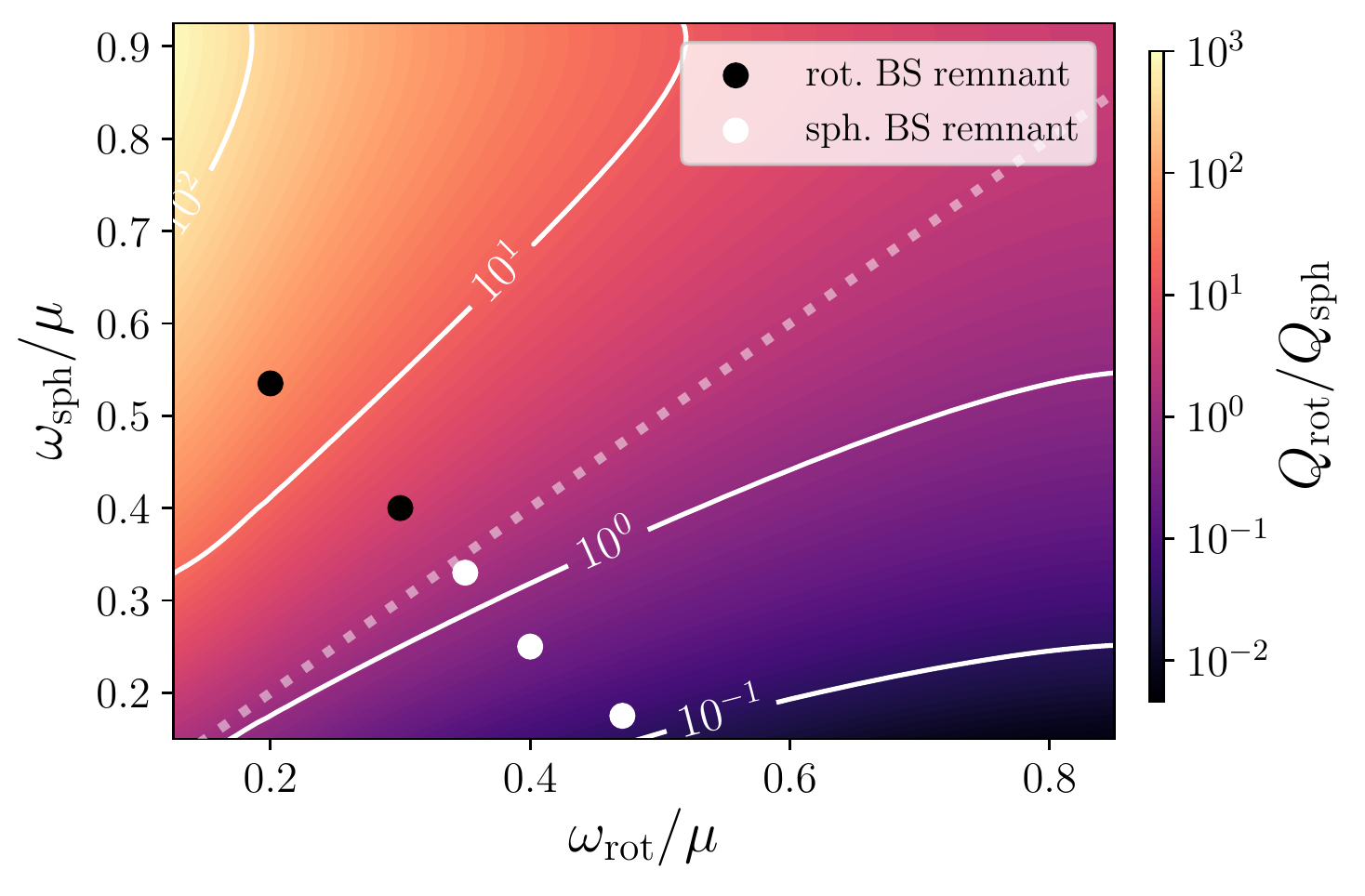}
\caption{We plot the mass ratio $\tilde{q}=M_\text{rot}/M_\text{sph}$ (white contour lines) and charge ratio $\tilde{\zeta}=Q_\text{rot}/Q_\text{sph}$ (color) across the parameter space of a (superposed) binary BS consisting of a non-spinning star with frequency $\omega_{\rm sph}$ and an $m=1$ rotating star with frequency $\omega_{\rm rot}$ in the $\sigma=0.05$ solitonic scalar model. The dashed white line indicates where $\omega_{\rm sph}=\omega_{\rm rot}$. The mergers of binaries with parameters indicated by the black (white) points result in a single rotating (non-rotating) BS remnant.}
\label{fig:emribbs}
\end{figure}

\begin{table}[b]
\begin{ruledtabular}
\begin{tabular}{c c c c}
    $\omega_{\rm rot}/\mu$ & $\omega_{\rm sph}/\mu$ & $\tilde{q}=M_{\rm rot}/M_{\rm sph}$ & $\tilde{\zeta}=Q_\text{rot}/Q_\text{sph}$ \\ \hline \hline 
$0.20$ & $0.535$ & $22$ & $63$ \\ \hline 
$0.30$ & $0.40$ & $3.9$ & $4.6$ \\ \hline 
$0.35$ & $0.33$ & $1.5$ & $0.95$ \\ \hline 
$0.40$ & $0.25$ & $0.51$ & $0.30$ \\ \hline 
$0.471$ & $0.175$ & $0.15$ & $0.06$ \\ 
\end{tabular} 
\end{ruledtabular}
\caption{The constituent BS frequencies, mass ratios, and charge ratios of the mixed spinning-non-spinning binary BS initial data discussed in the
main text. The two stars have frequencies $\omega_{\rm rot}$ and $\omega_{\rm
sph}$, (purely tangential) boost velocities $v_{\rm rot}=0.85 M_{\rm
sph}v_t/M_0$ and $v_{\rm sph}=-0.85 M_{\rm rot}v_t/M_0$ with
$v_t=\sqrt{M_0/D_0}$, and coordinate separation $D_0=10M_0$, in units of the ADM mass.}
\label{tab:partspinBBSidproperties}
\end{table}

To further understand the formation of a rotating BS remnant from the merger of
two isolated BSs, we now turn to mergers of binaries consisting of a
non-spinning and a $m=1$ spinning star. In this scenario, one can use the mass-ratio
$\tilde{q}=M_{\rm rot}/M_{\rm sph}$, or equivalently the charge-ratio
$\tilde{\zeta}=Q_\text{rot}/Q_\text{sph}$, to classify the system as an equal
mass-ratio, $\tilde{q}\sim 1$, or extreme mass-ratio, $\tilde{q}\ll 1$ and $\tilde{q}\gg 1$, system. 
The types of BSs we consider here are perturbatively stable against known instabilities.
Hence, in the $\tilde{q},\tilde{\zeta}\rightarrow \infty$ limit,
we expect the non-rotating star to be a small perturbation to the spinning
star, and thus for the remnant to be spinning (i.e., the only relevant vortex line 
is that of the spinning star). Conversely, in the $\tilde{q},\tilde{\zeta}\rightarrow 0$
limit, the remnant will be a non-spinning BS (hence, the vortex line of the spinning 
star is perturbative and neglected in the vortex condition 
\eqref{eq:vortexcondition}). This is entirely consistent with
the expectation based on the remnant map introduced in Sec.~\ref{sec:remnantmap}. 
In the intermediate regime, where
$\tilde{q},\tilde{\zeta}\sim 1$, the importance of the vortex from the spinning BS in ambiguous, 
and therefore, and the application of the remnant map is less clear.
Instead of assuming the accuracy of the remnant map, here we simply perform a series of
non-linear simulations of binary BSs in the solitonic scalar model with
coupling $\sigma=0.05$, covering the range of mass-ratios from $\tilde{q}=22$
to $\tilde{q}=0.15$. We summarize the properties of the constructed initial
data in \tablename{ \ref{tab:partspinBBSidproperties}}.

In \figurename{ \ref{fig:emribbs}}, we show the parameter space of such binary
BSs consisting of one rotating and one non-rotating constituent together with
the merger product of the binaries characterized in \tablename{
\ref{tab:partspinBBSidproperties}}. By construction, the binaries merge after
roughly one orbit in all cases considered.  During the merger of the three cases
with smallest mass-ratio, the vortex of the lighter spinning star is ejected
from the system at the point of contact or shortly after. 
The merger dynamics become more
violent with increasing initial mass-ratio, up to $\tilde{q}\sim 1$. The endstate
in all three cases is a single, non-rotating star, while in the $\tilde{q}= 1.5$
case, a small scalar clump is ejected from the system similar to what was found
in Ref.~\cite{Bezares:2022obu}. The two cases with largest mass- and
charge-ratios merge into a single rotating BS remnant, identified by the $q=1$
vortex at the center of the remnant star. The remnant star of the
$\tilde{q}=22$ binary persists for $t\gtrsim 600 M_0$ without sign of
instability, while the rotating remnant of the $\tilde{q}=3.9$ remains
non-perturbatively rotating for $t\approx 700 M_0$. In the latter case, after this time 
in our simulation the central vortex is ejected in the same way as
occurred at lower resolutions for the spinning remnant formed from the merger of non-spinning BSs
described in Sec.~\ref{sec:formationrobustness} (see
Appendix~\ref{app:vortexejinstability} for details). Hence, we speculate that this is an
artifact of the low resolution used to study this binary, and that at sufficiently
high resolution simulations the star will remain stable after merger. However, we have not
explicitly checked this for this case. In conclusion, we find that in the merger of a non-spinning
and a rotating BS, the remnant is non-rotating roughly when the mass (or charge) ratio
$\tilde{q}<1$, and rotating when $\tilde{q}>1$.

\section{Discussion and Conclusion}
\label{sec:conc}

In this work, we have studied the inspiral and merger dynamics of a large class of
scalar binary BS systems in representative repulsive and attractive
scalar potentials. To that end, we constructed constraint satisfying binary BS 
initial data using the conformal thin-sandwich approach and 
numerically evolved these data using the coupled Einstein-Klein-Gordon equations. We reviewed
important results on the non-gravitational interactions between two Q-balls,
and identified the impact of such interactions on the GW
phenomenology of an inspiraling binary BS. We pointed out
the pivotal role the scalar phase and vortex structure plays during the
inspiral and merger process of a binary BS. In particular, in the cases
we study we find that
\textit{(i)} the scalar interactions secularly accumulate throughout the late inspiral
and drive strong de-phasing and amplitude modulations of the 
gravitational radiation, \textit{(ii)} during the merger of a binary BS 
the GW signal is strongly affected by vortex dynamics, and that
\textit{(iii)} the nonlinear scalar interactions induce a prolonged
hypermassive BS phase, and can delay the collapse to a black hole by several hundred light
crossing times of the remnant, though do not strongly affect the final black hole's mass and spin. 
These findings demonstrate that, to have a 
consistent description of the late inspiral for such models that accurately predicts the emitted GWs
be taken into account.
Analogous to axion mediated forces in the 
inspiral of a binary neutron star considered in Refs.~\cite{Hook:2017psm,Huang:2018pbu,Zhang:2021mks},
the de-phasing and amplitude modulations from scalar interactions may enter at lower orders in a
perturbative post-Newtonian expansion of the binary dynamics compared to tidal Love number effects, and hence, be a
strong handle to efficiently distinguish binary BS mergers from binary
neutron star ones. 
Additionally, our results suggest that predicting the
GWs emitted during the nonlinear merger process (and of a
potential hypermassive BS remnant) for a large class of binary constituents
requires a substantial suite of nonlinear numerical evolutions, since the GW phenomenology is likely richer than even in the case of
binary neutron stars, due to the star's internal scalar phase degree of
freedom characterizing the merger. Lastly, if the final remnant of the merger
is a black hole, than the ringdown GW signal may be
indistinguishable from that of a binary black hole merger remnant (or a binary
neutron star merger resulting in black hole formation) in all but the least
compact cases. If the constituent stars of the binary are less compact than typical
neutron stars, then we show that accreting residual matter surrounding the
remnant black hole may alter the ringdown signal sufficiently to distinguish it
from the typical exponential black hole ringdown GW signals. On
the other hand, if the remnant is a BS (spinning or non-rotating), then
the ringdown signal is likely an efficient means of distinguishing the types of
binaries, particularly if the total mass of the system is $M_0\gtrsim
\mathcal{O}(10)M_\odot$ (positioning the ringdown GW signal in the most sensitive band of ground-based GW detectors).

Furthermore, in this work, we have constructed a remnant map, augmented by several 
conditions, which approximately predicts the remnant star that results from 
a binary BS merger. We illustrated the utility of
this mapping to qualitatively predict the outcome of the nonlinear dynamics
of merging BSs using a particularly peculiar
spinning binary BS inspiral. This inspiral results in the formation of a rotating 
dipolar BS solution (see, e.g., Refs.~\cite{Palenzuela:2006wp,Yoshida:1997nd}), as 
predicted by the remnant map. We also briefly commented on the
implications of the remnant map on PS dynamics. We emphasize the central
role of the scalar phase and vortex structure during the merger of binary BSs.
In other areas of physics, vortices also play an important role, leading
to phenomenology such as vortex reconnection and vortex lattices
\cite{PRL.71.1375,PRL.91.135301,PRB.38.2398,PRA.65.023603}. These, and similar
processes, may also arise in the context of BSs and are an
interesting direction for the future. While the predictions
of the remnant map are most accurate for equal frequency stars, we find it to
be a decent approximation even in more complex scenarios. However, in a few merger 
examples, we explicitly show the limitations of the remnant map. In particular, the 
classification of a vortex line as ``significant" within the vortex condition 
is ambiguous. For instance, in the inspiral of a light spinning star and a heavy 
non-rotating star, the vortex line of the former must be classified as insignificant
compared with the heavy companion in order for the vortex condition to be consistent
with the formation of non-rotating remnant. However, as the relevant
mass of the spinning star is increased, it is unclear, a priori, at exactly
what mass ratio the vortex becomes significant to the remnant, and therefore there is no
firm prediction for the exact threshold in the comparable mass regime where a rotating remnant will 
be formed.
In this work, we primarily focused on
the special case of a binary inspiral resulting in a single remnant star.
However, this map could be extended to include more than two stars taking part
in the merger and more than one remnant star.

Moreover, in this work we find, for the first time, examples where the
merger of non-rotating stars results in the
formation of a rotating BS. We achieve this by utilizing the
remnant map paired with, in particular, the vortex condition, which is a
crucial ingredient in understanding the formation of rotating BSs
in mergers. We investigated the robustness of this new formation mechanism to
changes in the binary's orbital angular momentum and initial scalar
phase-offset, finding that variations of up to a factor of two of the orbital angular
momentum and up to 12\% of the initial phase-offset still result in the
formation of a rotating remnant star. Hence, a large set of binary
configurations may form rotating remnant stars, rather than non-spinning BSs or
black holes. Furthermore, we find that quasi-circular orbits may
inhibit the formation of rotating remnant BSs, since these orbits
have either too little angular momentum (in the case of the solitonic scalar
potential), or too much angular momentum (in the case of PSs and for
stars in the repulsive scalar model considered). However, our results suggest that there may be limited regions of
parameter space where a quasi-circular orbit leads to the formation of a
rotating remnant star. Finally, regarding the merger remnant of a binary made up of a
spinning and a non-spinning star, the mass-ratio (or charge-ratio)
is a decent classifier of the remnant product: if the non-spinning star is
heavier, than the remnant is a non-rotating star, while if the spinning star
dominates the mass-ratio, then the remnant will be a rotating BS. We
also briefly comment on the implications of the formation mechanism on PS
dynamics and the formation of rotating PSs after a binary
inspiral. Previous attempts at forming a rotating BS or PS
remnant from the merger of two non-spinning stars have been unsuccessful, as
they violated one or several of the conditions found here. The mergers of binary
BSs considered in
Refs.~\cite{Palenzuela:2007dm,Palenzuela:2017kcg,Bezares:2017mzk,Bezares:2022obu}
violated the vortex condition or stability condition, while the orbital PS mergers considered in
Ref.~\cite{Sanchis-Gual:2018oui} violated, at the least, the kinematic and vortex
conditions. Lastly, if an astrophysical population of these stars existed,
then a subset of the binary inspirals may form rotating BS remnants, provided
the scalar self-interactions allow it. In fact, using our results, one could
quantify how many rotating boson star may be formed given the characteristics of a population of
non-spinning binary stars. This formation mechanism could be used to study the inspiral,
merger, and ringdown dynamics of black hole mimicking ultra compact objects
self-consistently in a nonlinear setting. In some cases, these rotating BS remnants are highly
relativistic, exhibiting stable light rings and ergoregions. The impact of
these features on the emitted gravitational waveform may be studied within
these scenarios and extrapolated to a larger set of ultra compact objects.

\begin{acknowledgments}
We would like to thank Luis Lehner for valuable discussions. We acknowledge support from an NSERC Discovery grant. Research at Perimeter Institute is supported in part by the Government of Canada through the Department of Innovation, Science and Economic Development Canada and by the Province of Ontario through the Ministry of Colleges and Universities. This research was undertaken thanks in part to funding from the Canada First Research Excellence Fund through the Arthur B. McDonald Canadian Astroparticle Physics Research Institute. This research was enabled in part by support provided by SciNet (www.scinethpc.ca), Compute Canada (www.computecanada.ca), and the Digital Research Alliance of Canada (www.alliancecan.ca). Simulations were performed on the Symmetry cluster at Perimeter Institute, the Niagara cluster at the University of Toronto, and the Narval cluster at the École de technologie supérieure in Montreal.
\end{acknowledgments}

\appendix

\section{Revisiting the non-axisymmetric instability of rotating boson stars} \label{app:revisitNAI}

In this appendix, we briefly revisit the non-axisymmetric instability (referred
to as NAI in the following) discovered in \cite{Sanchis-Gual:2019ljs}, focusing 
on the importance of the scalar phase and vortex structure in
these solutions. To gain intuition, we re-analyze two unstable BSs with $m=1$ and $2$,
in the solitonic and
repulsive scalar theories introduced in \eqref{eq:solitonic} and
\eqref{eq:repulsive}, respectively,
originally considered in Ref.~\cite{Siemonsen:2020hcg}, and discuss the instability in light of the
remnant map constructed in Sec.~\ref{sec:remnantmap}.

\subsection{Instability of a $m=2$ boson star}

\begin{figure}
\includegraphics[width=0.49\textwidth]{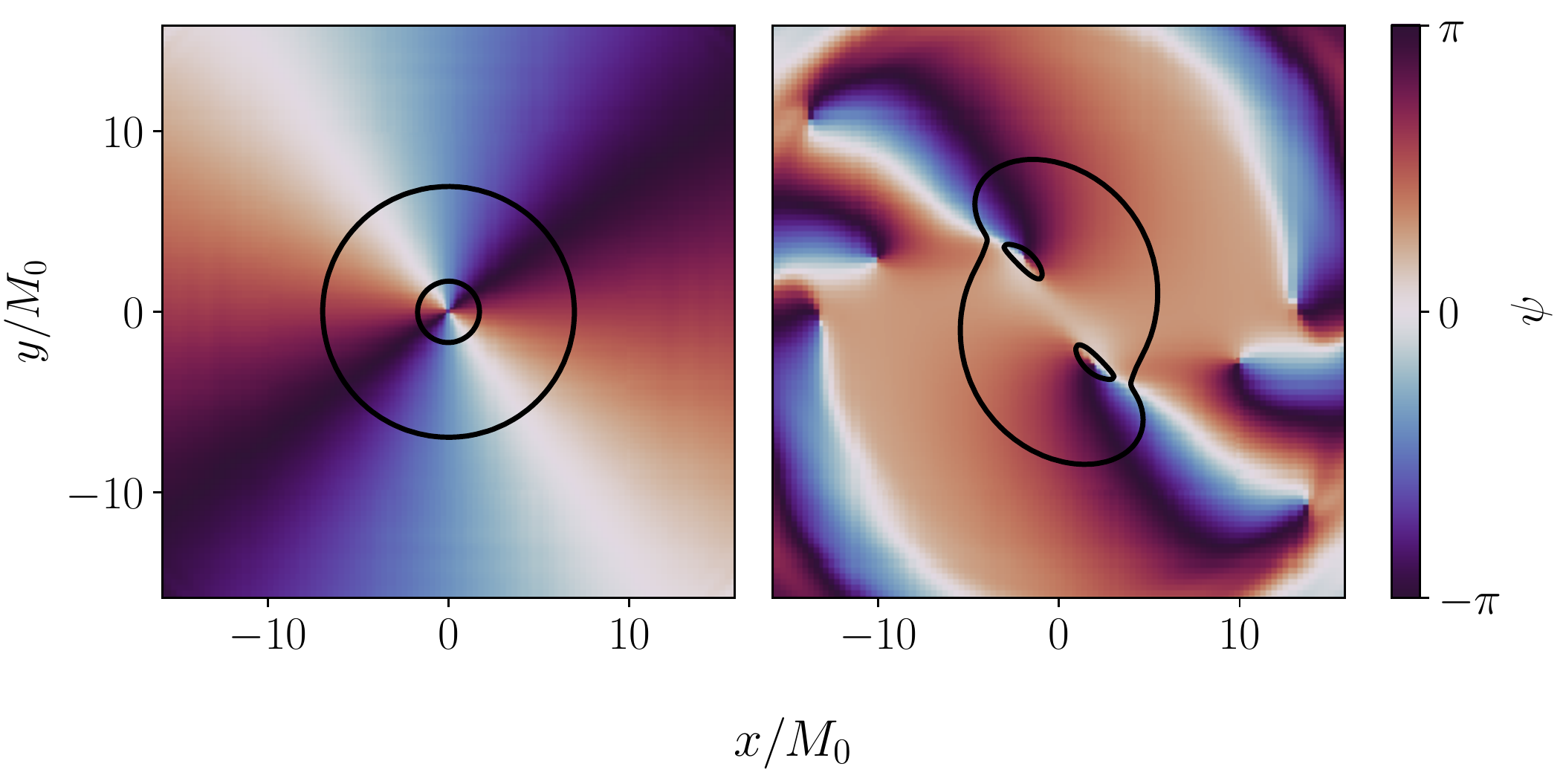}
\caption{We show the scalar phase in two equatorial slices of the $m=2$
rotating BS of frequency $\omega/\mu=0.4$ in the $\sigma=0.05$ solitonic scalar
model undergoing the NAI (first discussed in Ref.~\cite{Siemonsen:2020hcg}). The
left panel shows the scalar phase at $t/M_0=0$, while the right panel shows the
scalar phase during the nonlinear saturation of the NAI. The black lines
indicate arbitrarily chosen level surfaces of the scalar field magnitude within the
equatorial plane.}
\label{fig:m2NAIphase}
\end{figure}

We begin with the phase evolution during the development and saturation of the
NAI of a $m=2$ rotating BS solution. Specifically, we evolve the $m=2$ rotating
BS of frequency $\omega/\mu=0.4$ in the $\sigma=0.05$ solitonic scalar model in
\eqref{eq:solitonic}. The NAI of this stationary solution exhibits
characteristic growth timescales of $\tau_{\rm NAI}/M_0\approx 19$, leading to
the ejection of two blobs of scalar matter in opposite directions, leaving
behind a single non-spinning remnant star at the center of mass (see panel b
of Fig. 8 in Ref.~\cite{Siemonsen:2020hcg} for snapshots of the saturation of the
NAI in this star). In \figurename{ \ref{fig:m2NAIphase}}, we show the scalar
phase in the equatorial slice at the start of the evolution, and during the
non-linear saturation of the NAI. Initially, the vortex at the center of the
star is a $q=2$ vortex (as expected for a $m=2$ star). However, during the
non-linear saturation of the NAI shown on the right in \figurename{
\ref{fig:m2NAIphase}}, this $q=m=2$ vortex at the center of mass breaks up into
\textit{two} $q=1$ vortices, which are quickly ejected from the star, leaving
only a non-spinning star at late times (not shown in \figurename{
\ref{fig:m2NAIphase}}). This is reminiscent of, for instance, the break-up of a
$m=2$ string in the Abelian-Higgs model (see, e.g., Ref.~\cite{Thatcher_1997}).

\subsection{Instability of a $m=1$ boson star}

\begin{figure}[t]
\centering
\includegraphics[width=0.48\textwidth]{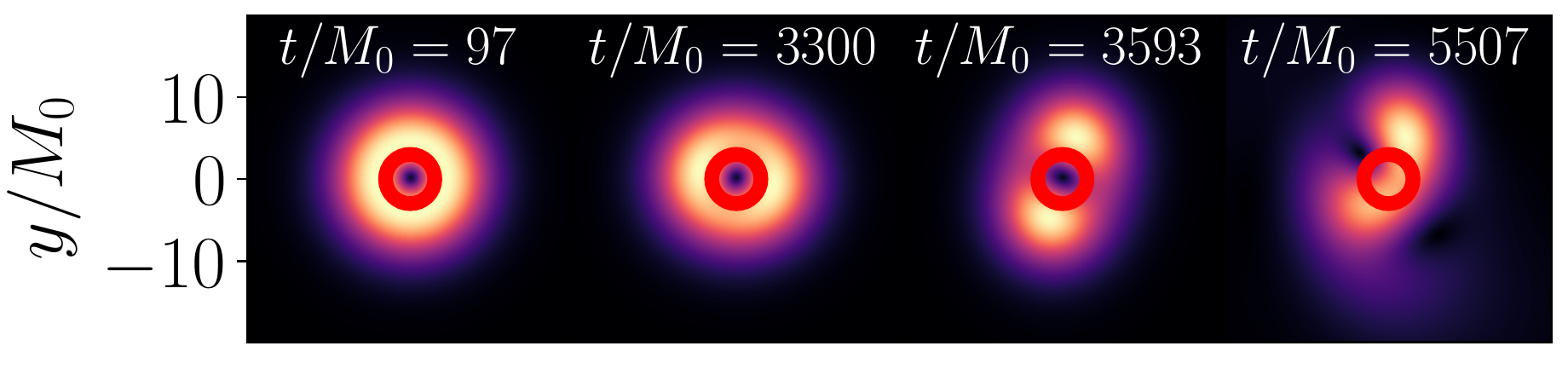}
\includegraphics[width=0.48\textwidth]{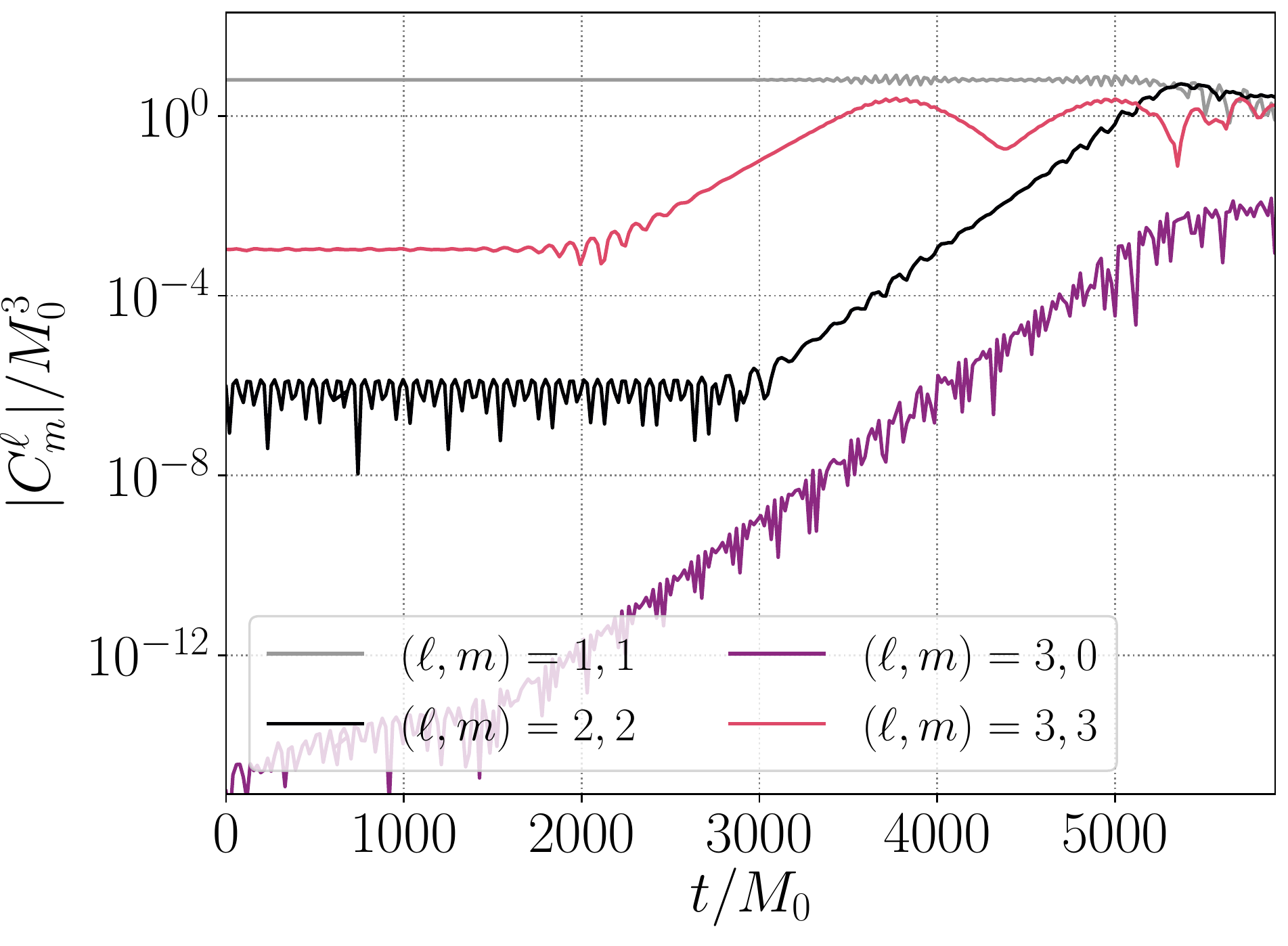}
\caption{\textit{(top)} We show the normalized scalar field magnitude $|\Phi|/|\Phi|^{t=0}_{\rm max}$ in four equatorial slices at different times during the evolution of an isolated $m=1$ rotating BS of mass $M_0$ in the repulsive model with $\lambda/\mu^2=10^2$ and frequency $\omega/\mu=0.9$. The star was shown to be unstable to the NAI in Ref.~\cite{Siemonsen:2020hcg}. Red circles indicate the coordinate location of the vortex at early times in all snapshots. \textit{(bottom)} The evolution of some of the spherical harmonic modes $C^\ell_m$, defined in \eqref{eq:sphericalharmonicmodes}, corresponding to the same star as in the top panel.}
\label{fig:NAIvsVEI}
\end{figure} 

We turn now to the NAI active in a $m=1$ rotating BS of frequency
$\omega/\mu=0.9$ in the repulsive scalar model with coupling
$\lambda/\mu^2=10^2$. In the top panel of \figurename{ \ref{fig:NAIvsVEI}}, we
show four snapshots of the evolution throughout the development and
saturation of the NAI. At early times, the scalar field magnitude morphology is
toroidal (first panel), while at intermediate times (second and third panel),
the scalar field magnitude exhibits a quadrupolar pattern, which is broken up
at late times (fourth panel) and eventually becomes a monopolar (i.e., perturbed
spherically symmetric) remnant BS. It is clear from the top panel of
\figurename{ \ref{fig:NAIvsVEI}} that the $q=1$ vortex of the stationary BS
solution remains at the center of mass even during the \textit{first}
fragmentation (first and second panel) until the \textit{second} fragmentation
phase, when this vortex is ejected from the system (last panel). 

In order to understand this two-staged fragmentation process, we decompose the scalar field $\Phi_R=\text{Re}(\Phi)$ into coordinate spherical-harmonic components $C^\ell_m$ as
\begin{align}
C^\ell_m=\int_D d^3x \Phi_R Y^\ell_m(\theta,\varphi),
\label{eq:sphericalharmonicmodes}
\end{align}
centered on the center of mass and aligned with the spin-axis of an
isolated star. (Here, the domain of integration $D$ is the ball of
coordinate radius $r/M_0=25$). In the bottom panel of \figurename{
\ref{fig:NAIvsVEI}}, we show the evolution of some of the spherical harmonic components during 
the development and saturation of the NAI. First, the $(\ell,m)=(1,1)$ component
dominates throughout the evolution, since this is a $m=1$ (toroidal) BS.
Secondly, the even-$m$ components are seeded roughly at the level of floating point roundoff at
$t=0$, while the odd-$m$ components (except for $m=1$) have amplitudes
set by the truncation error (orders of magnitude larger than the floating point
roundoff). While the even- and odd-$m$ components shown in the bottom panel of \figurename{
\ref{fig:NAIvsVEI}} all exhibit the same e-folding growth rate, suggesting that
they are all associated with the same unstable mode, the NAI has a much larger
overlap with the $(\ell,m)=(3,3)$ component than the other components.
 During the first fragmentation phase around $t/M_0\approx 3500$, the
quadrupolar patter of $|\Phi|$ can be understood by considering the bottom
panel of \figurename{ \ref{fig:NAIvsVEI}}. In this phase, the odd-$m$
components dominant in the initial non-linear phase. In particular, the
$(\ell,m)=(3,3)$ perturbation mixes with the $(\ell,m)=(1,1)$ background
solution into a $\tilde{m}=3\pm 1$ mode of $|\Phi|$, corresponding precisely to
the quadrupolar pattern observed in the top panel of \figurename{
\ref{fig:NAIvsVEI}}. During the second fragmentation around $t/M_0\approx 5000$,
the even-$m$ components also become significant and their presence
results in the ejection of the central vortex and the formation
of the non-spinning, i.e., $m=0$, BS remnant. The numerical convergence of this 
instability was checked in Ref.~\cite{Siemonsen:2020hcg}.

From this, we conclude that the NAI is dominated by odd-$m$ perturbations of
$\Phi_R$ and does not significantly affect the scalar vortex, even as the
unstable mode starts to become large (and nonlinear). However, in the subsequent
evolution in the nonlinear phase, even-$m$ perturbations continue to grow and
eventually lead to the ejection of the central vortex.

\subsection{The instability and the remnant map}

\begin{figure}
\includegraphics[width=0.49\textwidth]{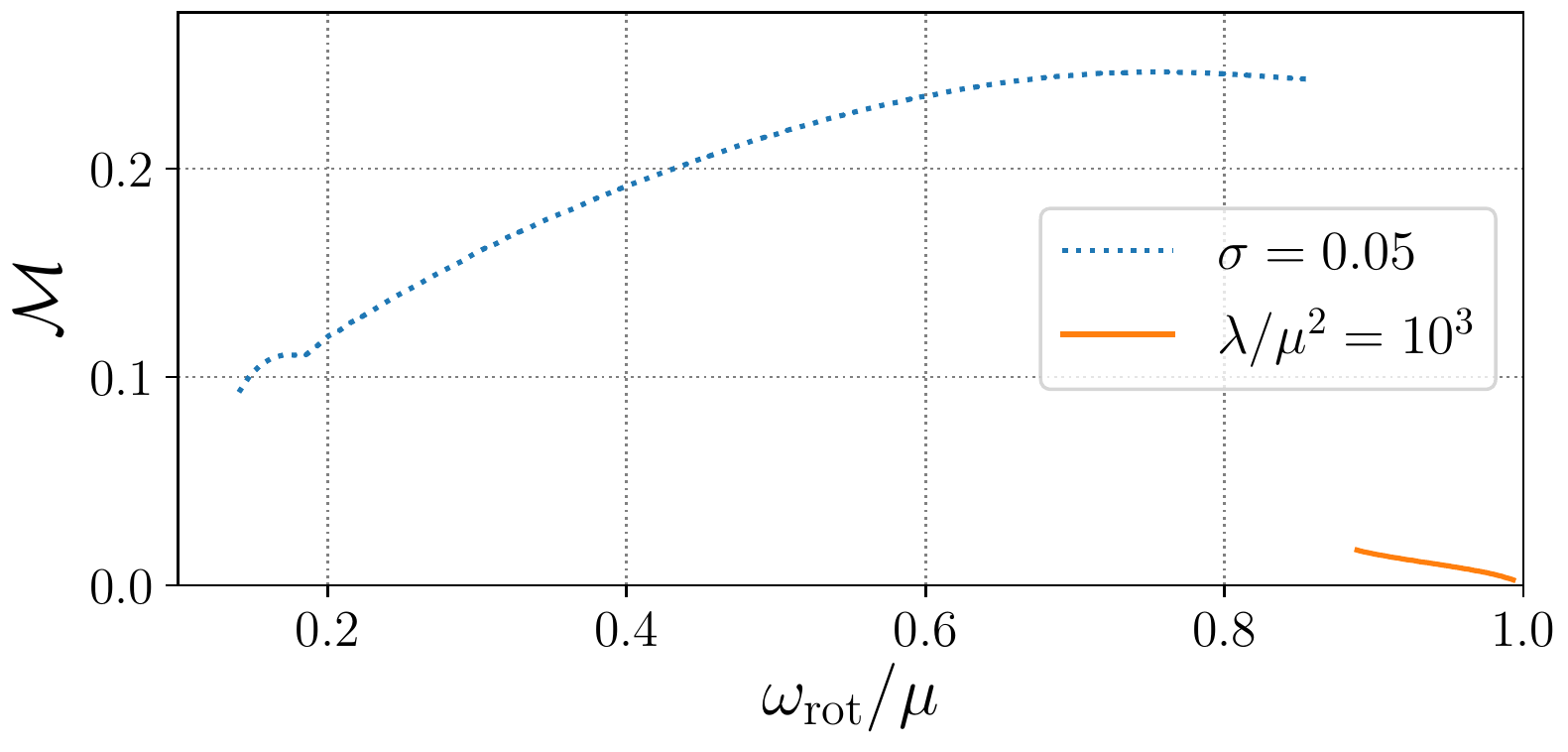}
\caption{We show the normalized mass difference $\mathcal{M}=(M_*-M_{\rm
rot})/M_*$ between the mass $M_{\rm rot}$ of an isolated $m=1$ spinning star of
frequency $\omega_{\rm rot}$ and a non-rotating star of mass $M_*$ in the same
models, when assuming the remnant map described in Sec.~\ref{sec:remnantmap}.
We do this for the solitonic and repulsive scalar models, with couplings
$\sigma$ and $\lambda$, respectively.  Therefore, for each $\omega_{\rm rot}$
shown, $\mathcal{M}$ indicates the energy gained by transitioning the $m=1$
rotating star to a non-rotating star of the same charge. We show only the
branches below the maximum mass of the families of solutions.}
\label{fig:RotToSphEnergy}
\end{figure}

For completeness, we briefly discuss the transition from a $m=1$ rotating BS to
a non-spinning star of the same scalar model with the remnant map described in
Sec.~\ref{sec:remnantmap}. This transition (similar to the NAI and its
non-linear saturation) corresponds to the decay of a rotating solution to a
non-rotating solution, which may be dynamically achieved by means of an
instability in the $m=1$ star. In \figurename{ \ref{fig:RotToSphEnergy}}, show
the energy gained by such a transition assuming the remnant map, i.e., assuming
no scalar radiation. Clearly, it is energetically favorable to decay
from a $m=1$ rotating star to a non-rotating star. However, whether this is a
hint of a linear instability everywhere in the parameter space shown in
\figurename{ \ref{fig:RotToSphEnergy}} is non-trivial. Furthermore, a dynamical
transition requires the ejection of all angular momentum from the system, which may
break the assumption of local charge conservation around the star. On the other
hand, this demonstrates that the transition of two identical $m=1$ BSs
into two identical non-spinning stars is also energetically favorable,
a scenario we explicitly found performing a nonlinear evolution in
Sec.~\ref{sec:remnantmap}.

\section{Angular momentum variation} \label{app:angularmomentumvariation}

\begin{figure*}
\includegraphics[width=0.99\textwidth]{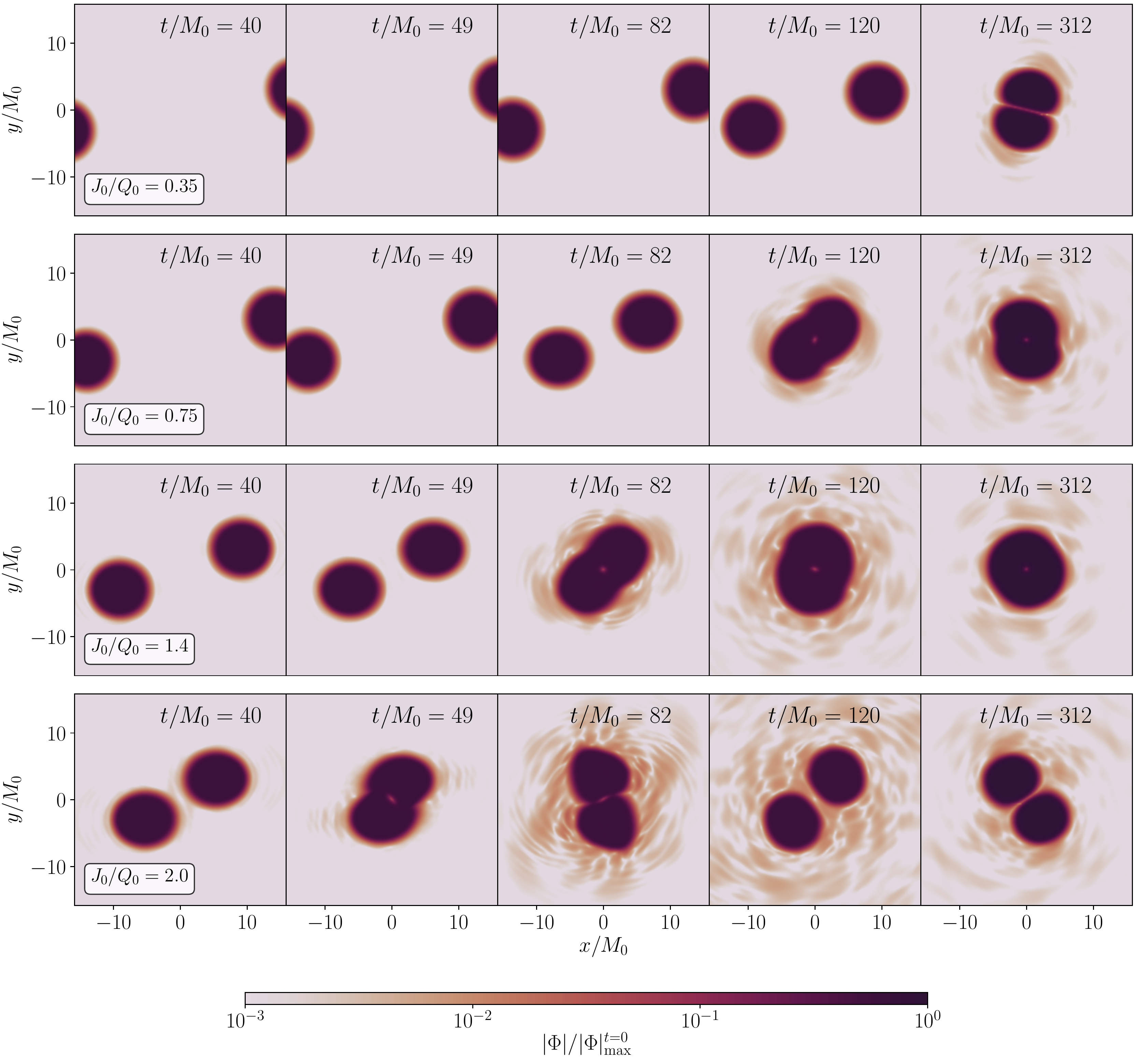}
\caption{We show the scalar field magnitude $|\Phi|$ (normalized by the maximum
in the initial time-slice) in a few snapshots of the equatorial slice of the
binary BS simulations discussed in \figurename{
\ref{fig:angularmomrobustness}}. We label each sequence of time-slices by the
initial angular momentum $J_0$ in units of initial charge $Q_0$. In the top and
bottom rows, the binary merges into a rotating DBS solution (i.e. two non-rotating 
BSs separated by scalar interactions, as discussed in
Sec.~\ref{sec:scalarinteractions}). Compared with the top row, the binary in
the bottom row rotates at high angular velocities around the center of mass at
late times, i.e., $t/M_0>300$. The case shown in the middle two rows merge to form
a remnant with $q=1$ vortex at the center of mass at late times. Notice, we
find that the $J_0/Q_0=0.75$ case relaxes to a rotating BS at late
times $t/M_0>300$.}
\label{fig:angmomvariationevolution}
\end{figure*}

In \figurename{ \ref{fig:angmomvariationevolution}}, we present snapshots of
the evolution of the sequence of non-spinning binary BSs of varying initial
orbital angular momentum discussed in Sec.~\ref{sec:formationrobustness}.

\section{Isolated Proca star solutions} \label{app:bssolutions}

The details of the construction of isolated scalar BSs are presented in
Ref.~\cite{Siemonsen:2020hcg}. Hence, here we focus on our approach to
constructing the PS solutions considered in this work. PSs are, analogous to
scalar BSs, solutions to a massive complex field---in this case a vector field
$A_\alpha$ of mass $\mu$---minimally coupled to general relativity as \cite{Brito:2015pxa}
\begin{align}
S=\int d^4x\sqrt{-g}\left[\frac{R}{16\pi}-\frac{1}{4}F_{\alpha\beta}\bar{F}^{\alpha\beta}-\frac{1}{2}\mu^2 A_\alpha\bar{A}^\alpha \right].
\label{eq:vectoraction}
\end{align}
Here  $F_{\alpha\beta} = \nabla_\alpha A_\beta - \nabla_\beta A_\alpha$ is the field strength,
while the overbar denotes complex conjugation. This theory is also invariant
under a global $U(1)$ transformation $A_\beta\rightarrow A_\beta e^{i\theta}$. The
associated Noether current and Noether charge are
\begin{align}
j^\mu = \frac{i}{2}(\bar{F}^\mu{}_\alpha A^\alpha-F^\mu{}_\alpha \bar{A}^\alpha), & & Q= - \int d^3x \sqrt{-g} j^0 ,
\end{align}
respectively. The Lorenz relation, $\nabla_\alpha A^\alpha=0$, is identically
satisfied due to the antisymmetry in the field-strength, assuming $\mu\neq 0$ (and no vector
self-interactions). The stress-energy associated with the action
\eqref{eq:vectoraction} is given by
\begin{align}
\begin{aligned}
T_{\mu\nu}= & \ -F_{\alpha (\mu}\bar{F}_{\nu)}{}^\alpha-\frac{1}{4}g_{\mu\nu}F^{\alpha\beta}\bar{F}_{\alpha\beta}\\
 & \ +\mu^2\bar{A}_{(\mu} A_{\nu)} -\frac{1}{2}\mu^2 g_{\mu\nu} A_\alpha \bar{A}^\alpha.
\end{aligned}
\end{align}
With this, the Einstein equations, sourced by the vector matter, and the vector field equations are given by 
\begin{align}
G_{\alpha\beta}=8\pi T_{\alpha\beta}, & & \nabla_\alpha F^{\alpha\beta}=\mu^2 A^\beta,
\end{align}
where $G_{\alpha\beta}$ is the Einstein tensor. Rotating PSs are asymptotically
flat, axisymmetric solutions to the above system of equations. Hence, we make
the ansatz
\begin{align}
\begin{aligned}
ds^2=-fdt^2 & \ +lf^{-1}\big\{ j(dr^2+r^2d\theta^2) \\
& \ +r^2\sin^2\theta(d\varphi-\Omega r^{-1} dt)^2\big\},
\end{aligned}
\end{align}
where $f,l,j$, and $\Omega$ are functions of $r$ and $\theta$. The corresponding ansatz for the vector 1-form is given by
\begin{align}
\textbf{A}=e^{i\omega t}(Vdt+iBdr)
\end{align}
for spherically symmetric solutions, and 
\begin{align}
\textbf{A}=e^{i\omega t}e^{im\varphi}\left(iV dt+\frac{H_1}{r}dr+H_2d\theta+i\sin\theta H_3 d\varphi\right)
\end{align}
for rotating solutions. Here, in direct analogy to scalar BSs, this
approach yields an infinite set of families of solution indexed by their
azimuthal mode $m$, and parameterized by their frequency $\omega$. The boundary
conditions for the vector field are $A_\mu\rightarrow 0$ for
$r\rightarrow\infty$. However, in order to obtain non-trivial solutions to the
field equations and these boundary conditions, we promote the frequency to a
field $\omega\rightarrow\omega_s(r,\theta)$ and introduce the second auxiliary
field $\rho(r,\theta)$, following Ref.~\cite{Kleihaus:2005me}, where this was
applied in the scalar case (see also Ref.~\cite{Siemonsen:2020hcg}). Both
fields follow $\square_g=g^{\mu\nu}\nabla_\mu \nabla_\nu$ of
the spacetime $g_{\mu\nu}$:
\begin{align}
\square_g \omega_s =0, & & \square_g \rho = \frac{j^t}{\omega_s}.
\end{align}
Through the boundary conditions $\omega_s(r\rightarrow\infty)\rightarrow \omega$, where $\omega$ is the solution's frequency, we ensure that the solution is non-trivial. 
Integrating over the entire three-volume of a time-slice, we see that $\rho$ encodes the total $U(1)$-charge in its boundary data:
\begin{align}
\lim_{r\rightarrow\infty} r^2 \partial_r\rho(r,\theta)=-\frac{Q}{4\pi\omega}.
\label{eq:chargeboundary}
\end{align}
Conversely, using \eqref{eq:chargeboundary}, we can impose $Q$ as a boundary
condition, and solve for $\omega_s$. Hence, within this formalism, either the
solution's frequency $\omega$, or the solution's charge $Q$ can be imposed as
a boundary condition. The list of boundary conditions for the metric variables
and auxiliary functions are given in Ref.~\cite{Siemonsen:2020hcg}, while the
boundary conditions of the vector components are listed in
Ref.~\cite{Brito:2015pxa}. One subtlety arises for $B(r)$ (in the spherically
symmetric case), which follows a first order differential equation. We promote
this equation trivially to a second order equation by
$B(r)\rightarrow\partial_r\tilde{B}(r)$, and then impose the boundary
condition $\lim_{r\rightarrow\infty}\tilde{B}(r)=\tilde{B}_\infty$ (the
solution is independent of $\tilde{B}_\infty$), while requiring $\partial_r
\tilde{B}(r)|_{r=0}=0$.

To numerically solve the system of equations introduced above subject to the
boundary conditions, we follow the implementation for scalar BSs in
Ref.~\cite{Siemonsen:2020hcg}. We compactify the radial coordinate
$r\rightarrow \bar{r}=r/(1+r)$, such that $\bar{r}\in(0,1)$, and restrict to the
upper-half plane, i.e., $\theta\in(0,\pi/2)$. We utilize fifth-order accurate
finite differences both in the radial and polar directions, and consider a
uniformly spaced grid in the compactified coordinates $(\bar{r},\theta)$. We
utilize a Newton-Raphson-type relaxation procedure that iteratively approaches
the true solution given a sufficiently close initial guess. The initial guess
is constructed from plots given in Ref.~\cite{Herdeiro:2019mbz}. Once a
solution is found at $\omega_1$, we explore the parameter space of each family
$m$, by imposing the boundary conditions $\omega_2=\omega_1+\delta\omega$ (or
analogously the vector charge), where $\delta\omega$ is chosen sufficiently
small such that the solution at $\omega_1$ is a good enough initial guess to
obtain the solution at $\omega_2$. For spherically symmetric stars, we found
that a resolution of $N_r=1000$ was sufficient for all considered cases, while
in the rotating case we typically use $N_r\times N_\theta=500\times 100$.

\section{Numerical setup} \label{app:idandnumevo}

\subsection{Initial data}

In this work, we numerically evolve only scalar BSs. Hence, in what follows, we
restrict ourselves entirely to the scalar case. In order to evolve a binary BS
solution in general relativity consistently, we solve the Hamiltonian and
momentum constraints of a superposed binary BS. Here, we only provide a brief
summary of our procedure, and defer further details to Ref.~\cite{inprep}. We
follow Ref.~\cite{East:2012zn} and consider the constraint equations in the
conformal thin-sandwich (CTS) formulation~\cite{York:1998hy}, where the spatial metric 
$\gamma_{ij}$ on the initial time-slice is related to the conformal metric
$\tilde{\gamma}_{ij}$ by the conformal factor $\Psi$:
\begin{align}
\gamma_{ij}=\Psi^4\tilde{\gamma}_{ij}.
\end{align}
Within the CTS framework, the extrinsic curvature
$K_{ij}=-\mathcal{L}_n\gamma_{ij}/2$, defined by the Lie-derivative along the
hypersurface normal $\mathcal{L}_n$, is decomposed into the trace-free part
$A^{ij}=\Psi^{-10}\hat{A}^{ij}$ and the trace $K=\gamma_{ij}K^{ij}$. With the
conformal lapse $\tilde{\alpha}=\Psi^{-6}\alpha$, Ricci scalar $\tilde{R}$
associated with the conformal metric, the conformal covariant derivative
$\tilde{D}_i$, and the matter energy and momentum densities $E$ and $p^i$,
the Hamiltonian and momentum constraint equations in the CTS formulation are
\begin{align}
\tilde{D}^i\tilde{D}_i\Psi-\frac{\tilde{R}}{8}\Psi+\frac{\hat{A}_{ij}\hat{A}^{ij}}{8}\Psi^ {-7}-\frac{K^2}{12}\Psi^5 = & \ -2\pi\Psi^5 E, \\
\tilde{D}_j\hat{A}^{ij}-\frac{2}{3}\Psi^6 \tilde{D}^iK= & \ 8\pi\Psi^{10}p^i.
\end{align}
We parameterize the energy and momentum densities as
\begin{align}
E= & \ \Psi^{-12}\tilde{\eta}\bar{\tilde{\eta}}+\Psi^{-4}\tilde{D}^i\Phi\tilde{D}_i\Phi+V(|\Phi|), \\
p^i= & \ -\Psi^{-10}(\tilde{\eta}\tilde{D}^i\bar{\Phi}+\bar{\tilde{\eta}}\tilde{D}^i\Phi),
\end{align}
where $\eta=\mathcal{L}_n\Phi$ and $\eta=\Psi^{6}\tilde{\eta}$, 
following what was done in Ref.~\cite{Corman:2022alv}. 
In addition to the usual conformal metric quantities that are free data in the CTS formalism, we specify
$\Phi$ and $\tilde{\eta}$ (as opposed to a conformally rescaled energy and momentum density)
as free data when solving the constraints.

Following Ref.~\cite{East:2012zn}, the free data for the CTS constraint
equations is constructed from the plain superposition of the metric variables
and scalar field degrees of freedom of two
isolated star solutions: 
\begin{align} \Phi= & \ \Phi_{(1)}+\Phi_{(2)}, & &
\tilde{\eta}=\tilde{\eta}_{(1)}+\tilde{\eta}_{(2)}.  
\end{align} 
We displace and boost the isolated solutions along vectors $\beta_i^{(1)}$ and
$\beta_i^{(2)}$ with coordinate velocity $v_{(1)}$ and $v_{(2)}$, respectively,
prior to superposing the stars' solutions. The numerical construction of the
isolated BS solutions used in this work is outlined in
Ref.~\cite{Siemonsen:2020hcg}.  The elliptic constraint equations, subject to
asymptotically flat boundary conditions, are then solved using Newton-Raphson
relaxation combined with the multigrid method~\cite{East:2012zn}.

\subsection{Numerical evolution}

Given initial data for a scalar binary BS, we evolve the Einstein-Klein-Gordon
equations (following from the scalar action \eqref{eq:action}) forward in time
employing the generalized harmonic formulation of the Einstein evolution
equations \cite{Pretorius:2004jg}. To that end, we utilize fourth-order
accurate finite difference stencils over a compactified Cartesian grid
containing spatial infinity. There we impose asymptotically flat boundary
conditions on the metric variables and set the scalar field to zero. This is
paired with a fourth-order accurate Runge-Kutta time-integration. Of particular
importance is the use of adaptive mesh refinement, with refinement ratio 2:1,
to track the stars as they move across the Cartesian grid (see
Ref.~\cite{East:2011aa} for details). The compactness of the stars sets the
number of levels required to resolve the stars sufficiently; for
low-compactness solutions [typically stars in the repulsive scalar model,
\eqref{eq:repulsive}], we require five to six refinement levels, while for the
high-compactness solutions [usually those in the solitonic scalar model,
\eqref{eq:solitonic}], we require six to seven levels. In the cases with black
hole formation, we add refinement levels dynamically to resolve the
gravitational collapse and apparent horizon (this requires seven to nine
levels). The resolution on the finest mesh refinement level for the binary
evolutions presented in Sec.~\ref{sec:gwimprints} is $\Delta x/M_0=0.15$. The
resolution for the solitonic cases shown in Sec.~\ref{sec:formationdynamics} is
$\Delta x/M_0=0.075$ on the finest refinement level, while for the binaries in the repulsive
model it is $\Delta x/M_0=0.2$. Throughout, we use the standard damped
harmonic gauge condition to set the generalized harmonic source functions
$H_\mu$ \cite{Choptuik:2009ww,Lindblom:2009tu}.

\subsection{Convergence tests}

\begin{figure}[t]
\includegraphics[width=0.485\textwidth]{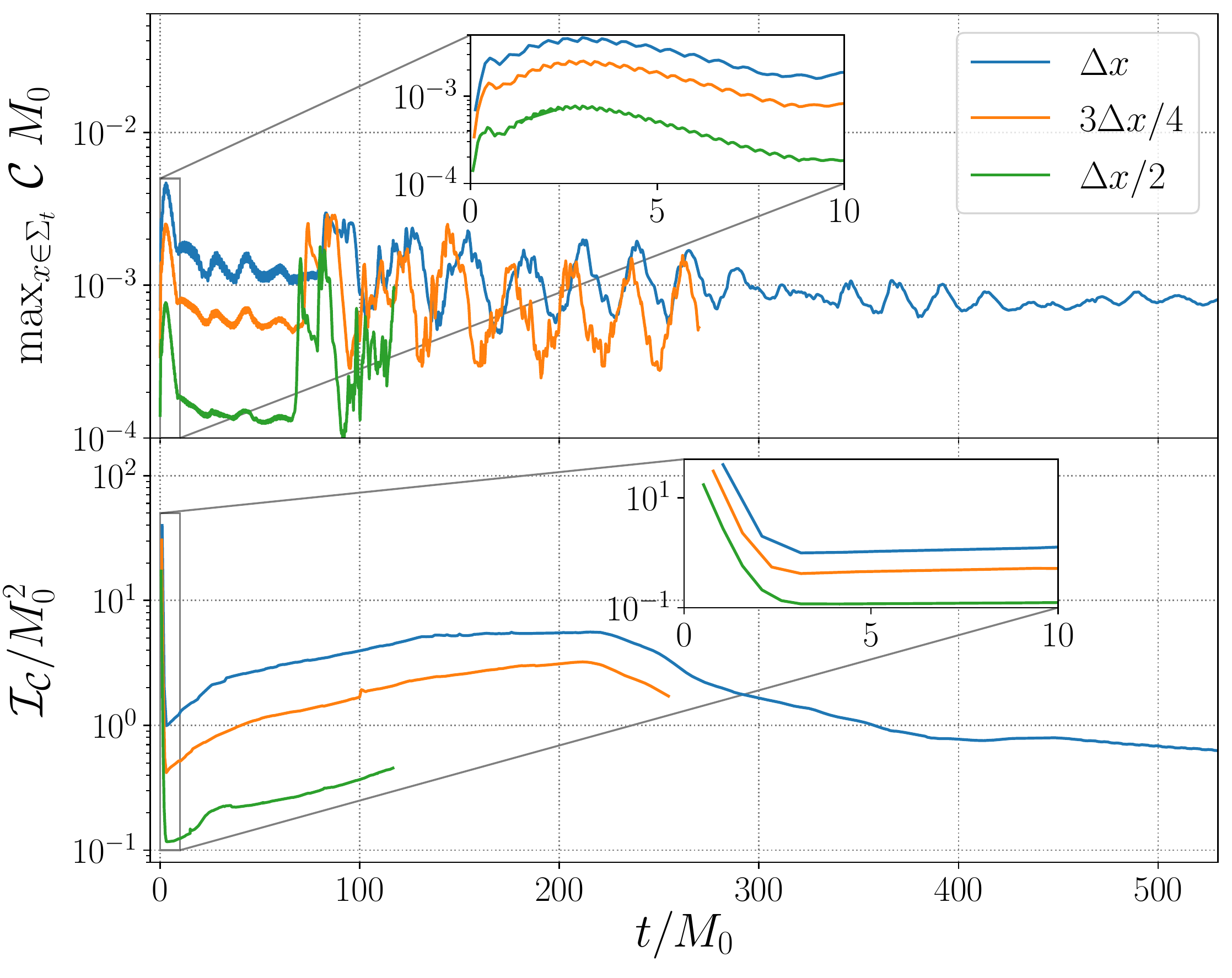}
\caption{Here we consider the convergence behavior of the binary BS in the
    $\sigma=0.05$ solitonic scalar model, with properties summarized in
    \tablename{ \ref{tab:rotBBSidproperties}}, with decreasing grid spacing.
    The quantities $\mathcal{C}$ and  $\mathcal{I}_\mathcal{C}$ (defined in the
    text) are a positive definite measure of the constraint violation, which
    we track throughout the simulation. The rapid variation of the constraints is driven by
    gauge dynamics at early times. The maximum of the constraint violation
    $\mathcal{C}$ occurs during the merger of the binary at around $t/M_0\approx
    75$. The binary merges earlier with increasing resolution, and only the medium
    and high resolutions capture small-scale features present in the remnant
    after merger. The quantity $\max\mathcal{C}$ converges to zero roughly at third order, as expected, since it is primarily set by the third-order accurate time interpolations on the mesh refinement boundaries. On the other hand, the integrated quantity $\mathcal{I}_\mathcal{C}$ converges at the expected forth order, as it is largely insensitive to the lower-order time interpolations.
    }
\label{fig:Sbbsconstraints}
\end{figure}

\begin{figure}[t]
\includegraphics[width=0.485\textwidth]{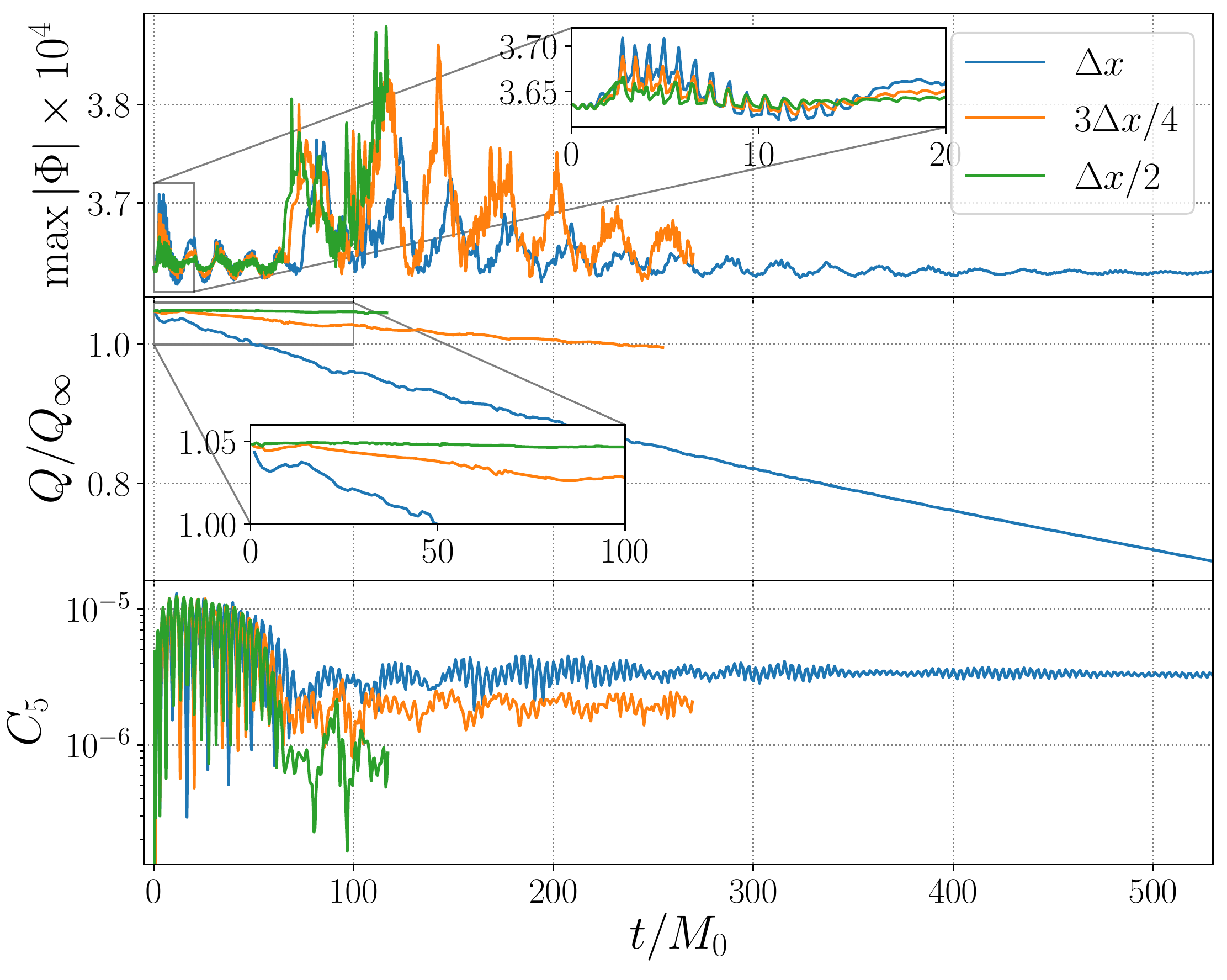}
\caption{We consider the convergence behavior of the global maximum of
    $|\Phi|$, the total $U(1)$-charge $Q$, and the azimuthal mode $C_5$ of the
    scalar field for the binary BS shown in \figurename{
        \ref{fig:Sbbsconstraints}}. The total charge $Q$ is calculated in a
    coordinate sphere of radius $100M_0$ around the center of mass of the
    system. We normalize $Q$ by $Q_\infty$, the sum of the BSs' isolated
    charges $Q_\infty=Q_1+Q_2$.  As the initial separation between the two
    stars increases, the total charge approaches the superposed charge
    \cite{inprep}: $Q\rightarrow Q_\infty$. Lastly, we also show the
    convergence behavior of the $C_5$ mode [defined in
    \eqref{eq:azimuthaldecomp}] during the binary evolution. The $m=5$
    perturbations remaining after the merger (and the formation of an $m=1$
    rotating remnant) at around $t/M_0\approx 75$ are converging towards zero
    with increasing resolution at roughly the expected fourth order.}
\label{fig:Sbbsproperties}
\end{figure}

We present resolution studies of two exemplary binary mergers. First, we focus
on the $\sigma=0.05$ solitonic scalar model and the binary with parameters
given in \tablename{ \ref{tab:rotBBSidproperties}}. We consider three
resolutions, corresponding to $\Delta x$, $3\Delta x/4$, and $\Delta x/2$,
where the lowest resolution corresponds to a grid spacing of $\Delta
x/M_0\approx0.1$ on the finest level, and the medium
resolution is the default resolution for all simulations discussed in
Sec.~\ref{sec:formationofrotremn}.
In order to explicitly demonstrate that we
are solving the Hamiltonian and momentum constraints, we track the violations
of the constraints, given by $C_\mu=H_\mu-\square x_\mu$, in time. In
\figurename{ \ref{fig:Sbbsconstraints}}, we plot the evolution of the
constraints at these different resolutions of the binary with parameters given
in \tablename{ \ref{tab:rotBBSidproperties}}. To track the constraint
violations, we define $\mathcal{C}=\sum_\mu|(C_\mu)^2|/4$, and consider the
global maximum $\max \mathcal{C}$ in each time-slice, as well as the integrated
norm $\mathcal{I}_\mathcal{C}= \int d^3x\sqrt{\gamma} \mathcal{C}$. In
\figurename{ \ref{fig:Sbbsproperties}}, we show the convergence behavior of the
total $U(1)$-charge of the system. Overall, the constraint violations converge
to zero at the expected forth order of our numerical methods. The violation of
the conservation of the $U(1)$ charge $Q$, shown in \figurename{
\ref{fig:Sbbsproperties}}, also converges towards zero. Likely, due to the
compactness ($C=0.13$) of the BSs, rapid exponential decay of the scalar field
outside the stars, i.e., $ \Phi\sim\exp(-\sqrt{\mu^2-\omega^2}r)$, with
$\omega/\mu=0.25$, and the large initial separation (of $D=40M_0$), the low and
medium resolutions exhibit relatively large drifts in the total conserved
charge. Hence, the scalar field gradients on the surface of the stars, as well
as the spatial scales of perturbations, require relatively high resolution.

\begin{figure}[t]
\includegraphics[width=0.49\textwidth]{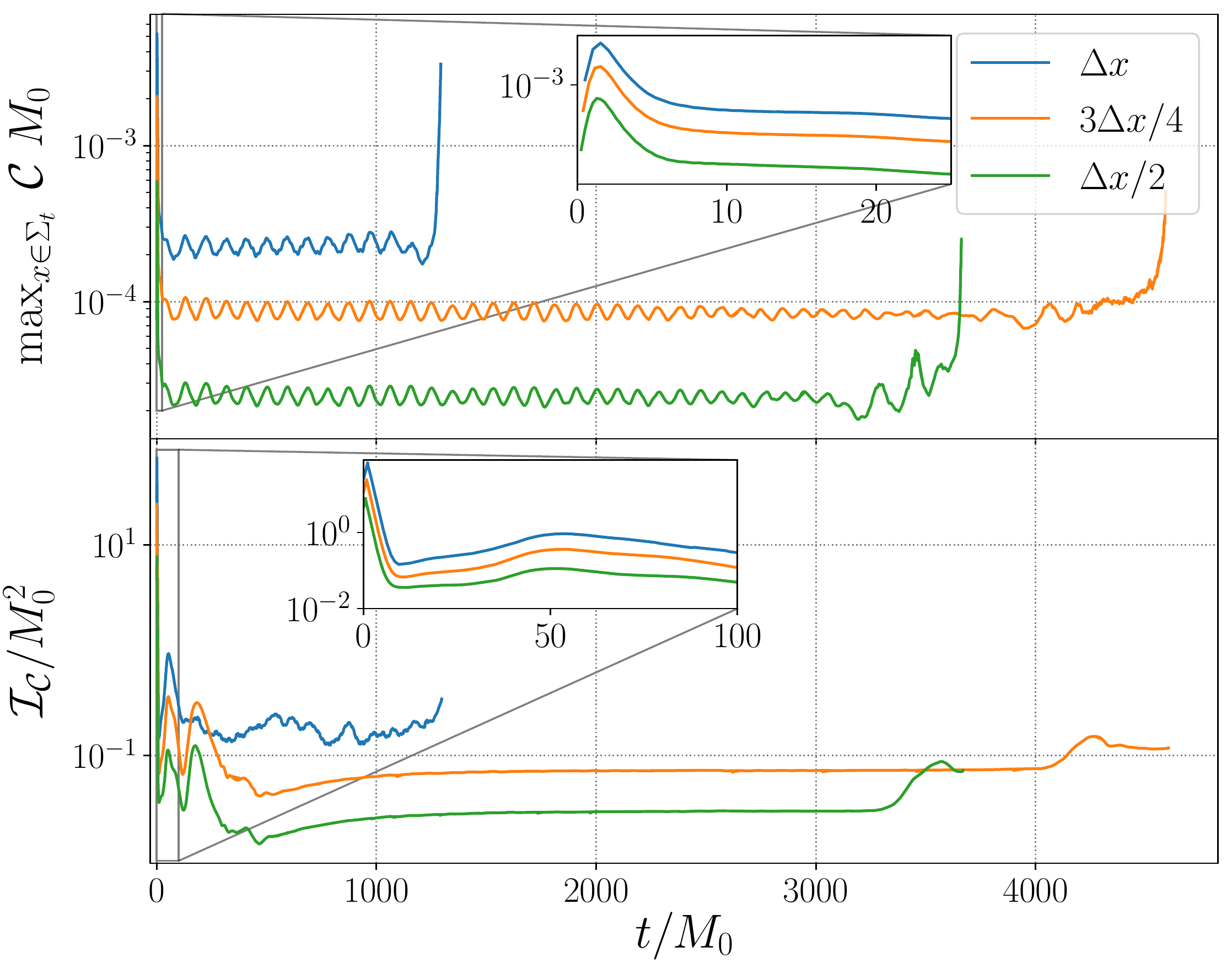}
\caption{We consider the convergence behavior of the $\alpha=\pi/2$ case of
    Sec.~\ref{sec:gwimprints} with decreasing grid spacing $\Delta x$. The
    quantities $\mathcal{C}$ and $\mathcal{I}_{\mathcal{C}}$ are defined in the
    text. The low resolution evolution is based on a different mesh-refinement
    layout (as discussed in the text,) and, hence, exhibits slightly different
    convergence behavior. At early times, the convergence orders of these quantities are the same as those discussed in the caption of \figurename{ \ref{fig:Sbbsconstraints}}.}
\label{fig:Rbbsconstraints}
\end{figure}

Secondly, we discuss the numerical convergence of one of the binaries considered
in Sec.~\ref{sec:gwimprints}. In particular, we focus on the $\alpha=\pi/2$
case, and compare its convergence behavior to that of the $\alpha=\pi$ binary
evolution. In \figurename{ \ref{fig:Rbbsconstraints}}, we present the
convergence of the constraint violations with increasing resolution of the
$\alpha=\pi/2$ evolution. Again, this demonstrates explicitly that we are
solving the Hamiltonian and momentum constraints consistently within the $t=0$
slice. In the subsequent evolution up to $t/M_0=100$, the constraints converge
at the expected orders. 
For numerical stability purposes, we have to increase the size of the
second coarsest mesh-refinement level in the lowest resolution run, moving the
outer boundary of this level from $|x_i|/M_0=100$ to $|x_i|/M_0\approx 241$.
This explains the disagreement between the $\Delta x$ and the $3\Delta x/4$ as
well as $\Delta x/2$ resolutions in \figurename{ \ref{fig:Rbbsconstraints}},
\textit{after} $t/M_0\approx 100$ (as at this time constraint violations
propagating outward reach the mesh-refinement boundary in the medium and high
resolution runs, but not yet in the low-resolution case). Furthermore, this
different mesh-refinement layout in the low resolution case alters the
convergence behavior, such that this case mergers much earlier compared with
the medium and high resolution runs. However, we have checked explicitly that the
merger delay between the $\alpha=\pi/2$ and $\alpha=\pi$ cases
\textit{increases} from low (of $\Delta t/M_0\approx 43$) to medium resolution
evolutions (of $\Delta t/M_0\approx 262$). Hence, the dephasing, delayed merger
and black hole collapse discussed in Sec.~\ref{sec:gwimprints} are physical,
and we are likely \textit{underestimating} their impact on the GWs. Notice
also, identical numerical setups were used for \textit{all} cases presented
Sec.~\ref{sec:gwimprints}, both for the initial data construction and
evolution. Therefore, while \textit{absolute} differences are not resolved,
this is suggestive that the \textit{relative} difference in amplitude in the GW
waveform between the $\alpha$-cases are driven by the scalar interactions,
rather than numerical truncation error.

\section{Vortex ejection as an artifact of numerical resolution} \label{app:vortexejinstability}

\begin{figure}
\includegraphics[width=0.485\textwidth]{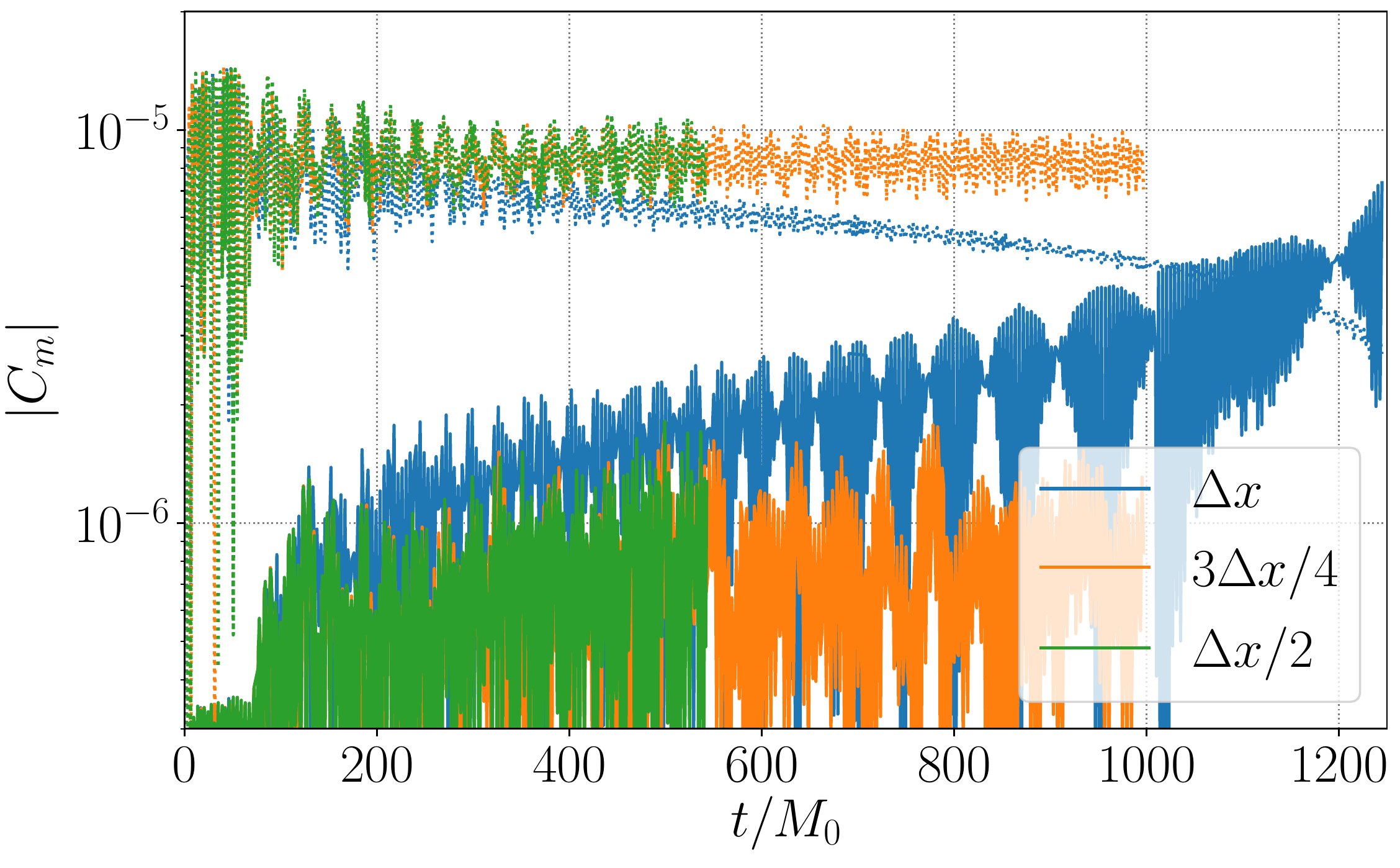}
\caption{The evolution of the scalar field modes $C_m$ (dotted and solid lines 
    corresponding to $m=1$ and 2, respectively) 
    defined in \eqref{eq:azimuthaldecomp} for the binary
    BS merger specified in \tablename{ \ref{tab:rotBBSidproperties}} with
    phase variation $\alpha/\pi=63/64$. The merger occurs roughly at
    $t/M_0\approx 75$, after which the even-$m$ modes promptly begin to grow
    exponentially in the evolution with the \textit{lowest} resolution (the $m=0$
    mode is representative of all even-$m$ modes). This apparent  
    instability is an artifact of low numerical resolution, and disappears with increasing resolution.
    }
\label{fig:modedecomposition}
\end{figure}

We find that in our simulations of the rotating BS formed from the merger of two non-rotating BS
with a phase variation of
$63/64\geq \alpha/\pi\geq 7/8$ exhibit a growing perturbation leading to vortex ejection
at low resolutions, but that this behavior disappears at sufficiently
high resolution. In order to understand this behavior, it is
instructive to consider an azimuthal mode decomposition of the real part of the
scalar field, $\Phi_R=\text{Re}(\Phi)$, defined in \eqref{eq:azimuthaldecomp}.
In \figurename{ \ref{fig:modedecomposition}}, we show the scalar field modes
$C_m$ during the merger of the binary BS specified in \tablename{
    \ref{tab:rotBBSidproperties}} with initial phase variation
$\alpha/\pi=63/64$. During, and shortly after, the merger around $t/M_0=75$, the
$m=1$ mode is the most dominant mode representing the formation of a $m=1$
rotating BS, and indicating the formation of a $q=1$ central vortex.
Additionally, the amplitude of the even-$m$ modes right after merger is
consistent across resolutions. On the other hand, the even-$m$ modes begin to
grow exponentially right after formation of the rotating remnant 
(the representative $m=0$ mode is shown in \figurename{ \ref{fig:modedecomposition}})
in the evolution with \textit{lowest} resolution. Furthermore, we find that
with increasing $\alpha$, the amplitude of the even-$m$ modes after merger
decreases, but in all cases the artificial instability appears at lowest
resolution; in fact, even in the $\alpha=\pi$ case, where the even-$m$
modes are seeded at amplitudes consistent with floating point roundoff, we find
this behavior. In all cases considered, this growing perturbation at low resolution
saturates in the vortex ejection of the solution. 
However, we performed higher resolution evolutions in the binaries with
$\alpha/ \pi\in\{63/64,31/32,7/8\}$ and explicitly checked that the unstable behavior disappears.
This is illustrated for $\alpha/\pi=63/64$ in \figurename{ \ref{fig:modedecomposition}}.

\bibliography{bib.bib,bib2.bib}

\begin{thebibliography}{99}%
\makeatletter
\providecommand \@ifxundefined [1]{%
 \@ifx{#1\undefined}
}%
\providecommand \@ifnum [1]{%
 \ifnum #1\expandafter \@firstoftwo
 \else \expandafter \@secondoftwo
 \fi
}%
\providecommand \@ifx [1]{%
 \ifx #1\expandafter \@firstoftwo
 \else \expandafter \@secondoftwo
 \fi
}%
\providecommand \natexlab [1]{#1}%
\providecommand \enquote  [1]{``#1''}%
\providecommand \bibnamefont  [1]{#1}%
\providecommand \bibfnamefont [1]{#1}%
\providecommand \citenamefont [1]{#1}%
\providecommand \href@noop [0]{\@secondoftwo}%
\providecommand \href [0]{\begingroup \@sanitize@url \@href}%
\providecommand \@href[1]{\@@startlink{#1}\@@href}%
\providecommand \@@href[1]{\endgroup#1\@@endlink}%
\providecommand \@sanitize@url [0]{\catcode `\\12\catcode `\$12\catcode
  `\&12\catcode `\#12\catcode `\^12\catcode `\_12\catcode `\%12\relax}%
\providecommand \@@startlink[1]{}%
\providecommand \@@endlink[0]{}%
\providecommand \url  [0]{\begingroup\@sanitize@url \@url }%
\providecommand \@url [1]{\endgroup\@href {#1}{\urlprefix }}%
\providecommand \urlprefix  [0]{URL }%
\providecommand \Eprint [0]{\href }%
\providecommand \doibase [0]{http://dx.doi.org/}%
\providecommand \selectlanguage [0]{\@gobble}%
\providecommand \bibinfo  [0]{\@secondoftwo}%
\providecommand \bibfield  [0]{\@secondoftwo}%
\providecommand \translation [1]{[#1]}%
\providecommand \BibitemOpen [0]{}%
\providecommand \bibitemStop [0]{}%
\providecommand \bibitemNoStop [0]{.\EOS\space}%
\providecommand \EOS [0]{\spacefactor3000\relax}%
\providecommand \BibitemShut  [1]{\csname bibitem#1\endcsname}%
\let\auto@bib@innerbib\@empty
\bibitem [{\citenamefont {Cardoso}\ and\ \citenamefont
  {Pani}(2019)}]{Cardoso:2019rvt}%
  \BibitemOpen
  \bibfield  {author} {\bibinfo {author} {\bibfnamefont {V.}~\bibnamefont
  {Cardoso}}\ and\ \bibinfo {author} {\bibfnamefont {P.}~\bibnamefont {Pani}},\
  }\href {\doibase 10.1007/s41114-019-0020-4} {\bibfield  {journal} {\bibinfo
  {journal} {Living Rev. Rel.}\ }\textbf {\bibinfo {volume} {22}},\ \bibinfo
  {pages} {4} (\bibinfo {year} {2019})},\ \Eprint
  {http://arxiv.org/abs/1904.05363} {arXiv:1904.05363 [gr-qc]} \BibitemShut
  {NoStop}%
\bibitem [{\citenamefont {{Bowers}}\ and\ \citenamefont
  {{Liang}}(1974)}]{1974ApJ...188..657B}%
  \BibitemOpen
  \bibfield  {author} {\bibinfo {author} {\bibfnamefont {R.~L.}\ \bibnamefont
  {{Bowers}}}\ and\ \bibinfo {author} {\bibfnamefont {E.~P.~T.}\ \bibnamefont
  {{Liang}}},\ }\href {\doibase 10.1086/152760} {\bibfield  {journal} {\bibinfo
   {journal} {\apj}\ }\textbf {\bibinfo {volume} {188}},\ \bibinfo {pages}
  {657} (\bibinfo {year} {1974})}\BibitemShut {NoStop}%
\bibitem [{\citenamefont {Letelier}(1980)}]{PhysRevD.22.807}%
  \BibitemOpen
  \bibfield  {author} {\bibinfo {author} {\bibfnamefont {P.~S.}\ \bibnamefont
  {Letelier}},\ }\href {\doibase 10.1103/PhysRevD.22.807} {\bibfield  {journal}
  {\bibinfo  {journal} {Phys. Rev. D}\ }\textbf {\bibinfo {volume} {22}},\
  \bibinfo {pages} {807} (\bibinfo {year} {1980})}\BibitemShut {NoStop}%
\bibitem [{\citenamefont {Herrera}\ \emph {et~al.}(2004)\citenamefont
  {Herrera}, \citenamefont {Di~Prisco}, \citenamefont {Martin}, \citenamefont
  {Ospino}, \citenamefont {Santos},\ and\ \citenamefont
  {Troconis}}]{Herrera:2004xc}%
  \BibitemOpen
  \bibfield  {author} {\bibinfo {author} {\bibfnamefont {L.}~\bibnamefont
  {Herrera}}, \bibinfo {author} {\bibfnamefont {A.}~\bibnamefont {Di~Prisco}},
  \bibinfo {author} {\bibfnamefont {J.}~\bibnamefont {Martin}}, \bibinfo
  {author} {\bibfnamefont {J.}~\bibnamefont {Ospino}}, \bibinfo {author}
  {\bibfnamefont {N.}~\bibnamefont {Santos}}, \ and\ \bibinfo {author}
  {\bibfnamefont {O.}~\bibnamefont {Troconis}},\ }\href {\doibase
  10.1103/PhysRevD.69.084026} {\bibfield  {journal} {\bibinfo  {journal} {Phys.
  Rev. D}\ }\textbf {\bibinfo {volume} {69}},\ \bibinfo {pages} {084026}
  (\bibinfo {year} {2004})},\ \Eprint {http://arxiv.org/abs/gr-qc/0403006}
  {arXiv:gr-qc/0403006} \BibitemShut {NoStop}%
\bibitem [{\citenamefont {Mathur}(2005)}]{Mathur:2005zp}%
  \BibitemOpen
  \bibfield  {author} {\bibinfo {author} {\bibfnamefont {S.~D.}\ \bibnamefont
  {Mathur}},\ }\href {\doibase 10.1002/prop.200410203} {\bibfield  {journal}
  {\bibinfo  {journal} {Fortsch. Phys.}\ }\textbf {\bibinfo {volume} {53}},\
  \bibinfo {pages} {793} (\bibinfo {year} {2005})},\ \Eprint
  {http://arxiv.org/abs/hep-th/0502050} {arXiv:hep-th/0502050} \BibitemShut
  {NoStop}%
\bibitem [{\citenamefont {Bena}\ and\ \citenamefont
  {Warner}(2008)}]{Bena:2007kg}%
  \BibitemOpen
  \bibfield  {author} {\bibinfo {author} {\bibfnamefont {I.}~\bibnamefont
  {Bena}}\ and\ \bibinfo {author} {\bibfnamefont {N.~P.}\ \bibnamefont
  {Warner}},\ }\href {\doibase 10.1007/978-3-540-79523-0_1} {\bibfield
  {journal} {\bibinfo  {journal} {Lect. Notes Phys.}\ }\textbf {\bibinfo
  {volume} {755}},\ \bibinfo {pages} {1} (\bibinfo {year} {2008})},\ \Eprint
  {http://arxiv.org/abs/hep-th/0701216} {arXiv:hep-th/0701216} \BibitemShut
  {NoStop}%
\bibitem [{\citenamefont {Balasubramanian}\ \emph {et~al.}(2008)\citenamefont
  {Balasubramanian}, \citenamefont {de~Boer}, \citenamefont {El-Showk},\ and\
  \citenamefont {Messamah}}]{Balasubramanian:2008da}%
  \BibitemOpen
  \bibfield  {author} {\bibinfo {author} {\bibfnamefont {V.}~\bibnamefont
  {Balasubramanian}}, \bibinfo {author} {\bibfnamefont {J.}~\bibnamefont
  {de~Boer}}, \bibinfo {author} {\bibfnamefont {S.}~\bibnamefont {El-Showk}}, \
  and\ \bibinfo {author} {\bibfnamefont {I.}~\bibnamefont {Messamah}},\ }\href
  {\doibase 10.1088/0264-9381/25/21/214004} {\bibfield  {journal} {\bibinfo
  {journal} {Class. Quant. Grav.}\ }\textbf {\bibinfo {volume} {25}},\ \bibinfo
  {pages} {214004} (\bibinfo {year} {2008})},\ \Eprint
  {http://arxiv.org/abs/0811.0263} {arXiv:0811.0263 [hep-th]} \BibitemShut
  {NoStop}%
\bibitem [{\citenamefont {Cardoso}\ \emph {et~al.}(2016)\citenamefont
  {Cardoso}, \citenamefont {Hopper}, \citenamefont {Macedo}, \citenamefont
  {Palenzuela},\ and\ \citenamefont {Pani}}]{Cardoso:2016oxy}%
  \BibitemOpen
  \bibfield  {author} {\bibinfo {author} {\bibfnamefont {V.}~\bibnamefont
  {Cardoso}}, \bibinfo {author} {\bibfnamefont {S.}~\bibnamefont {Hopper}},
  \bibinfo {author} {\bibfnamefont {C.~F.~B.}\ \bibnamefont {Macedo}}, \bibinfo
  {author} {\bibfnamefont {C.}~\bibnamefont {Palenzuela}}, \ and\ \bibinfo
  {author} {\bibfnamefont {P.}~\bibnamefont {Pani}},\ }\href {\doibase
  10.1103/PhysRevD.94.084031} {\bibfield  {journal} {\bibinfo  {journal} {Phys.
  Rev. D}\ }\textbf {\bibinfo {volume} {94}},\ \bibinfo {pages} {084031}
  (\bibinfo {year} {2016})},\ \Eprint {http://arxiv.org/abs/1608.08637}
  {arXiv:1608.08637 [gr-qc]} \BibitemShut {NoStop}%
\bibitem [{\citenamefont {Friedman}(1978)}]{friedman1978}%
  \BibitemOpen
  \bibfield  {author} {\bibinfo {author} {\bibfnamefont {J.~L.}\ \bibnamefont
  {Friedman}},\ }\href@noop {} {\bibfield  {journal} {\bibinfo  {journal}
  {Comm. Math. Phys.}\ }\textbf {\bibinfo {volume} {63}},\ \bibinfo {pages}
  {243} (\bibinfo {year} {1978})}\BibitemShut {NoStop}%
\bibitem [{\citenamefont {Kaup}(1968)}]{Kaup:1968zz}%
  \BibitemOpen
  \bibfield  {author} {\bibinfo {author} {\bibfnamefont {D.~J.}\ \bibnamefont
  {Kaup}},\ }\href {\doibase 10.1103/PhysRev.172.1331} {\bibfield  {journal}
  {\bibinfo  {journal} {Phys. Rev.}\ }\textbf {\bibinfo {volume} {172}},\
  \bibinfo {pages} {1331} (\bibinfo {year} {1968})}\BibitemShut {NoStop}%
\bibitem [{\citenamefont {Ruffini}\ and\ \citenamefont
  {Bonazzola}(1969)}]{Ruffini:1969qy}%
  \BibitemOpen
  \bibfield  {author} {\bibinfo {author} {\bibfnamefont {R.}~\bibnamefont
  {Ruffini}}\ and\ \bibinfo {author} {\bibfnamefont {S.}~\bibnamefont
  {Bonazzola}},\ }\href {\doibase 10.1103/PhysRev.187.1767} {\bibfield
  {journal} {\bibinfo  {journal} {Phys. Rev.}\ }\textbf {\bibinfo {volume}
  {187}},\ \bibinfo {pages} {1767} (\bibinfo {year} {1969})}\BibitemShut
  {NoStop}%
\bibitem [{\citenamefont {Seidel}\ and\ \citenamefont
  {Suen}(1994)}]{Seidel:1993zk}%
  \BibitemOpen
  \bibfield  {author} {\bibinfo {author} {\bibfnamefont {E.}~\bibnamefont
  {Seidel}}\ and\ \bibinfo {author} {\bibfnamefont {W.-M.}\ \bibnamefont
  {Suen}},\ }\href {\doibase 10.1103/PhysRevLett.72.2516} {\bibfield  {journal}
  {\bibinfo  {journal} {Phys. Rev. Lett.}\ }\textbf {\bibinfo {volume} {72}},\
  \bibinfo {pages} {2516} (\bibinfo {year} {1994})},\ \Eprint
  {http://arxiv.org/abs/gr-qc/9309015} {arXiv:gr-qc/9309015} \BibitemShut
  {NoStop}%
\bibitem [{\citenamefont {Seidel}\ and\ \citenamefont
  {Suen}(1991)}]{Seidel:1991zh}%
  \BibitemOpen
  \bibfield  {author} {\bibinfo {author} {\bibfnamefont {E.}~\bibnamefont
  {Seidel}}\ and\ \bibinfo {author} {\bibfnamefont {W.~M.}\ \bibnamefont
  {Suen}},\ }\href {\doibase 10.1103/PhysRevLett.66.1659} {\bibfield  {journal}
  {\bibinfo  {journal} {Phys. Rev. Lett.}\ }\textbf {\bibinfo {volume} {66}},\
  \bibinfo {pages} {1659} (\bibinfo {year} {1991})}\BibitemShut {NoStop}%
\bibitem [{\citenamefont {Alcubierre}\ \emph {et~al.}(2003)\citenamefont
  {Alcubierre}, \citenamefont {Becerril}, \citenamefont {Guzman}, \citenamefont
  {Matos}, \citenamefont {Nunez},\ and\ \citenamefont
  {Urena-Lopez}}]{Alcubierre:2003sx}%
  \BibitemOpen
  \bibfield  {author} {\bibinfo {author} {\bibfnamefont {M.}~\bibnamefont
  {Alcubierre}}, \bibinfo {author} {\bibfnamefont {R.}~\bibnamefont
  {Becerril}}, \bibinfo {author} {\bibfnamefont {S.~F.}\ \bibnamefont
  {Guzman}}, \bibinfo {author} {\bibfnamefont {T.}~\bibnamefont {Matos}},
  \bibinfo {author} {\bibfnamefont {D.}~\bibnamefont {Nunez}}, \ and\ \bibinfo
  {author} {\bibfnamefont {L.~A.}\ \bibnamefont {Urena-Lopez}},\ }\href
  {\doibase 10.1088/0264-9381/20/13/332} {\bibfield  {journal} {\bibinfo
  {journal} {Class. Quant. Grav.}\ }\textbf {\bibinfo {volume} {20}},\ \bibinfo
  {pages} {2883} (\bibinfo {year} {2003})},\ \Eprint
  {http://arxiv.org/abs/gr-qc/0301105} {arXiv:gr-qc/0301105} \BibitemShut
  {NoStop}%
\bibitem [{\citenamefont {Schunck}\ and\ \citenamefont
  {Mielke}(2003)}]{Schunck:2003kk}%
  \BibitemOpen
  \bibfield  {author} {\bibinfo {author} {\bibfnamefont {F.~E.}\ \bibnamefont
  {Schunck}}\ and\ \bibinfo {author} {\bibfnamefont {E.~W.}\ \bibnamefont
  {Mielke}},\ }\href {\doibase 10.1088/0264-9381/20/20/201} {\bibfield
  {journal} {\bibinfo  {journal} {Class. Quant. Grav.}\ }\textbf {\bibinfo
  {volume} {20}},\ \bibinfo {pages} {R301} (\bibinfo {year} {2003})},\ \Eprint
  {http://arxiv.org/abs/0801.0307} {arXiv:0801.0307 [astro-ph]} \BibitemShut
  {NoStop}%
\bibitem [{\citenamefont {Liebling}\ and\ \citenamefont
  {Palenzuela}(2017)}]{Liebling:2012fv}%
  \BibitemOpen
  \bibfield  {author} {\bibinfo {author} {\bibfnamefont {S.~L.}\ \bibnamefont
  {Liebling}}\ and\ \bibinfo {author} {\bibfnamefont {C.}~\bibnamefont
  {Palenzuela}},\ }\href {\doibase 10.12942/lrr-2012-6} {\bibfield  {journal}
  {\bibinfo  {journal} {Living Rev. Rel.}\ }\textbf {\bibinfo {volume} {20}},\
  \bibinfo {pages} {5} (\bibinfo {year} {2017})},\ \Eprint
  {http://arxiv.org/abs/1202.5809} {arXiv:1202.5809 [gr-qc]} \BibitemShut
  {NoStop}%
\bibitem [{\citenamefont {Visinelli}(2021)}]{Visinelli:2021uve}%
  \BibitemOpen
  \bibfield  {author} {\bibinfo {author} {\bibfnamefont {L.}~\bibnamefont
  {Visinelli}},\ }\href {\doibase 10.1142/S0218271821300068} {\bibfield
  {journal} {\bibinfo  {journal} {Int. J. Mod. Phys. D}\ }\textbf {\bibinfo
  {volume} {30}},\ \bibinfo {pages} {2130006} (\bibinfo {year} {2021})},\
  \Eprint {http://arxiv.org/abs/2109.05481} {arXiv:2109.05481 [gr-qc]}
  \BibitemShut {NoStop}%
\bibitem [{\citenamefont {Friedberg}\ \emph {et~al.}(1987)\citenamefont
  {Friedberg}, \citenamefont {Lee},\ and\ \citenamefont
  {Pang}}]{Friedberg:1986tq}%
  \BibitemOpen
  \bibfield  {author} {\bibinfo {author} {\bibfnamefont {R.}~\bibnamefont
  {Friedberg}}, \bibinfo {author} {\bibfnamefont {T.}~\bibnamefont {Lee}}, \
  and\ \bibinfo {author} {\bibfnamefont {Y.}~\bibnamefont {Pang}},\ }\href
  {\doibase 10.1103/PhysRevD.35.3658} {\bibfield  {journal} {\bibinfo
  {journal} {Phys. Rev. D}\ }\textbf {\bibinfo {volume} {35}},\ \bibinfo
  {pages} {3658} (\bibinfo {year} {1987})}\BibitemShut {NoStop}%
\bibitem [{\citenamefont {Balakrishna}\ \emph {et~al.}(1998)\citenamefont
  {Balakrishna}, \citenamefont {Seidel},\ and\ \citenamefont
  {Suen}}]{Balakrishna:1997ej}%
  \BibitemOpen
  \bibfield  {author} {\bibinfo {author} {\bibfnamefont {J.}~\bibnamefont
  {Balakrishna}}, \bibinfo {author} {\bibfnamefont {E.}~\bibnamefont {Seidel}},
  \ and\ \bibinfo {author} {\bibfnamefont {W.-M.}\ \bibnamefont {Suen}},\
  }\href {\doibase 10.1103/PhysRevD.58.104004} {\bibfield  {journal} {\bibinfo
  {journal} {Phys. Rev. D}\ }\textbf {\bibinfo {volume} {58}},\ \bibinfo
  {pages} {104004} (\bibinfo {year} {1998})},\ \Eprint
  {http://arxiv.org/abs/gr-qc/9712064} {arXiv:gr-qc/9712064} \BibitemShut
  {NoStop}%
\bibitem [{\citenamefont {Schunck}\ and\ \citenamefont
  {Torres}(2000)}]{Schunck:1999zu}%
  \BibitemOpen
  \bibfield  {author} {\bibinfo {author} {\bibfnamefont {F.~E.}\ \bibnamefont
  {Schunck}}\ and\ \bibinfo {author} {\bibfnamefont {D.~F.}\ \bibnamefont
  {Torres}},\ }\href {\doibase 10.1142/S0218271800000608} {\bibfield  {journal}
  {\bibinfo  {journal} {Int. J. Mod. Phys. D}\ }\textbf {\bibinfo {volume}
  {9}},\ \bibinfo {pages} {601} (\bibinfo {year} {2000})},\ \Eprint
  {http://arxiv.org/abs/gr-qc/9911038} {arXiv:gr-qc/9911038} \BibitemShut
  {NoStop}%
\bibitem [{\citenamefont {Sorkin}(1981)}]{Sorkin:1981jc}%
  \BibitemOpen
  \bibfield  {author} {\bibinfo {author} {\bibfnamefont {R.}~\bibnamefont
  {Sorkin}},\ }\href {\doibase 10.1086/159282} {\bibfield  {journal} {\bibinfo
  {journal} {Astrophys. J.}\ }\textbf {\bibinfo {volume} {249}},\ \bibinfo
  {pages} {254} (\bibinfo {year} {1981})}\BibitemShut {NoStop}%
\bibitem [{\citenamefont {Gleiser}(1988)}]{Gleiser:1988rq}%
  \BibitemOpen
  \bibfield  {author} {\bibinfo {author} {\bibfnamefont {M.}~\bibnamefont
  {Gleiser}},\ }\href {\doibase 10.1103/PhysRevD.38.2376} {\bibfield  {journal}
  {\bibinfo  {journal} {Phys. Rev. D}\ }\textbf {\bibinfo {volume} {38}},\
  \bibinfo {pages} {2376} (\bibinfo {year} {1988})},\ \bibinfo {note}
  {[Erratum: Phys.Rev.D 39, 1257 (1989)]}\BibitemShut {NoStop}%
\bibitem [{\citenamefont {Gleiser}\ and\ \citenamefont
  {Watkins}(1989)}]{Gleiser:1988ih}%
  \BibitemOpen
  \bibfield  {author} {\bibinfo {author} {\bibfnamefont {M.}~\bibnamefont
  {Gleiser}}\ and\ \bibinfo {author} {\bibfnamefont {R.}~\bibnamefont
  {Watkins}},\ }\href {\doibase 10.1016/0550-3213(89)90627-5} {\bibfield
  {journal} {\bibinfo  {journal} {Nucl. Phys. B}\ }\textbf {\bibinfo {volume}
  {319}},\ \bibinfo {pages} {733} (\bibinfo {year} {1989})}\BibitemShut
  {NoStop}%
\bibitem [{\citenamefont {Lee}\ and\ \citenamefont {Pang}(1989)}]{Lee:1988av}%
  \BibitemOpen
  \bibfield  {author} {\bibinfo {author} {\bibfnamefont {T.}~\bibnamefont
  {Lee}}\ and\ \bibinfo {author} {\bibfnamefont {Y.}~\bibnamefont {Pang}},\
  }\href {\doibase 10.1016/0550-3213(89)90365-9} {\bibfield  {journal}
  {\bibinfo  {journal} {Nucl. Phys. B}\ }\textbf {\bibinfo {volume} {315}},\
  \bibinfo {pages} {477} (\bibinfo {year} {1989})}\BibitemShut {NoStop}%
\bibitem [{\citenamefont {Kusmartsev}\ \emph {et~al.}(1991)\citenamefont
  {Kusmartsev}, \citenamefont {Mielke},\ and\ \citenamefont
  {Schunck}}]{Kusmartsev:1990cr}%
  \BibitemOpen
  \bibfield  {author} {\bibinfo {author} {\bibfnamefont {F.~V.}\ \bibnamefont
  {Kusmartsev}}, \bibinfo {author} {\bibfnamefont {E.~W.}\ \bibnamefont
  {Mielke}}, \ and\ \bibinfo {author} {\bibfnamefont {F.~E.}\ \bibnamefont
  {Schunck}},\ }\href {\doibase 10.1103/PhysRevD.43.3895} {\bibfield  {journal}
  {\bibinfo  {journal} {Phys. Rev. D}\ }\textbf {\bibinfo {volume} {43}},\
  \bibinfo {pages} {3895} (\bibinfo {year} {1991})},\ \Eprint
  {http://arxiv.org/abs/0810.0696} {arXiv:0810.0696 [astro-ph]} \BibitemShut
  {NoStop}%
\bibitem [{\citenamefont {Guzman}(2004)}]{Guzman:2004jw}%
  \BibitemOpen
  \bibfield  {author} {\bibinfo {author} {\bibfnamefont {F.}~\bibnamefont
  {Guzman}},\ }\href {\doibase 10.1103/PhysRevD.70.044033} {\bibfield
  {journal} {\bibinfo  {journal} {Phys. Rev. D}\ }\textbf {\bibinfo {volume}
  {70}},\ \bibinfo {pages} {044033} (\bibinfo {year} {2004})},\ \Eprint
  {http://arxiv.org/abs/gr-qc/0407054} {arXiv:gr-qc/0407054} \BibitemShut
  {NoStop}%
\bibitem [{\citenamefont {Sanchis-Gual}\ \emph
  {et~al.}(2022{\natexlab{a}})\citenamefont {Sanchis-Gual}, \citenamefont
  {Herdeiro},\ and\ \citenamefont {Radu}}]{Sanchis-Gual:2021phr}%
  \BibitemOpen
  \bibfield  {author} {\bibinfo {author} {\bibfnamefont {N.}~\bibnamefont
  {Sanchis-Gual}}, \bibinfo {author} {\bibfnamefont {C.}~\bibnamefont
  {Herdeiro}}, \ and\ \bibinfo {author} {\bibfnamefont {E.}~\bibnamefont
  {Radu}},\ }\href {\doibase 10.1088/1361-6382/ac4b9b} {\bibfield  {journal}
  {\bibinfo  {journal} {Class. Quant. Grav.}\ }\textbf {\bibinfo {volume}
  {39}},\ \bibinfo {pages} {064001} (\bibinfo {year} {2022}{\natexlab{a}})},\
  \Eprint {http://arxiv.org/abs/2110.03000} {arXiv:2110.03000 [gr-qc]}
  \BibitemShut {NoStop}%
\bibitem [{\citenamefont {Guzman}\ and\ \citenamefont
  {Urena-Lopez}(2003)}]{Guzman:2003kt}%
  \BibitemOpen
  \bibfield  {author} {\bibinfo {author} {\bibfnamefont {F.~S.}\ \bibnamefont
  {Guzman}}\ and\ \bibinfo {author} {\bibfnamefont {L.~A.}\ \bibnamefont
  {Urena-Lopez}},\ }\href {\doibase 10.1103/PhysRevD.68.024023} {\bibfield
  {journal} {\bibinfo  {journal} {Phys. Rev. D}\ }\textbf {\bibinfo {volume}
  {68}},\ \bibinfo {pages} {024023} (\bibinfo {year} {2003})},\ \Eprint
  {http://arxiv.org/abs/astro-ph/0303440} {arXiv:astro-ph/0303440} \BibitemShut
  {NoStop}%
\bibitem [{\citenamefont {Amin}\ and\ \citenamefont
  {Mocz}(2019)}]{Amin:2019ums}%
  \BibitemOpen
  \bibfield  {author} {\bibinfo {author} {\bibfnamefont {M.~A.}\ \bibnamefont
  {Amin}}\ and\ \bibinfo {author} {\bibfnamefont {P.}~\bibnamefont {Mocz}},\
  }\href {\doibase 10.1103/PhysRevD.100.063507} {\bibfield  {journal} {\bibinfo
   {journal} {Phys. Rev. D}\ }\textbf {\bibinfo {volume} {100}},\ \bibinfo
  {pages} {063507} (\bibinfo {year} {2019})},\ \Eprint
  {http://arxiv.org/abs/1902.07261} {arXiv:1902.07261 [astro-ph.CO]}
  \BibitemShut {NoStop}%
\bibitem [{\citenamefont {Levkov}\ \emph {et~al.}(2018)\citenamefont {Levkov},
  \citenamefont {Panin},\ and\ \citenamefont {Tkachev}}]{Levkov:2018kau}%
  \BibitemOpen
  \bibfield  {author} {\bibinfo {author} {\bibfnamefont {D.~G.}\ \bibnamefont
  {Levkov}}, \bibinfo {author} {\bibfnamefont {A.~G.}\ \bibnamefont {Panin}}, \
  and\ \bibinfo {author} {\bibfnamefont {I.~I.}\ \bibnamefont {Tkachev}},\
  }\href {\doibase 10.1103/PhysRevLett.121.151301} {\bibfield  {journal}
  {\bibinfo  {journal} {Phys. Rev. Lett.}\ }\textbf {\bibinfo {volume} {121}},\
  \bibinfo {pages} {151301} (\bibinfo {year} {2018})},\ \Eprint
  {http://arxiv.org/abs/1804.05857} {arXiv:1804.05857 [astro-ph.CO]}
  \BibitemShut {NoStop}%
\bibitem [{\citenamefont {Veltmaat}\ \emph {et~al.}(2018)\citenamefont
  {Veltmaat}, \citenamefont {Niemeyer},\ and\ \citenamefont
  {Schwabe}}]{Veltmaat:2018dfz}%
  \BibitemOpen
  \bibfield  {author} {\bibinfo {author} {\bibfnamefont {J.}~\bibnamefont
  {Veltmaat}}, \bibinfo {author} {\bibfnamefont {J.~C.}\ \bibnamefont
  {Niemeyer}}, \ and\ \bibinfo {author} {\bibfnamefont {B.}~\bibnamefont
  {Schwabe}},\ }\href {\doibase 10.1103/PhysRevD.98.043509} {\bibfield
  {journal} {\bibinfo  {journal} {Phys. Rev. D}\ }\textbf {\bibinfo {volume}
  {98}},\ \bibinfo {pages} {043509} (\bibinfo {year} {2018})},\ \Eprint
  {http://arxiv.org/abs/1804.09647} {arXiv:1804.09647 [astro-ph.CO]}
  \BibitemShut {NoStop}%
\bibitem [{\citenamefont {Arvanitaki}\ \emph {et~al.}(2020)\citenamefont
  {Arvanitaki}, \citenamefont {Dimopoulos}, \citenamefont {Galanis},
  \citenamefont {Lehner}, \citenamefont {Thompson},\ and\ \citenamefont
  {Van~Tilburg}}]{Arvanitaki:2019rax}%
  \BibitemOpen
  \bibfield  {author} {\bibinfo {author} {\bibfnamefont {A.}~\bibnamefont
  {Arvanitaki}}, \bibinfo {author} {\bibfnamefont {S.}~\bibnamefont
  {Dimopoulos}}, \bibinfo {author} {\bibfnamefont {M.}~\bibnamefont {Galanis}},
  \bibinfo {author} {\bibfnamefont {L.}~\bibnamefont {Lehner}}, \bibinfo
  {author} {\bibfnamefont {J.~O.}\ \bibnamefont {Thompson}}, \ and\ \bibinfo
  {author} {\bibfnamefont {K.}~\bibnamefont {Van~Tilburg}},\ }\href {\doibase
  10.1103/PhysRevD.101.083014} {\bibfield  {journal} {\bibinfo  {journal}
  {Phys. Rev. D}\ }\textbf {\bibinfo {volume} {101}},\ \bibinfo {pages}
  {083014} (\bibinfo {year} {2020})},\ \Eprint
  {http://arxiv.org/abs/1909.11665} {arXiv:1909.11665 [astro-ph.CO]}
  \BibitemShut {NoStop}%
\bibitem [{\citenamefont {Lai}(2004)}]{Lai:2004fw}%
  \BibitemOpen
  \bibfield  {author} {\bibinfo {author} {\bibfnamefont {C.-W.}\ \bibnamefont
  {Lai}},\ }\emph {\bibinfo {title} {{A Numerical study of boson stars}}},\
  \href@noop {} {\bibinfo {type} {Other thesis}} (\bibinfo {year} {2004}),\
  \Eprint {http://arxiv.org/abs/gr-qc/0410040} {arXiv:gr-qc/0410040}
  \BibitemShut {NoStop}%
\bibitem [{\citenamefont {Choptuik}\ and\ \citenamefont
  {Pretorius}(2010)}]{Choptuik:2009ww}%
  \BibitemOpen
  \bibfield  {author} {\bibinfo {author} {\bibfnamefont {M.~W.}\ \bibnamefont
  {Choptuik}}\ and\ \bibinfo {author} {\bibfnamefont {F.}~\bibnamefont
  {Pretorius}},\ }\href {\doibase 10.1103/PhysRevLett.104.111101} {\bibfield
  {journal} {\bibinfo  {journal} {Phys. Rev. Lett.}\ }\textbf {\bibinfo
  {volume} {104}},\ \bibinfo {pages} {111101} (\bibinfo {year} {2010})},\
  \Eprint {http://arxiv.org/abs/0908.1780} {arXiv:0908.1780 [gr-qc]}
  \BibitemShut {NoStop}%
\bibitem [{\citenamefont {Paredes}\ and\ \citenamefont
  {Michinel}(2016)}]{Paredes:2015wga}%
  \BibitemOpen
  \bibfield  {author} {\bibinfo {author} {\bibfnamefont {A.}~\bibnamefont
  {Paredes}}\ and\ \bibinfo {author} {\bibfnamefont {H.}~\bibnamefont
  {Michinel}},\ }\href {\doibase 10.1016/j.dark.2016.02.003} {\bibfield
  {journal} {\bibinfo  {journal} {Phys. Dark Univ.}\ }\textbf {\bibinfo
  {volume} {12}},\ \bibinfo {pages} {50} (\bibinfo {year} {2016})},\ \Eprint
  {http://arxiv.org/abs/1512.05121} {arXiv:1512.05121 [astro-ph.CO]}
  \BibitemShut {NoStop}%
\bibitem [{\citenamefont {Bernal}\ and\ \citenamefont
  {Siddhartha~Guzman}(2006)}]{Bernal:2006ci}%
  \BibitemOpen
  \bibfield  {author} {\bibinfo {author} {\bibfnamefont {A.}~\bibnamefont
  {Bernal}}\ and\ \bibinfo {author} {\bibfnamefont {F.}~\bibnamefont
  {Siddhartha~Guzman}},\ }\href {\doibase 10.1103/PhysRevD.74.103002}
  {\bibfield  {journal} {\bibinfo  {journal} {Phys. Rev. D}\ }\textbf {\bibinfo
  {volume} {74}},\ \bibinfo {pages} {103002} (\bibinfo {year} {2006})},\
  \Eprint {http://arxiv.org/abs/astro-ph/0610682} {arXiv:astro-ph/0610682}
  \BibitemShut {NoStop}%
\bibitem [{\citenamefont {Schwabe}\ \emph {et~al.}(2016)\citenamefont
  {Schwabe}, \citenamefont {Niemeyer},\ and\ \citenamefont
  {Engels}}]{Schwabe:2016rze}%
  \BibitemOpen
  \bibfield  {author} {\bibinfo {author} {\bibfnamefont {B.}~\bibnamefont
  {Schwabe}}, \bibinfo {author} {\bibfnamefont {J.~C.}\ \bibnamefont
  {Niemeyer}}, \ and\ \bibinfo {author} {\bibfnamefont {J.~F.}\ \bibnamefont
  {Engels}},\ }\href {\doibase 10.1103/PhysRevD.94.043513} {\bibfield
  {journal} {\bibinfo  {journal} {Phys. Rev. D}\ }\textbf {\bibinfo {volume}
  {94}},\ \bibinfo {pages} {043513} (\bibinfo {year} {2016})},\ \Eprint
  {http://arxiv.org/abs/1606.05151} {arXiv:1606.05151 [astro-ph.CO]}
  \BibitemShut {NoStop}%
\bibitem [{\citenamefont {Palenzuela}\ \emph {et~al.}(2007)\citenamefont
  {Palenzuela}, \citenamefont {Olabarrieta}, \citenamefont {Lehner},\ and\
  \citenamefont {Liebling}}]{Palenzuela:2006wp}%
  \BibitemOpen
  \bibfield  {author} {\bibinfo {author} {\bibfnamefont {C.}~\bibnamefont
  {Palenzuela}}, \bibinfo {author} {\bibfnamefont {I.}~\bibnamefont
  {Olabarrieta}}, \bibinfo {author} {\bibfnamefont {L.}~\bibnamefont {Lehner}},
  \ and\ \bibinfo {author} {\bibfnamefont {S.~L.}\ \bibnamefont {Liebling}},\
  }\href {\doibase 10.1103/PhysRevD.75.064005} {\bibfield  {journal} {\bibinfo
  {journal} {Phys. Rev. D}\ }\textbf {\bibinfo {volume} {75}},\ \bibinfo
  {pages} {064005} (\bibinfo {year} {2007})},\ \Eprint
  {http://arxiv.org/abs/gr-qc/0612067} {arXiv:gr-qc/0612067} \BibitemShut
  {NoStop}%
\bibitem [{\citenamefont {Mundim}(2010)}]{Mundim:2010hi}%
  \BibitemOpen
  \bibfield  {author} {\bibinfo {author} {\bibfnamefont {B.~C.}\ \bibnamefont
  {Mundim}},\ }\emph {\bibinfo {title} {{A Numerical Study of Boson Star
  Binaries}}},\ \href@noop {} {Ph.D. thesis},\ \bibinfo  {school} {British
  Columbia U.} (\bibinfo {year} {2010}),\ \Eprint
  {http://arxiv.org/abs/1003.0239} {arXiv:1003.0239 [gr-qc]} \BibitemShut
  {NoStop}%
\bibitem [{\citenamefont {Bezares}\ \emph {et~al.}(2017)\citenamefont
  {Bezares}, \citenamefont {Palenzuela},\ and\ \citenamefont
  {Bona}}]{Bezares:2017mzk}%
  \BibitemOpen
  \bibfield  {author} {\bibinfo {author} {\bibfnamefont {M.}~\bibnamefont
  {Bezares}}, \bibinfo {author} {\bibfnamefont {C.}~\bibnamefont {Palenzuela}},
  \ and\ \bibinfo {author} {\bibfnamefont {C.}~\bibnamefont {Bona}},\ }\href
  {\doibase 10.1103/PhysRevD.95.124005} {\bibfield  {journal} {\bibinfo
  {journal} {Phys. Rev. D}\ }\textbf {\bibinfo {volume} {95}},\ \bibinfo
  {pages} {124005} (\bibinfo {year} {2017})},\ \Eprint
  {http://arxiv.org/abs/1705.01071} {arXiv:1705.01071 [gr-qc]} \BibitemShut
  {NoStop}%
\bibitem [{\citenamefont {Helfer}\ \emph {et~al.}(2021)\citenamefont {Helfer},
  \citenamefont {Sperhake}, \citenamefont {Croft}, \citenamefont {Radia},
  \citenamefont {Ge},\ and\ \citenamefont {Lim}}]{Helfer:2021brt}%
  \BibitemOpen
  \bibfield  {author} {\bibinfo {author} {\bibfnamefont {T.}~\bibnamefont
  {Helfer}}, \bibinfo {author} {\bibfnamefont {U.}~\bibnamefont {Sperhake}},
  \bibinfo {author} {\bibfnamefont {R.}~\bibnamefont {Croft}}, \bibinfo
  {author} {\bibfnamefont {M.}~\bibnamefont {Radia}}, \bibinfo {author}
  {\bibfnamefont {B.-X.}\ \bibnamefont {Ge}}, \ and\ \bibinfo {author}
  {\bibfnamefont {E.~A.}\ \bibnamefont {Lim}},\ }\href@noop {} {\  (\bibinfo
  {year} {2021})},\ \Eprint {http://arxiv.org/abs/2108.11995} {arXiv:2108.11995
  [gr-qc]} \BibitemShut {NoStop}%
\bibitem [{\citenamefont {Palenzuela}\ \emph {et~al.}(2008)\citenamefont
  {Palenzuela}, \citenamefont {Lehner},\ and\ \citenamefont
  {Liebling}}]{Palenzuela:2007dm}%
  \BibitemOpen
  \bibfield  {author} {\bibinfo {author} {\bibfnamefont {C.}~\bibnamefont
  {Palenzuela}}, \bibinfo {author} {\bibfnamefont {L.}~\bibnamefont {Lehner}},
  \ and\ \bibinfo {author} {\bibfnamefont {S.~L.}\ \bibnamefont {Liebling}},\
  }\href {\doibase 10.1103/PhysRevD.77.044036} {\bibfield  {journal} {\bibinfo
  {journal} {Phys. Rev. D}\ }\textbf {\bibinfo {volume} {77}},\ \bibinfo
  {pages} {044036} (\bibinfo {year} {2008})},\ \Eprint
  {http://arxiv.org/abs/0706.2435} {arXiv:0706.2435 [gr-qc]} \BibitemShut
  {NoStop}%
\bibitem [{\citenamefont {Palenzuela}\ \emph {et~al.}(2017)\citenamefont
  {Palenzuela}, \citenamefont {Pani}, \citenamefont {Bezares}, \citenamefont
  {Cardoso}, \citenamefont {Lehner},\ and\ \citenamefont
  {Liebling}}]{Palenzuela:2017kcg}%
  \BibitemOpen
  \bibfield  {author} {\bibinfo {author} {\bibfnamefont {C.}~\bibnamefont
  {Palenzuela}}, \bibinfo {author} {\bibfnamefont {P.}~\bibnamefont {Pani}},
  \bibinfo {author} {\bibfnamefont {M.}~\bibnamefont {Bezares}}, \bibinfo
  {author} {\bibfnamefont {V.}~\bibnamefont {Cardoso}}, \bibinfo {author}
  {\bibfnamefont {L.}~\bibnamefont {Lehner}}, \ and\ \bibinfo {author}
  {\bibfnamefont {S.}~\bibnamefont {Liebling}},\ }\href {\doibase
  10.1103/PhysRevD.96.104058} {\bibfield  {journal} {\bibinfo  {journal} {Phys.
  Rev. D}\ }\textbf {\bibinfo {volume} {96}},\ \bibinfo {pages} {104058}
  (\bibinfo {year} {2017})},\ \Eprint {http://arxiv.org/abs/1710.09432}
  {arXiv:1710.09432 [gr-qc]} \BibitemShut {NoStop}%
\bibitem [{\citenamefont {Bezares}\ \emph {et~al.}(2022)\citenamefont
  {Bezares}, \citenamefont {Bo\v{s}kovi\'c}, \citenamefont {Liebling},
  \citenamefont {Palenzuela}, \citenamefont {Pani},\ and\ \citenamefont
  {Barausse}}]{Bezares:2022obu}%
  \BibitemOpen
  \bibfield  {author} {\bibinfo {author} {\bibfnamefont {M.}~\bibnamefont
  {Bezares}}, \bibinfo {author} {\bibfnamefont {M.}~\bibnamefont
  {Bo\v{s}kovi\'c}}, \bibinfo {author} {\bibfnamefont {S.}~\bibnamefont
  {Liebling}}, \bibinfo {author} {\bibfnamefont {C.}~\bibnamefont
  {Palenzuela}}, \bibinfo {author} {\bibfnamefont {P.}~\bibnamefont {Pani}}, \
  and\ \bibinfo {author} {\bibfnamefont {E.}~\bibnamefont {Barausse}},\ }\href
  {\doibase 10.1103/PhysRevD.105.064067} {\bibfield  {journal} {\bibinfo
  {journal} {Phys. Rev. D}\ }\textbf {\bibinfo {volume} {105}},\ \bibinfo
  {pages} {064067} (\bibinfo {year} {2022})},\ \Eprint
  {http://arxiv.org/abs/2201.06113} {arXiv:2201.06113 [gr-qc]} \BibitemShut
  {NoStop}%
\bibitem [{\citenamefont {Bezares}\ and\ \citenamefont
  {Palenzuela}(2018)}]{Bezares:2018qwa}%
  \BibitemOpen
  \bibfield  {author} {\bibinfo {author} {\bibfnamefont {M.}~\bibnamefont
  {Bezares}}\ and\ \bibinfo {author} {\bibfnamefont {C.}~\bibnamefont
  {Palenzuela}},\ }\href {\doibase 10.1088/1361-6382/aae87c} {\bibfield
  {journal} {\bibinfo  {journal} {Class. Quant. Grav.}\ }\textbf {\bibinfo
  {volume} {35}},\ \bibinfo {pages} {234002} (\bibinfo {year} {2018})},\
  \Eprint {http://arxiv.org/abs/1808.10732} {arXiv:1808.10732 [gr-qc]}
  \BibitemShut {NoStop}%
\bibitem [{\citenamefont {Brito}\ \emph {et~al.}(2016)\citenamefont {Brito},
  \citenamefont {Cardoso}, \citenamefont {Herdeiro},\ and\ \citenamefont
  {Radu}}]{Brito:2015pxa}%
  \BibitemOpen
  \bibfield  {author} {\bibinfo {author} {\bibfnamefont {R.}~\bibnamefont
  {Brito}}, \bibinfo {author} {\bibfnamefont {V.}~\bibnamefont {Cardoso}},
  \bibinfo {author} {\bibfnamefont {C.~A.~R.}\ \bibnamefont {Herdeiro}}, \ and\
  \bibinfo {author} {\bibfnamefont {E.}~\bibnamefont {Radu}},\ }\href {\doibase
  10.1016/j.physletb.2015.11.051} {\bibfield  {journal} {\bibinfo  {journal}
  {Phys. Lett. B}\ }\textbf {\bibinfo {volume} {752}},\ \bibinfo {pages} {291}
  (\bibinfo {year} {2016})},\ \Eprint {http://arxiv.org/abs/1508.05395}
  {arXiv:1508.05395 [gr-qc]} \BibitemShut {NoStop}%
\bibitem [{\citenamefont {Sanchis-Gual}\ \emph
  {et~al.}(2022{\natexlab{b}})\citenamefont {Sanchis-Gual}, \citenamefont
  {Calder\'on~Bustillo}, \citenamefont {Herdeiro}, \citenamefont {Radu},
  \citenamefont {Font}, \citenamefont {Leong},\ and\ \citenamefont
  {Torres-Forn\'e}}]{Sanchis-Gual:2022mkk}%
  \BibitemOpen
  \bibfield  {author} {\bibinfo {author} {\bibfnamefont {N.}~\bibnamefont
  {Sanchis-Gual}}, \bibinfo {author} {\bibfnamefont {J.}~\bibnamefont
  {Calder\'on~Bustillo}}, \bibinfo {author} {\bibfnamefont {C.}~\bibnamefont
  {Herdeiro}}, \bibinfo {author} {\bibfnamefont {E.}~\bibnamefont {Radu}},
  \bibinfo {author} {\bibfnamefont {J.~A.}\ \bibnamefont {Font}}, \bibinfo
  {author} {\bibfnamefont {S.~H.~W.}\ \bibnamefont {Leong}}, \ and\ \bibinfo
  {author} {\bibfnamefont {A.}~\bibnamefont {Torres-Forn\'e}},\ }\href@noop {}
  {\  (\bibinfo {year} {2022}{\natexlab{b}})},\ \Eprint
  {http://arxiv.org/abs/2208.11717} {arXiv:2208.11717 [gr-qc]} \BibitemShut
  {NoStop}%
\bibitem [{\citenamefont {Sanchis-Gual}\ \emph
  {et~al.}(2019{\natexlab{a}})\citenamefont {Sanchis-Gual}, \citenamefont
  {Herdeiro}, \citenamefont {Font}, \citenamefont {Radu},\ and\ \citenamefont
  {Di~Giovanni}}]{Sanchis-Gual:2018oui}%
  \BibitemOpen
  \bibfield  {author} {\bibinfo {author} {\bibfnamefont {N.}~\bibnamefont
  {Sanchis-Gual}}, \bibinfo {author} {\bibfnamefont {C.}~\bibnamefont
  {Herdeiro}}, \bibinfo {author} {\bibfnamefont {J.~A.}\ \bibnamefont {Font}},
  \bibinfo {author} {\bibfnamefont {E.}~\bibnamefont {Radu}}, \ and\ \bibinfo
  {author} {\bibfnamefont {F.}~\bibnamefont {Di~Giovanni}},\ }\href {\doibase
  10.1103/PhysRevD.99.024017} {\bibfield  {journal} {\bibinfo  {journal} {Phys.
  Rev. D}\ }\textbf {\bibinfo {volume} {99}},\ \bibinfo {pages} {024017}
  (\bibinfo {year} {2019}{\natexlab{a}})},\ \Eprint
  {http://arxiv.org/abs/1806.07779} {arXiv:1806.07779 [gr-qc]} \BibitemShut
  {NoStop}%
\bibitem [{\citenamefont {Di~Giovanni}\ \emph {et~al.}(2018)\citenamefont
  {Di~Giovanni}, \citenamefont {Sanchis-Gual}, \citenamefont {Herdeiro},\ and\
  \citenamefont {Font}}]{DiGiovanni:2018bvo}%
  \BibitemOpen
  \bibfield  {author} {\bibinfo {author} {\bibfnamefont {F.}~\bibnamefont
  {Di~Giovanni}}, \bibinfo {author} {\bibfnamefont {N.}~\bibnamefont
  {Sanchis-Gual}}, \bibinfo {author} {\bibfnamefont {C.~A.~R.}\ \bibnamefont
  {Herdeiro}}, \ and\ \bibinfo {author} {\bibfnamefont {J.~A.}\ \bibnamefont
  {Font}},\ }\href {\doibase 10.1103/PhysRevD.98.064044} {\bibfield  {journal}
  {\bibinfo  {journal} {Phys. Rev. D}\ }\textbf {\bibinfo {volume} {98}},\
  \bibinfo {pages} {064044} (\bibinfo {year} {2018})},\ \Eprint
  {http://arxiv.org/abs/1803.04802} {arXiv:1803.04802 [gr-qc]} \BibitemShut
  {NoStop}%
\bibitem [{\citenamefont {Sanchis-Gual}\ \emph {et~al.}(2017)\citenamefont
  {Sanchis-Gual}, \citenamefont {Herdeiro}, \citenamefont {Radu}, \citenamefont
  {Degollado},\ and\ \citenamefont {Font}}]{Sanchis-Gual:2017bhw}%
  \BibitemOpen
  \bibfield  {author} {\bibinfo {author} {\bibfnamefont {N.}~\bibnamefont
  {Sanchis-Gual}}, \bibinfo {author} {\bibfnamefont {C.}~\bibnamefont
  {Herdeiro}}, \bibinfo {author} {\bibfnamefont {E.}~\bibnamefont {Radu}},
  \bibinfo {author} {\bibfnamefont {J.~C.}\ \bibnamefont {Degollado}}, \ and\
  \bibinfo {author} {\bibfnamefont {J.~A.}\ \bibnamefont {Font}},\ }\href
  {\doibase 10.1103/PhysRevD.95.104028} {\bibfield  {journal} {\bibinfo
  {journal} {Phys. Rev. D}\ }\textbf {\bibinfo {volume} {95}},\ \bibinfo
  {pages} {104028} (\bibinfo {year} {2017})},\ \Eprint
  {http://arxiv.org/abs/1702.04532} {arXiv:1702.04532 [gr-qc]} \BibitemShut
  {NoStop}%
\bibitem [{\citenamefont {Kobayashi}\ \emph {et~al.}(1994)\citenamefont
  {Kobayashi}, \citenamefont {Kasai},\ and\ \citenamefont
  {Futamase}}]{Kobayashi:1994qi}%
  \BibitemOpen
  \bibfield  {author} {\bibinfo {author} {\bibfnamefont {Y.}~\bibnamefont
  {Kobayashi}}, \bibinfo {author} {\bibfnamefont {M.}~\bibnamefont {Kasai}}, \
  and\ \bibinfo {author} {\bibfnamefont {T.}~\bibnamefont {Futamase}},\ }\href
  {\doibase 10.1103/PhysRevD.50.7721} {\bibfield  {journal} {\bibinfo
  {journal} {Phys. Rev. D}\ }\textbf {\bibinfo {volume} {50}},\ \bibinfo
  {pages} {7721} (\bibinfo {year} {1994})}\BibitemShut {NoStop}%
\bibitem [{\citenamefont {Kling}\ \emph {et~al.}(2021)\citenamefont {Kling},
  \citenamefont {Rajaraman},\ and\ \citenamefont {Rivera}}]{Kling:2020xjj}%
  \BibitemOpen
  \bibfield  {author} {\bibinfo {author} {\bibfnamefont {F.}~\bibnamefont
  {Kling}}, \bibinfo {author} {\bibfnamefont {A.}~\bibnamefont {Rajaraman}}, \
  and\ \bibinfo {author} {\bibfnamefont {F.~L.}\ \bibnamefont {Rivera}},\
  }\href {\doibase 10.1103/PhysRevD.103.075020} {\bibfield  {journal} {\bibinfo
   {journal} {Phys. Rev. D}\ }\textbf {\bibinfo {volume} {103}},\ \bibinfo
  {pages} {075020} (\bibinfo {year} {2021})},\ \Eprint
  {http://arxiv.org/abs/2010.09880} {arXiv:2010.09880 [hep-th]} \BibitemShut
  {NoStop}%
\bibitem [{\citenamefont {Kleihaus}\ \emph {et~al.}(2005)\citenamefont
  {Kleihaus}, \citenamefont {Kunz},\ and\ \citenamefont
  {List}}]{Kleihaus:2005me}%
  \BibitemOpen
  \bibfield  {author} {\bibinfo {author} {\bibfnamefont {B.}~\bibnamefont
  {Kleihaus}}, \bibinfo {author} {\bibfnamefont {J.}~\bibnamefont {Kunz}}, \
  and\ \bibinfo {author} {\bibfnamefont {M.}~\bibnamefont {List}},\ }\href
  {\doibase 10.1103/PhysRevD.72.064002} {\bibfield  {journal} {\bibinfo
  {journal} {Phys. Rev. D}\ }\textbf {\bibinfo {volume} {72}},\ \bibinfo
  {pages} {064002} (\bibinfo {year} {2005})},\ \Eprint
  {http://arxiv.org/abs/gr-qc/0505143} {arXiv:gr-qc/0505143} \BibitemShut
  {NoStop}%
\bibitem [{\citenamefont {Kleihaus}\ \emph {et~al.}(2008)\citenamefont
  {Kleihaus}, \citenamefont {Kunz}, \citenamefont {List},\ and\ \citenamefont
  {Schaffer}}]{Kleihaus:2007vk}%
  \BibitemOpen
  \bibfield  {author} {\bibinfo {author} {\bibfnamefont {B.}~\bibnamefont
  {Kleihaus}}, \bibinfo {author} {\bibfnamefont {J.}~\bibnamefont {Kunz}},
  \bibinfo {author} {\bibfnamefont {M.}~\bibnamefont {List}}, \ and\ \bibinfo
  {author} {\bibfnamefont {I.}~\bibnamefont {Schaffer}},\ }\href {\doibase
  10.1103/PhysRevD.77.064025} {\bibfield  {journal} {\bibinfo  {journal} {Phys.
  Rev. D}\ }\textbf {\bibinfo {volume} {77}},\ \bibinfo {pages} {064025}
  (\bibinfo {year} {2008})},\ \Eprint {http://arxiv.org/abs/0712.3742}
  {arXiv:0712.3742 [gr-qc]} \BibitemShut {NoStop}%
\bibitem [{\citenamefont {Sanchis-Gual}\ \emph
  {et~al.}(2019{\natexlab{b}})\citenamefont {Sanchis-Gual}, \citenamefont
  {Di~Giovanni}, \citenamefont {Zilhão}, \citenamefont {Herdeiro},
  \citenamefont {Cerdá-Durán}, \citenamefont {Font},\ and\ \citenamefont
  {Radu}}]{Sanchis-Gual:2019ljs}%
  \BibitemOpen
  \bibfield  {author} {\bibinfo {author} {\bibfnamefont {N.}~\bibnamefont
  {Sanchis-Gual}}, \bibinfo {author} {\bibfnamefont {F.}~\bibnamefont
  {Di~Giovanni}}, \bibinfo {author} {\bibfnamefont {M.}~\bibnamefont
  {Zilhão}}, \bibinfo {author} {\bibfnamefont {C.}~\bibnamefont {Herdeiro}},
  \bibinfo {author} {\bibfnamefont {P.}~\bibnamefont {Cerdá-Durán}}, \bibinfo
  {author} {\bibfnamefont {J.}~\bibnamefont {Font}}, \ and\ \bibinfo {author}
  {\bibfnamefont {E.}~\bibnamefont {Radu}},\ }\href {\doibase
  10.1103/PhysRevLett.123.221101} {\bibfield  {journal} {\bibinfo  {journal}
  {Phys. Rev. Lett.}\ }\textbf {\bibinfo {volume} {123}},\ \bibinfo {pages}
  {221101} (\bibinfo {year} {2019}{\natexlab{b}})},\ \Eprint
  {http://arxiv.org/abs/1907.12565} {arXiv:1907.12565 [gr-qc]} \BibitemShut
  {NoStop}%
\bibitem [{\citenamefont {Di~Giovanni}\ \emph {et~al.}(2020)\citenamefont
  {Di~Giovanni}, \citenamefont {Sanchis-Gual}, \citenamefont {Cerd\'a-Dur\'an},
  \citenamefont {Zilh\~ao}, \citenamefont {Herdeiro}, \citenamefont {Font},\
  and\ \citenamefont {Radu}}]{DiGiovanni:2020ror}%
  \BibitemOpen
  \bibfield  {author} {\bibinfo {author} {\bibfnamefont {F.}~\bibnamefont
  {Di~Giovanni}}, \bibinfo {author} {\bibfnamefont {N.}~\bibnamefont
  {Sanchis-Gual}}, \bibinfo {author} {\bibfnamefont {P.}~\bibnamefont
  {Cerd\'a-Dur\'an}}, \bibinfo {author} {\bibfnamefont {M.}~\bibnamefont
  {Zilh\~ao}}, \bibinfo {author} {\bibfnamefont {C.}~\bibnamefont {Herdeiro}},
  \bibinfo {author} {\bibfnamefont {J.}~\bibnamefont {Font}}, \ and\ \bibinfo
  {author} {\bibfnamefont {E.}~\bibnamefont {Radu}},\ }\href@noop {} {\
  (\bibinfo {year} {2020})},\ \Eprint {http://arxiv.org/abs/2010.05845}
  {arXiv:2010.05845 [gr-qc]} \BibitemShut {NoStop}%
\bibitem [{\citenamefont {Siemonsen}\ and\ \citenamefont
  {East}(2021)}]{Siemonsen:2020hcg}%
  \BibitemOpen
  \bibfield  {author} {\bibinfo {author} {\bibfnamefont {N.}~\bibnamefont
  {Siemonsen}}\ and\ \bibinfo {author} {\bibfnamefont {W.~E.}\ \bibnamefont
  {East}},\ }\href {\doibase 10.1103/PhysRevD.103.044022} {\bibfield  {journal}
  {\bibinfo  {journal} {Phys. Rev. D}\ }\textbf {\bibinfo {volume} {103}},\
  \bibinfo {pages} {044022} (\bibinfo {year} {2021})},\ \Eprint
  {http://arxiv.org/abs/2011.08247} {arXiv:2011.08247 [gr-qc]} \BibitemShut
  {NoStop}%
\bibitem [{\citenamefont {Dmitriev}\ \emph {et~al.}(2021)\citenamefont
  {Dmitriev}, \citenamefont {Levkov}, \citenamefont {Panin}, \citenamefont
  {Pushnaya},\ and\ \citenamefont {Tkachev}}]{Dmitriev:2021utv}%
  \BibitemOpen
  \bibfield  {author} {\bibinfo {author} {\bibfnamefont {A.~S.}\ \bibnamefont
  {Dmitriev}}, \bibinfo {author} {\bibfnamefont {D.~G.}\ \bibnamefont
  {Levkov}}, \bibinfo {author} {\bibfnamefont {A.~G.}\ \bibnamefont {Panin}},
  \bibinfo {author} {\bibfnamefont {E.~K.}\ \bibnamefont {Pushnaya}}, \ and\
  \bibinfo {author} {\bibfnamefont {I.~I.}\ \bibnamefont {Tkachev}},\ }\href
  {\doibase 10.1103/PhysRevD.104.023504} {\bibfield  {journal} {\bibinfo
  {journal} {Phys. Rev. D}\ }\textbf {\bibinfo {volume} {104}},\ \bibinfo
  {pages} {023504} (\bibinfo {year} {2021})},\ \Eprint
  {http://arxiv.org/abs/2104.00962} {arXiv:2104.00962 [gr-qc]} \BibitemShut
  {NoStop}%
\bibitem [{\citenamefont {Croft}\ \emph {et~al.}(2022)\citenamefont {Croft},
  \citenamefont {Helfer}, \citenamefont {Ge}, \citenamefont {Radia},
  \citenamefont {Evstafyeva}, \citenamefont {Lim}, \citenamefont {Sperhake},\
  and\ \citenamefont {Clough}}]{Croft:2022bxq}%
  \BibitemOpen
  \bibfield  {author} {\bibinfo {author} {\bibfnamefont {R.}~\bibnamefont
  {Croft}}, \bibinfo {author} {\bibfnamefont {T.}~\bibnamefont {Helfer}},
  \bibinfo {author} {\bibfnamefont {B.-X.}\ \bibnamefont {Ge}}, \bibinfo
  {author} {\bibfnamefont {M.}~\bibnamefont {Radia}}, \bibinfo {author}
  {\bibfnamefont {T.}~\bibnamefont {Evstafyeva}}, \bibinfo {author}
  {\bibfnamefont {E.~A.}\ \bibnamefont {Lim}}, \bibinfo {author} {\bibfnamefont
  {U.}~\bibnamefont {Sperhake}}, \ and\ \bibinfo {author} {\bibfnamefont
  {K.}~\bibnamefont {Clough}},\ }\href@noop {} {\  (\bibinfo {year} {2022})},\
  \Eprint {http://arxiv.org/abs/2207.05690} {arXiv:2207.05690 [gr-qc]}
  \BibitemShut {NoStop}%
\bibitem [{\citenamefont {Friedberg}\ \emph {et~al.}(1976)\citenamefont
  {Friedberg}, \citenamefont {Lee},\ and\ \citenamefont
  {Sirlin}}]{Friedberg:1976me}%
  \BibitemOpen
  \bibfield  {author} {\bibinfo {author} {\bibfnamefont {R.}~\bibnamefont
  {Friedberg}}, \bibinfo {author} {\bibfnamefont {T.~D.}\ \bibnamefont {Lee}},
  \ and\ \bibinfo {author} {\bibfnamefont {A.}~\bibnamefont {Sirlin}},\ }\href
  {\doibase 10.1103/PhysRevD.13.2739} {\bibfield  {journal} {\bibinfo
  {journal} {Phys. Rev. D}\ }\textbf {\bibinfo {volume} {13}},\ \bibinfo
  {pages} {2739} (\bibinfo {year} {1976})}\BibitemShut {NoStop}%
\bibitem [{\citenamefont {Coleman}(1985)}]{Coleman:1985ki}%
  \BibitemOpen
  \bibfield  {author} {\bibinfo {author} {\bibfnamefont {S.~R.}\ \bibnamefont
  {Coleman}},\ }\href {\doibase 10.1016/0550-3213(86)90520-1} {\bibfield
  {journal} {\bibinfo  {journal} {Nucl. Phys. B}\ }\textbf {\bibinfo {volume}
  {262}},\ \bibinfo {pages} {263} (\bibinfo {year} {1985})},\ \bibinfo {note}
  {[Addendum: Nucl.Phys.B 269, 744 (1986)]}\BibitemShut {NoStop}%
\bibitem [{\citenamefont {Tsubota}\ \emph {et~al.}(2002)\citenamefont
  {Tsubota}, \citenamefont {Kasamatsu},\ and\ \citenamefont
  {Ueda}}]{PRA.65.023603}%
  \BibitemOpen
  \bibfield  {author} {\bibinfo {author} {\bibfnamefont {M.}~\bibnamefont
  {Tsubota}}, \bibinfo {author} {\bibfnamefont {K.}~\bibnamefont {Kasamatsu}},
  \ and\ \bibinfo {author} {\bibfnamefont {M.}~\bibnamefont {Ueda}},\ }\href
  {\doibase 10.1103/PhysRevA.65.023603} {\bibfield  {journal} {\bibinfo
  {journal} {Phys. Rev. A}\ }\textbf {\bibinfo {volume} {65}},\ \bibinfo
  {pages} {023603} (\bibinfo {year} {2002})}\BibitemShut {NoStop}%
\bibitem [{\citenamefont {Koplik}\ and\ \citenamefont
  {Levine}(1993)}]{PRL.71.1375}%
  \BibitemOpen
  \bibfield  {author} {\bibinfo {author} {\bibfnamefont {J.}~\bibnamefont
  {Koplik}}\ and\ \bibinfo {author} {\bibfnamefont {H.}~\bibnamefont
  {Levine}},\ }\href {\doibase 10.1103/PhysRevLett.71.1375} {\bibfield
  {journal} {\bibinfo  {journal} {Phys. Rev. Lett.}\ }\textbf {\bibinfo
  {volume} {71}},\ \bibinfo {pages} {1375} (\bibinfo {year}
  {1993})}\BibitemShut {NoStop}%
\bibitem [{\citenamefont {Vinen}\ \emph {et~al.}(2003)\citenamefont {Vinen},
  \citenamefont {Tsubota},\ and\ \citenamefont {Mitani}}]{PRL.91.135301}%
  \BibitemOpen
  \bibfield  {author} {\bibinfo {author} {\bibfnamefont {W.~F.}\ \bibnamefont
  {Vinen}}, \bibinfo {author} {\bibfnamefont {M.}~\bibnamefont {Tsubota}}, \
  and\ \bibinfo {author} {\bibfnamefont {A.}~\bibnamefont {Mitani}},\ }\href
  {\doibase 10.1103/PhysRevLett.91.135301} {\bibfield  {journal} {\bibinfo
  {journal} {Phys. Rev. Lett.}\ }\textbf {\bibinfo {volume} {91}},\ \bibinfo
  {pages} {135301} (\bibinfo {year} {2003})}\BibitemShut {NoStop}%
\bibitem [{\citenamefont {Schwarz}(1988)}]{PRB.38.2398}%
  \BibitemOpen
  \bibfield  {author} {\bibinfo {author} {\bibfnamefont {K.~W.}\ \bibnamefont
  {Schwarz}},\ }\href {\doibase 10.1103/PhysRevB.38.2398} {\bibfield  {journal}
  {\bibinfo  {journal} {Phys. Rev. B}\ }\textbf {\bibinfo {volume} {38}},\
  \bibinfo {pages} {2398} (\bibinfo {year} {1988})}\BibitemShut {NoStop}%
\bibitem [{\citenamefont {Yu}\ and\ \citenamefont {Morgan}(2002)}]{Yu:2002sz}%
  \BibitemOpen
  \bibfield  {author} {\bibinfo {author} {\bibfnamefont {R.~P.}\ \bibnamefont
  {Yu}}\ and\ \bibinfo {author} {\bibfnamefont {M.~J.}\ \bibnamefont
  {Morgan}},\ }\href {\doibase 10.1088/0264-9381/19/17/101} {\bibfield
  {journal} {\bibinfo  {journal} {Class. Quant. Grav.}\ }\textbf {\bibinfo
  {volume} {19}},\ \bibinfo {pages} {L157} (\bibinfo {year}
  {2002})}\BibitemShut {NoStop}%
\bibitem [{\citenamefont {Sikivie}\ and\ \citenamefont
  {Yang}(2009)}]{Sikivie:2009qn}%
  \BibitemOpen
  \bibfield  {author} {\bibinfo {author} {\bibfnamefont {P.}~\bibnamefont
  {Sikivie}}\ and\ \bibinfo {author} {\bibfnamefont {Q.}~\bibnamefont {Yang}},\
  }\href {\doibase 10.1103/PhysRevLett.103.111301} {\bibfield  {journal}
  {\bibinfo  {journal} {Phys. Rev. Lett.}\ }\textbf {\bibinfo {volume} {103}},\
  \bibinfo {pages} {111301} (\bibinfo {year} {2009})},\ \Eprint
  {http://arxiv.org/abs/0901.1106} {arXiv:0901.1106 [hep-ph]} \BibitemShut
  {NoStop}%
\bibitem [{\citenamefont {Kain}\ and\ \citenamefont
  {Ling}(2010)}]{Kain:2010rb}%
  \BibitemOpen
  \bibfield  {author} {\bibinfo {author} {\bibfnamefont {B.}~\bibnamefont
  {Kain}}\ and\ \bibinfo {author} {\bibfnamefont {H.~Y.}\ \bibnamefont
  {Ling}},\ }\href {\doibase 10.1103/PhysRevD.82.064042} {\bibfield  {journal}
  {\bibinfo  {journal} {Phys. Rev. D}\ }\textbf {\bibinfo {volume} {82}},\
  \bibinfo {pages} {064042} (\bibinfo {year} {2010})},\ \Eprint
  {http://arxiv.org/abs/1004.4692} {arXiv:1004.4692 [hep-ph]} \BibitemShut
  {NoStop}%
\bibitem [{\citenamefont {Rindler-Daller}\ and\ \citenamefont
  {Shapiro}(2012)}]{Rindler-Daller:2011afd}%
  \BibitemOpen
  \bibfield  {author} {\bibinfo {author} {\bibfnamefont {T.}~\bibnamefont
  {Rindler-Daller}}\ and\ \bibinfo {author} {\bibfnamefont {P.~R.}\
  \bibnamefont {Shapiro}},\ }\href {\doibase 10.1111/j.1365-2966.2012.20588.x}
  {\bibfield  {journal} {\bibinfo  {journal} {Mon. Not. Roy. Astron. Soc.}\
  }\textbf {\bibinfo {volume} {422}},\ \bibinfo {pages} {135} (\bibinfo {year}
  {2012})},\ \Eprint {http://arxiv.org/abs/1106.1256} {arXiv:1106.1256
  [astro-ph.CO]} \BibitemShut {NoStop}%
\bibitem [{\citenamefont {Kibble}(1976)}]{Kibble:1976sj}%
  \BibitemOpen
  \bibfield  {author} {\bibinfo {author} {\bibfnamefont {T.~W.~B.}\
  \bibnamefont {Kibble}},\ }\href {\doibase 10.1088/0305-4470/9/8/029}
  {\bibfield  {journal} {\bibinfo  {journal} {J. Phys. A}\ }\textbf {\bibinfo
  {volume} {9}},\ \bibinfo {pages} {1387} (\bibinfo {year} {1976})}\BibitemShut
  {NoStop}%
\bibitem [{\citenamefont {Zurek}(1985)}]{Zurek:1985qw}%
  \BibitemOpen
  \bibfield  {author} {\bibinfo {author} {\bibfnamefont {W.~H.}\ \bibnamefont
  {Zurek}},\ }\href {\doibase 10.1038/317505a0} {\bibfield  {journal} {\bibinfo
   {journal} {Nature}\ }\textbf {\bibinfo {volume} {317}},\ \bibinfo {pages}
  {505} (\bibinfo {year} {1985})}\BibitemShut {NoStop}%
\bibitem [{\citenamefont {del Campo}\ and\ \citenamefont
  {Zurek}(2014)}]{delCampo:2013nla}%
  \BibitemOpen
  \bibfield  {author} {\bibinfo {author} {\bibfnamefont {A.}~\bibnamefont {del
  Campo}}\ and\ \bibinfo {author} {\bibfnamefont {W.~H.}\ \bibnamefont
  {Zurek}},\ }\href {\doibase 10.1142/S0217751X1430018X} {\bibfield  {journal}
  {\bibinfo  {journal} {Int. J. Mod. Phys. A}\ }\textbf {\bibinfo {volume}
  {29}},\ \bibinfo {pages} {1430018} (\bibinfo {year} {2014})},\ \Eprint
  {http://arxiv.org/abs/1310.1600} {arXiv:1310.1600 [cond-mat.stat-mech]}
  \BibitemShut {NoStop}%
\bibitem [{\citenamefont {Bo\v{s}kovi\'c}\ and\ \citenamefont
  {Barausse}(2022)}]{Boskovic:2021nfs}%
  \BibitemOpen
  \bibfield  {author} {\bibinfo {author} {\bibfnamefont {M.}~\bibnamefont
  {Bo\v{s}kovi\'c}}\ and\ \bibinfo {author} {\bibfnamefont {E.}~\bibnamefont
  {Barausse}},\ }\href {\doibase 10.1088/1475-7516/2022/02/032} {\bibfield
  {journal} {\bibinfo  {journal} {JCAP}\ }\textbf {\bibinfo {volume} {02}},\
  \bibinfo {pages} {032} (\bibinfo {year} {2022})},\ \Eprint
  {http://arxiv.org/abs/2111.03870} {arXiv:2111.03870 [gr-qc]} \BibitemShut
  {NoStop}%
\bibitem [{\citenamefont {Axenides}\ \emph {et~al.}(2000)\citenamefont
  {Axenides}, \citenamefont {Komineas}, \citenamefont {Perivolaropoulos},\ and\
  \citenamefont {Floratos}}]{Axenides:1999hs}%
  \BibitemOpen
  \bibfield  {author} {\bibinfo {author} {\bibfnamefont {M.}~\bibnamefont
  {Axenides}}, \bibinfo {author} {\bibfnamefont {S.}~\bibnamefont {Komineas}},
  \bibinfo {author} {\bibfnamefont {L.}~\bibnamefont {Perivolaropoulos}}, \
  and\ \bibinfo {author} {\bibfnamefont {M.}~\bibnamefont {Floratos}},\ }\href
  {\doibase 10.1103/PhysRevD.61.085006} {\bibfield  {journal} {\bibinfo
  {journal} {Phys. Rev. D}\ }\textbf {\bibinfo {volume} {61}},\ \bibinfo
  {pages} {085006} (\bibinfo {year} {2000})},\ \Eprint
  {http://arxiv.org/abs/hep-ph/9910388} {arXiv:hep-ph/9910388} \BibitemShut
  {NoStop}%
\bibitem [{\citenamefont {Battye}\ and\ \citenamefont
  {Sutcliffe}(2000)}]{Battye:2000qj}%
  \BibitemOpen
  \bibfield  {author} {\bibinfo {author} {\bibfnamefont {R.}~\bibnamefont
  {Battye}}\ and\ \bibinfo {author} {\bibfnamefont {P.}~\bibnamefont
  {Sutcliffe}},\ }\href {\doibase 10.1016/S0550-3213(00)00506-X} {\bibfield
  {journal} {\bibinfo  {journal} {Nucl. Phys. B}\ }\textbf {\bibinfo {volume}
  {590}},\ \bibinfo {pages} {329} (\bibinfo {year} {2000})},\ \Eprint
  {http://arxiv.org/abs/hep-th/0003252} {arXiv:hep-th/0003252} \BibitemShut
  {NoStop}%
\bibitem [{\citenamefont {Bowcock}\ \emph {et~al.}(2009)\citenamefont
  {Bowcock}, \citenamefont {Foster},\ and\ \citenamefont
  {Sutcliffe}}]{Bowcock:2008dn}%
  \BibitemOpen
  \bibfield  {author} {\bibinfo {author} {\bibfnamefont {P.}~\bibnamefont
  {Bowcock}}, \bibinfo {author} {\bibfnamefont {D.}~\bibnamefont {Foster}}, \
  and\ \bibinfo {author} {\bibfnamefont {P.}~\bibnamefont {Sutcliffe}},\ }\href
  {\doibase 10.1088/1751-8113/42/8/085403} {\bibfield  {journal} {\bibinfo
  {journal} {J. Phys. A}\ }\textbf {\bibinfo {volume} {42}},\ \bibinfo {pages}
  {085403} (\bibinfo {year} {2009})},\ \Eprint {http://arxiv.org/abs/0809.3895}
  {arXiv:0809.3895 [hep-th]} \BibitemShut {NoStop}%
\bibitem [{\citenamefont {Yoshida}\ and\ \citenamefont
  {Eriguchi}(1997)}]{Yoshida:1997nd}%
  \BibitemOpen
  \bibfield  {author} {\bibinfo {author} {\bibfnamefont {S.}~\bibnamefont
  {Yoshida}}\ and\ \bibinfo {author} {\bibfnamefont {Y.}~\bibnamefont
  {Eriguchi}},\ }\href {\doibase 10.1103/PhysRevD.55.1994} {\bibfield
  {journal} {\bibinfo  {journal} {Phys. Rev. D}\ }\textbf {\bibinfo {volume}
  {55}},\ \bibinfo {pages} {1994} (\bibinfo {year} {1997})}\BibitemShut
  {NoStop}%
\bibitem [{\citenamefont {Herdeiro}\ \emph {et~al.}(2021)\citenamefont
  {Herdeiro}, \citenamefont {Kunz}, \citenamefont {Perapechka}, \citenamefont
  {Radu},\ and\ \citenamefont {Shnir}}]{Herdeiro:2021mol}%
  \BibitemOpen
  \bibfield  {author} {\bibinfo {author} {\bibfnamefont {C.~A.~R.}\
  \bibnamefont {Herdeiro}}, \bibinfo {author} {\bibfnamefont {J.}~\bibnamefont
  {Kunz}}, \bibinfo {author} {\bibfnamefont {I.}~\bibnamefont {Perapechka}},
  \bibinfo {author} {\bibfnamefont {E.}~\bibnamefont {Radu}}, \ and\ \bibinfo
  {author} {\bibfnamefont {Y.}~\bibnamefont {Shnir}},\ }\href {\doibase
  10.1103/PhysRevD.103.065009} {\bibfield  {journal} {\bibinfo  {journal}
  {Phys. Rev. D}\ }\textbf {\bibinfo {volume} {103}},\ \bibinfo {pages}
  {065009} (\bibinfo {year} {2021})},\ \Eprint
  {http://arxiv.org/abs/2101.06442} {arXiv:2101.06442 [gr-qc]} \BibitemShut
  {NoStop}%
\bibitem [{\citenamefont {Cunha}\ \emph {et~al.}(2022)\citenamefont {Cunha},
  \citenamefont {Herdeiro}, \citenamefont {Radu},\ and\ \citenamefont
  {Shnir}}]{Cunha:2022tvk}%
  \BibitemOpen
  \bibfield  {author} {\bibinfo {author} {\bibfnamefont {P.}~\bibnamefont
  {Cunha}}, \bibinfo {author} {\bibfnamefont {C.}~\bibnamefont {Herdeiro}},
  \bibinfo {author} {\bibfnamefont {E.}~\bibnamefont {Radu}}, \ and\ \bibinfo
  {author} {\bibfnamefont {Y.}~\bibnamefont {Shnir}},\ }\href@noop {} {\
  (\bibinfo {year} {2022})},\ \Eprint {http://arxiv.org/abs/2210.01833}
  {arXiv:2210.01833 [gr-qc]} \BibitemShut {NoStop}%
\bibitem [{\citenamefont {Siemonsen}\ and\ \citenamefont
  {East}(prep)}]{inprep}%
  \BibitemOpen
  \bibfield  {author} {\bibinfo {author} {\bibfnamefont {N.}~\bibnamefont
  {Siemonsen}}\ and\ \bibinfo {author} {\bibfnamefont {W.~E.}\ \bibnamefont
  {East}},\ }\href@noop {} {\  (\bibinfo {year} {\textit{in
  prep}})}\BibitemShut {NoStop}%
\bibitem [{\citenamefont {Gourgoulhon}(2007)}]{Gourgoulhon:2007ue}%
  \BibitemOpen
  \bibfield  {author} {\bibinfo {author} {\bibfnamefont {E.}~\bibnamefont
  {Gourgoulhon}},\ }\href@noop {} {\  (\bibinfo {year} {2007})},\ \Eprint
  {http://arxiv.org/abs/gr-qc/0703035} {arXiv:gr-qc/0703035} \BibitemShut
  {NoStop}%
\bibitem [{\citenamefont {Clough}(2021)}]{Clough:2021qlv}%
  \BibitemOpen
  \bibfield  {author} {\bibinfo {author} {\bibfnamefont {K.}~\bibnamefont
  {Clough}},\ }\href {\doibase 10.1088/1361-6382/ac10ee} {\bibfield  {journal}
  {\bibinfo  {journal} {Class. Quant. Grav.}\ }\textbf {\bibinfo {volume}
  {38}},\ \bibinfo {pages} {167001} (\bibinfo {year} {2021})},\ \Eprint
  {http://arxiv.org/abs/2104.13420} {arXiv:2104.13420 [gr-qc]} \BibitemShut
  {NoStop}%
\bibitem [{\citenamefont {Croft}(2022)}]{Croft:2022gks}%
  \BibitemOpen
  \bibfield  {author} {\bibinfo {author} {\bibfnamefont {R.}~\bibnamefont
  {Croft}},\ }\href@noop {} {\  (\bibinfo {year} {2022})},\ \Eprint
  {http://arxiv.org/abs/2203.13845} {arXiv:2203.13845 [gr-qc]} \BibitemShut
  {NoStop}%
\bibitem [{\citenamefont {Evstafyeva}\ \emph {et~al.}(2022)\citenamefont
  {Evstafyeva}, \citenamefont {Sperhake}, \citenamefont {Helfer}, \citenamefont
  {Croft}, \citenamefont {Radia}, \citenamefont {Ge},\ and\ \citenamefont
  {Lim}}]{Evstafyeva:2022bpr}%
  \BibitemOpen
  \bibfield  {author} {\bibinfo {author} {\bibfnamefont {T.}~\bibnamefont
  {Evstafyeva}}, \bibinfo {author} {\bibfnamefont {U.}~\bibnamefont
  {Sperhake}}, \bibinfo {author} {\bibfnamefont {T.}~\bibnamefont {Helfer}},
  \bibinfo {author} {\bibfnamefont {R.}~\bibnamefont {Croft}}, \bibinfo
  {author} {\bibfnamefont {M.}~\bibnamefont {Radia}}, \bibinfo {author}
  {\bibfnamefont {B.-X.}\ \bibnamefont {Ge}}, \ and\ \bibinfo {author}
  {\bibfnamefont {E.~A.}\ \bibnamefont {Lim}},\ }\href@noop {} {\  (\bibinfo
  {year} {2022})},\ \Eprint {http://arxiv.org/abs/2212.08023} {arXiv:2212.08023
  [gr-qc]} \BibitemShut {NoStop}%
\bibitem [{\citenamefont {Baumgarte}\ \emph {et~al.}(2000)\citenamefont
  {Baumgarte}, \citenamefont {Shapiro},\ and\ \citenamefont
  {Shibata}}]{Baumgarte:1999cq}%
  \BibitemOpen
  \bibfield  {author} {\bibinfo {author} {\bibfnamefont {T.~W.}\ \bibnamefont
  {Baumgarte}}, \bibinfo {author} {\bibfnamefont {S.~L.}\ \bibnamefont
  {Shapiro}}, \ and\ \bibinfo {author} {\bibfnamefont {M.}~\bibnamefont
  {Shibata}},\ }\href {\doibase 10.1086/312425} {\bibfield  {journal} {\bibinfo
   {journal} {Astrophys. J. Lett.}\ }\textbf {\bibinfo {volume} {528}},\
  \bibinfo {pages} {L29} (\bibinfo {year} {2000})},\ \Eprint
  {http://arxiv.org/abs/astro-ph/9910565} {arXiv:astro-ph/9910565} \BibitemShut
  {NoStop}%
\bibitem [{\citenamefont {Bernuzzi}\ \emph {et~al.}(2014)\citenamefont
  {Bernuzzi}, \citenamefont {Dietrich}, \citenamefont {Tichy},\ and\
  \citenamefont {Br\"ugmann}}]{Bernuzzi:2013rza}%
  \BibitemOpen
  \bibfield  {author} {\bibinfo {author} {\bibfnamefont {S.}~\bibnamefont
  {Bernuzzi}}, \bibinfo {author} {\bibfnamefont {T.}~\bibnamefont {Dietrich}},
  \bibinfo {author} {\bibfnamefont {W.}~\bibnamefont {Tichy}}, \ and\ \bibinfo
  {author} {\bibfnamefont {B.}~\bibnamefont {Br\"ugmann}},\ }\href {\doibase
  10.1103/PhysRevD.89.104021} {\bibfield  {journal} {\bibinfo  {journal} {Phys.
  Rev. D}\ }\textbf {\bibinfo {volume} {89}},\ \bibinfo {pages} {104021}
  (\bibinfo {year} {2014})},\ \Eprint {http://arxiv.org/abs/1311.4443}
  {arXiv:1311.4443 [gr-qc]} \BibitemShut {NoStop}%
\bibitem [{\citenamefont {\c{C}okluk}\ \emph {et~al.}(2023)\citenamefont
  {\c{C}okluk}, \citenamefont {Yakut},\ and\ \citenamefont
  {Giacomazzo}}]{Cokluk:2023xio}%
  \BibitemOpen
  \bibfield  {author} {\bibinfo {author} {\bibfnamefont {K.~A.}\ \bibnamefont
  {\c{C}okluk}}, \bibinfo {author} {\bibfnamefont {K.}~\bibnamefont {Yakut}}, \
  and\ \bibinfo {author} {\bibfnamefont {B.}~\bibnamefont {Giacomazzo}},\
  }\href@noop {} {\  (\bibinfo {year} {2023})},\ \Eprint
  {http://arxiv.org/abs/2301.09635} {arXiv:2301.09635 [astro-ph.HE]}
  \BibitemShut {NoStop}%
\bibitem [{\citenamefont {Moschidis}(2018)}]{Moschidis:2016zjy}%
  \BibitemOpen
  \bibfield  {author} {\bibinfo {author} {\bibfnamefont {G.}~\bibnamefont
  {Moschidis}},\ }\href {\doibase 10.1007/s00220-017-3010-y} {\bibfield
  {journal} {\bibinfo  {journal} {Commun. Math. Phys.}\ }\textbf {\bibinfo
  {volume} {358}},\ \bibinfo {pages} {437} (\bibinfo {year} {2018})},\ \Eprint
  {http://arxiv.org/abs/1608.02035} {arXiv:1608.02035 [math.AP]} \BibitemShut
  {NoStop}%
\bibitem [{\citenamefont {Hook}\ and\ \citenamefont
  {Huang}(2018)}]{Hook:2017psm}%
  \BibitemOpen
  \bibfield  {author} {\bibinfo {author} {\bibfnamefont {A.}~\bibnamefont
  {Hook}}\ and\ \bibinfo {author} {\bibfnamefont {J.}~\bibnamefont {Huang}},\
  }\href {\doibase 10.1007/JHEP06(2018)036} {\bibfield  {journal} {\bibinfo
  {journal} {JHEP}\ }\textbf {\bibinfo {volume} {06}},\ \bibinfo {pages} {036}
  (\bibinfo {year} {2018})},\ \Eprint {http://arxiv.org/abs/1708.08464}
  {arXiv:1708.08464 [hep-ph]} \BibitemShut {NoStop}%
\bibitem [{\citenamefont {Huang}\ \emph {et~al.}(2019)\citenamefont {Huang},
  \citenamefont {Johnson}, \citenamefont {Sagunski}, \citenamefont
  {Sakellariadou},\ and\ \citenamefont {Zhang}}]{Huang:2018pbu}%
  \BibitemOpen
  \bibfield  {author} {\bibinfo {author} {\bibfnamefont {J.}~\bibnamefont
  {Huang}}, \bibinfo {author} {\bibfnamefont {M.~C.}\ \bibnamefont {Johnson}},
  \bibinfo {author} {\bibfnamefont {L.}~\bibnamefont {Sagunski}}, \bibinfo
  {author} {\bibfnamefont {M.}~\bibnamefont {Sakellariadou}}, \ and\ \bibinfo
  {author} {\bibfnamefont {J.}~\bibnamefont {Zhang}},\ }\href {\doibase
  10.1103/PhysRevD.99.063013} {\bibfield  {journal} {\bibinfo  {journal} {Phys.
  Rev. D}\ }\textbf {\bibinfo {volume} {99}},\ \bibinfo {pages} {063013}
  (\bibinfo {year} {2019})},\ \Eprint {http://arxiv.org/abs/1807.02133}
  {arXiv:1807.02133 [hep-ph]} \BibitemShut {NoStop}%
\bibitem [{\citenamefont {Zhang}\ \emph {et~al.}(2021)\citenamefont {Zhang},
  \citenamefont {Lyu}, \citenamefont {Huang}, \citenamefont {Johnson},
  \citenamefont {Sagunski}, \citenamefont {Sakellariadou},\ and\ \citenamefont
  {Yang}}]{Zhang:2021mks}%
  \BibitemOpen
  \bibfield  {author} {\bibinfo {author} {\bibfnamefont {J.}~\bibnamefont
  {Zhang}}, \bibinfo {author} {\bibfnamefont {Z.}~\bibnamefont {Lyu}}, \bibinfo
  {author} {\bibfnamefont {J.}~\bibnamefont {Huang}}, \bibinfo {author}
  {\bibfnamefont {M.~C.}\ \bibnamefont {Johnson}}, \bibinfo {author}
  {\bibfnamefont {L.}~\bibnamefont {Sagunski}}, \bibinfo {author}
  {\bibfnamefont {M.}~\bibnamefont {Sakellariadou}}, \ and\ \bibinfo {author}
  {\bibfnamefont {H.}~\bibnamefont {Yang}},\ }\href {\doibase
  10.1103/PhysRevLett.127.161101} {\bibfield  {journal} {\bibinfo  {journal}
  {Phys. Rev. Lett.}\ }\textbf {\bibinfo {volume} {127}},\ \bibinfo {pages}
  {161101} (\bibinfo {year} {2021})},\ \Eprint
  {http://arxiv.org/abs/2105.13963} {arXiv:2105.13963 [hep-ph]} \BibitemShut
  {NoStop}%
\bibitem [{\citenamefont {Thatcher}\ and\ \citenamefont
  {Morgan}(1997)}]{Thatcher_1997}%
  \BibitemOpen
  \bibfield  {author} {\bibinfo {author} {\bibfnamefont {M.~J.}\ \bibnamefont
  {Thatcher}}\ and\ \bibinfo {author} {\bibfnamefont {M.~J.}\ \bibnamefont
  {Morgan}},\ }\href {\doibase 10.1088/0264-9381/14/11/016} {\bibfield
  {journal} {\bibinfo  {journal} {Classical and Quantum Gravity}\ }\textbf
  {\bibinfo {volume} {14}},\ \bibinfo {pages} {3161} (\bibinfo {year}
  {1997})}\BibitemShut {NoStop}%
\bibitem [{\citenamefont {Herdeiro}\ \emph {et~al.}(2019)\citenamefont
  {Herdeiro}, \citenamefont {Perapechka}, \citenamefont {Radu},\ and\
  \citenamefont {Shnir}}]{Herdeiro:2019mbz}%
  \BibitemOpen
  \bibfield  {author} {\bibinfo {author} {\bibfnamefont {C.}~\bibnamefont
  {Herdeiro}}, \bibinfo {author} {\bibfnamefont {I.}~\bibnamefont
  {Perapechka}}, \bibinfo {author} {\bibfnamefont {E.}~\bibnamefont {Radu}}, \
  and\ \bibinfo {author} {\bibfnamefont {Y.}~\bibnamefont {Shnir}},\ }\href
  {\doibase 10.1016/j.physletb.2019.134845} {\bibfield  {journal} {\bibinfo
  {journal} {Phys. Lett. B}\ }\textbf {\bibinfo {volume} {797}},\ \bibinfo
  {pages} {134845} (\bibinfo {year} {2019})},\ \Eprint
  {http://arxiv.org/abs/1906.05386} {arXiv:1906.05386 [gr-qc]} \BibitemShut
  {NoStop}%
\bibitem [{\citenamefont {East}\ \emph
  {et~al.}(2012{\natexlab{a}})\citenamefont {East}, \citenamefont
  {Ramazanoglu},\ and\ \citenamefont {Pretorius}}]{East:2012zn}%
  \BibitemOpen
  \bibfield  {author} {\bibinfo {author} {\bibfnamefont {W.~E.}\ \bibnamefont
  {East}}, \bibinfo {author} {\bibfnamefont {F.~M.}\ \bibnamefont
  {Ramazanoglu}}, \ and\ \bibinfo {author} {\bibfnamefont {F.}~\bibnamefont
  {Pretorius}},\ }\href {\doibase 10.1103/PhysRevD.86.104053} {\bibfield
  {journal} {\bibinfo  {journal} {Phys. Rev. D}\ }\textbf {\bibinfo {volume}
  {86}},\ \bibinfo {pages} {104053} (\bibinfo {year} {2012}{\natexlab{a}})},\
  \Eprint {http://arxiv.org/abs/1208.3473} {arXiv:1208.3473 [gr-qc]}
  \BibitemShut {NoStop}%
\bibitem [{\citenamefont {York}(1999)}]{York:1998hy}%
  \BibitemOpen
  \bibfield  {author} {\bibinfo {author} {\bibfnamefont {J.~W.}\ \bibnamefont
  {York}, \bibfnamefont {Jr.}},\ }\href {\doibase 10.1103/PhysRevLett.82.1350}
  {\bibfield  {journal} {\bibinfo  {journal} {Phys. Rev. Lett.}\ }\textbf
  {\bibinfo {volume} {82}},\ \bibinfo {pages} {1350} (\bibinfo {year}
  {1999})},\ \Eprint {http://arxiv.org/abs/gr-qc/9810051} {arXiv:gr-qc/9810051}
  \BibitemShut {NoStop}%
\bibitem [{\citenamefont {Corman}\ and\ \citenamefont
  {East}(2022)}]{Corman:2022alv}%
  \BibitemOpen
  \bibfield  {author} {\bibinfo {author} {\bibfnamefont {M.}~\bibnamefont
  {Corman}}\ and\ \bibinfo {author} {\bibfnamefont {W.~E.}\ \bibnamefont
  {East}},\ }\href@noop {} {\  (\bibinfo {year} {2022})},\ \Eprint
  {http://arxiv.org/abs/2212.04479} {arXiv:2212.04479 [gr-qc]} \BibitemShut
  {NoStop}%
\bibitem [{\citenamefont {Pretorius}(2005)}]{Pretorius:2004jg}%
  \BibitemOpen
  \bibfield  {author} {\bibinfo {author} {\bibfnamefont {F.}~\bibnamefont
  {Pretorius}},\ }\href {\doibase 10.1088/0264-9381/22/2/014} {\bibfield
  {journal} {\bibinfo  {journal} {Class. Quant. Grav.}\ }\textbf {\bibinfo
  {volume} {22}},\ \bibinfo {pages} {425} (\bibinfo {year} {2005})},\ \Eprint
  {http://arxiv.org/abs/gr-qc/0407110} {arXiv:gr-qc/0407110 [gr-qc]}
  \BibitemShut {NoStop}%
\bibitem [{\citenamefont {East}\ \emph
  {et~al.}(2012{\natexlab{b}})\citenamefont {East}, \citenamefont {Pretorius},\
  and\ \citenamefont {Stephens}}]{East:2011aa}%
  \BibitemOpen
  \bibfield  {author} {\bibinfo {author} {\bibfnamefont {W.~E.}\ \bibnamefont
  {East}}, \bibinfo {author} {\bibfnamefont {F.}~\bibnamefont {Pretorius}}, \
  and\ \bibinfo {author} {\bibfnamefont {B.~C.}\ \bibnamefont {Stephens}},\
  }\href {\doibase 10.1103/PhysRevD.85.124010} {\bibfield  {journal} {\bibinfo
  {journal} {Phys. Rev. D}\ }\textbf {\bibinfo {volume} {85}},\ \bibinfo
  {pages} {124010} (\bibinfo {year} {2012}{\natexlab{b}})},\ \Eprint
  {http://arxiv.org/abs/1112.3094} {arXiv:1112.3094 [gr-qc]} \BibitemShut
  {NoStop}%
\bibitem [{\citenamefont {Lindblom}\ and\ \citenamefont
  {Szilagyi}(2009)}]{Lindblom:2009tu}%
  \BibitemOpen
  \bibfield  {author} {\bibinfo {author} {\bibfnamefont {L.}~\bibnamefont
  {Lindblom}}\ and\ \bibinfo {author} {\bibfnamefont {B.}~\bibnamefont
  {Szilagyi}},\ }\href {\doibase 10.1103/PhysRevD.80.084019} {\bibfield
  {journal} {\bibinfo  {journal} {Phys. Rev.}\ }\textbf {\bibinfo {volume}
  {D80}},\ \bibinfo {pages} {084019} (\bibinfo {year} {2009})},\ \Eprint
  {http://arxiv.org/abs/0904.4873} {arXiv:0904.4873 [gr-qc]} \BibitemShut
  {NoStop}%
\end{thebibliography}%

\end{document}